\documentclass[fleqn,usenatbib]{mnras}

\usepackage{newtxtext,newtxmath}
\usepackage[T1]{fontenc}
\usepackage{ae,aecompl}
\usepackage{amsmath}	% Advanced maths commands
\usepackage{amssymb}	% Extra maths symbols  
\usepackage{graphicx,changepage}
\usepackage{rotating}
\usepackage{url}
\usepackage{natbib}
\usepackage{booktabs}
\usepackage{lscape}
\usepackage{longtable}
\usepackage{titlesec}
\usepackage{chngcntr}
\usepackage{threeparttable}

\newcommand{\etal}{\mbox{et\ al.\ }}

\newcommand{\msun}{\,\mbox{$\mbox{M}_{\odot}$}}

\title[SPLOT]
{SPLOT: a Snapshot survey for Polarised Light in Optical Transients}

\author[A. B. Higgins \etal]
{A. B. Higgins,$^1$ \thanks{E-mail: abh13@le.ac.uk} K. Wiersema,$^{1,2}$ S. Covino,$^3$ R. L. C. Starling,$^1$ H. F. Stevance,$^4$ \and \L. Wyrzykowski,$^5$ S. T. Hodgkin,$^6$ J. R. Maund,$^4$ P. T. O'Brien,$^1$ \and and N. R. Tanvir$^1$\\
$^1$Department of Physics and Astronomy, University of Leicester, University Road, Leicester LE1 7RH, UK\\
$^2$Department of Physics, University of Warwick, Coventry CV4 7AL, UK\\
$^3$INAF/Brera Astronomical Observatory, via Bianchi 46, I-23807, Merate (LC), Italy\\
$^4$Department of Physics and Astronomy, University of Sheffield, Hounsfield Rd, Sheffield S3 7RH, UK\\
$^5$Warsaw University Astronomical Observatory, Al. Ujazdowskie 4, PL-00-478 Warszawa, Poland\\
$^6$Institute of Astronomy, University of Cambridge, Madingley Road,
Cambridge CB3 0HA, UK\\}
\begin{document}
\date{Accepted . Received ; in original form .}
%\bigskip

\pagerange{\pageref{firstpage}--\pageref{lastpage}} \pubyear{}

\maketitle

\label{firstpage}

%%***************************************************************************
%%  ABSTRACT
%%***************************************************************************

\begin{abstract}
We present SPLOT, a small scale pilot survey to test the potential of snapshot (single epoch) linear imaging polarimetry as a supplementary tool to traditional transient follow-up. Transients exist in a vast volume of observational parameter space and polarimetry has the potential to highlight sources of scientific interest and add value to near real-time transient survey streams. We observed a sample of $\sim 50$ randomly selected optical transients with the EFOSC2 and SofI instruments, on the 3.6m New Technology Telescope (NTT) to test the feasibility of the survey. Our sample contained a number of interesting individual sources: a variety of supernovae, X-ray binaries, a tidal disruption event, blazar outbursts, and, by design, numerous transients of unknown nature. We discuss the results, both for the individual sources and the survey in detail. We provide an overview on the success and limitations of SPLOT and also describe a novel calibration method for removing instrumental polarisation effects from Nasymth-mounted telescopes. We find that a SPLOT-like survey would be a benefit to the large scale future transient survey streams such as LSST. The polarimetric measurements have added scientific value to a significant number of the sources and, most importantly, has shown the potential to highlight unclassified transient sources of scientific interest for further study.
\end{abstract}

\begin{keywords}
polarization; supernovae: general; galaxies: active;
\end{keywords}

% Format titles and footnote numbers for main body of text
\setcounter{footnote}{0} 
\titleformat*{\section}{\Large\bfseries}
\titleformat*{\subsection}{\large\bfseries}

\titleformat{\subsubsection}
   {\bfseries\normalsize}{\thesubsubsection}{1em}{}

\titleformat{\paragraph}
{\normalfont\normalsize\itshape}{\theparagraph}{1em}{}
\titlespacing*{\paragraph}
{0pt}{2.5ex plus 1ex minus .2ex}{1.25ex plus .2ex}

\renewcommand{\thefootnote}{\textsuperscript{\arabic{footnote}}}

%%%%% Intro
\section{Introduction} \label{sec:intro}
The discovery space of transients now spans an unprecedented range of wavelengths and timescales, continuously pushed by new campaigns and software and, as a result, the rate of transient candidate discovery has increased dramatically in recent years. There are a number of current facilities whose aim it is to detect a variety of transient phenomena including the Mobile Astronomical System of Telescope-Robots (MASTER; \citealt{Lipunov2004}), the All Sky Automated Survey for SuperNovae (ASAS-SN; \citealt{Shappee2014}), the {\it Gaia} satellite \citep{Gaia2016}, the Panoramic Survey Telescope and Rapid Response System (\citealt{Chambers2016i}; Pan-STARRS) and Optical Gravitational Lensing Experiment (OGLE) IV Transient Detection System \citep{Wyrzykowski2014} to name a few. The current number of detections from optical transient surveys lies at $\sim 1-10$ transients per night. At optical wavelengths, large additional increases in discovery rates are expected from the arrival of new surveys such as the Large Synoptic Survey Telescope (LSST; \citealt{Ivezic2008}) and the Gravitational-Wave Optical Transient Observer (GOTO)\footnote{https://goto-observatory.org/}. Moreover, the Zwicky Transient Facility (ZTF; \citealt{Kulkarni2016}) has also recently become operational and has been distributing alerts to the transient community since 2018 June. In many cases, the discovery data and subsequent photometry provided by these surveys alone does not provide enough information to accurately filter the targets of highest astrophysical interest from the streams and follow-up data are required. Traditionally, the key follow-up resource is spectroscopy, but spectroscopic observations are usually time expensive and cannot feasibly be used on large volumes of transients.

An important primary step is therefore the ability to filter and choose interesting transient sources in near real-time directly from incoming data streams.
The classification of new transient sources via follow-up spectroscopic observations is well studied by large programmes (i.e. PESSTO, \citealt{Smartt2015}). However, there is a large number of potentially interesting, transient-enabled astrophysics that does not map cleanly onto selection functions based on multi-wavelength flux(ratios), astrometric position, morphology or low resolution spectroscopic features - particularly with selection functions that are available early after alert. Linear polarimetry may go some way towards providing an additional observational parameter axis for large numbers of transients, with the potential to flag up astrophysics of interest.  

Many high energy astrophysical phenomena have complex internal geometry. Intrinsic linear polarisation of the order of several percent can help decipher the complex geometry and magnetic field configuration of regions with optical emission. Optical linear polarisation can arise from a number of mechanisms. The presence of non-thermal emission in the form of synchrotron emission, produced by relativistic electrons gyrating around magnetic field lines and thought to arise in a host of transient phenomena, exhibits a significant level of polarisation. This emission mechanism is thought to dominate the low energy (optical to radio) photon production in AGN/Blazars \citep{Trippe2014}, the emission from X-ray to radio wavelengths in GRB afterglows \citep{Wiersema2012b,Wiersema2014,Covino2016} and X-ray binary (XRB) jets \citep{Russell2008} to name a few. For core-collapse Supernovae (SNe) a non-zero measurement of polarisation arises from an asymmetric explosion ejecta \citep{Shapiro1982,Wang1997,Wang2008} and additionally can arise from inhomogeneous ejecta in novae outbursts \citep{Evans2002}. Type Ia SNe observations have shown that the intrinsic continuum light shows no significant levels of polarisation \citep{Wang1996,Wang1997,Wang2008}. However, multiple detections of significant polarisation have been detected in broad band optical filters (i.e. SN2014J; \citealt{Kawabata2014}). Polarisation of this nature has been attributed to line-of-sight dust, potentially in the SN host, giving us a window into the immediate environment of the source.

We have undertaken a pilot study measuring the optical linear polarisation of a variety of high energy transients and variables through single epoch polarimetry. 
We highlight the aims, use and justification of undertaking a polarimetric survey and introduce our observations in section \ref{sec:obs}. We discuss our polarimetric data analysis, calibration efforts and measurements in section \ref{sec:polmethod} and our photometry in section \ref{sec:photometry}. Sections \ref{sec:sourceresults} and \ref{sec:surveyresults} showcase the results of SPLOT - both the individual sources and as a sample. We discuss 
the impact of the survey and the shape of future polarimetric surveys in section \ref{sec:conclusion}. 

%%%% Observations
\section{Observations} \label{sec:obs}
\begin{figure*}
\includegraphics[width=\linewidth]{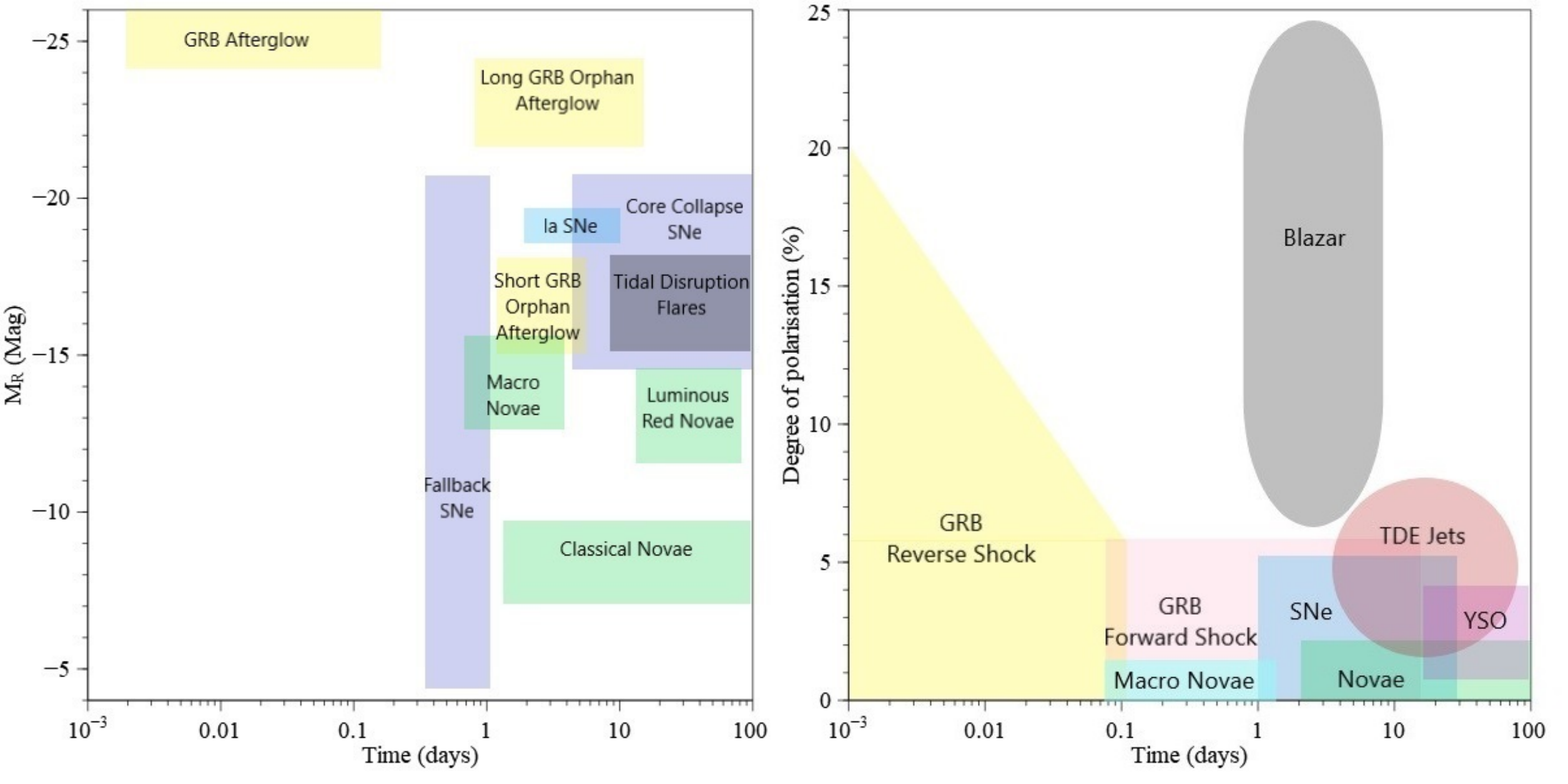}
\caption{Left: Absolute magnitude and characteristic timescales for a range of optical transients, demonstrating the large area of discovery space of optically selected transient searches in the lightcurve domain - similar layout to figure 8.1 in \citealt{LSST2009}. Right: Discovery space of SPLOT, in the optical polarimetric domain. The x-axis represents a characteristic timescale. Indicated are approximate regions where some polarimetric detections exist - not necessarily intrinsic polarisation. Current statistics are very poor for some of the discovery space: several of the source classes have just one or two polarimetric measurements.} 
\label{fig:discoveryspace}
\end{figure*}

\subsection{Survey rationale} \label{sec:survey}
The main aim of this survey was to investigate the opportunities and practicalities of using a snapshot linear polarisation measurement survey as a tool to add value to streams of optical transients. In particular we aimed to explore whether polarisation alone could allow us to highlight transient sources of high scientific interest, independent of the traditional classification tools of light curves and spectra. As discussed in section \ref{sec:intro}, astrophysical transients can exhibit significant and varying levels of intrinsic linear polarisation based on their internal structure and and equally wide range due to dust in the environment of the source. These transient events cover a large range of both absolute magnitudes and physical time scales (see Figure \ref{fig:discoveryspace}) resulting in a large polarimetric parameter space. If you include additional observational parameters such as multi-wavelength follow-up, colours and potential host information, transients cover a vast multi-dimensional space.
Value can be added onto survey transient streams by mapping out where sources fit into this multi-dimensional parameter set and hence highlight any sources of scientific interest.
Spectral classification, while crucially important to many aspects of transient science, may not highlight all sources of interest and we therefore want to test linear optical polarimetry as an independent aid of large scale transient streams. 

Linear optical polarimetry has been a fairly standard tool in the follow-up of some transients, in particular SN \citep{Wang2008} where optical spectropolarimetry has provided constraints on SNe geometry (i.e. \citealt{Maund2009,Reilly2017,Stevance2017}). SN rates are high enough that such a pre-selection can be made well, and a reasonable number of sources are available for spectropolarimetry.
For many other transient classes, only a very small number of sources have follow-up polarimetry (i.e Macronova; \citealt{Covino2017}). These uncommon transients typically have a low rate of detection and may be considerably fainter.
As we also aimed to observe a relatively large sample of sources we therefore opted for broadband imaging polarimetry which requires substantially shorter exposure times than spectropolarimetry.

To investigate the feasibility of our survey we required a relatively large sample of sources to:
\begin{itemize}
{\item[(a)] Sample both the contents of transient survey streams and a broad area of the discussed parameter space.}
{\item[(b)] Cover the effects of Galactic dust induced polarisation.}
{\item[(c)] Investigate the effect of practical constraints such as weather, instrument calibration and ease of access to transient alerts by surveys.}
{\item[(d)] Obtain results to sufficient precision that to enable scientific conclusions on individual sources ($\sigma_{P}\sim0.2\%$).}
\end{itemize}
To achieve this we chose a snapshot approach where the majority of sources are observed just once, in a single broadband filter. Detailed studies of some source classes in the literature can then be used to place selected single sources into context. A small subset of sources is observed more than once, generally as a test of calibration fidelity and occasionally to assess polarimetric variability over short time scales or multi-wavelength behaviour. 

\subsection{Source selection, exposure times and impact of conditions} \label{sec:sample}

\subsubsection{Telescope, instrumental set-up and filter choice} \label{sec:setup}
The rate of transients is currently sufficiently high that it is feasible to use $`$visitor-mode' observing to perform a survey like SPLOT, as demonstrated by the success of the ePESSTO\footnote{http://www.pessto.org} (\citealt{Smartt2015}) supernova survey.

% Choice of telescope and instrument and filter
We required the use of a medium sized telescope ($\sim4$m) with an execution time of 1 hour or less per target to fulfil the following criteria:
\begin{itemize}
{\item Cover a magnitude range down to $\sim20$\,mag in $V$ band - where more uncommon (extragalactic) transients typically appear (see Figure \ref{fig:discoveryspace}; \citealt{Rau2009}).}
{\item Aim for polarimetric uncertainties of $\sim0.2\%$ with $\sim0.5\%$ for the faintest sources. In reality the dominant source of uncertainty will be weather conditions and instrumental effects.}
{\item Observe a fairly large ($\sim50$) sample of transients.}
\end{itemize}
For the survey we used the ESO 3.6m New Technology Telescope (NTT) at La Silla, Chile, primarily with the ESO Faint Object Spectrograph and Camera v2 (EFOSC2; \citealt{Buzzoni1984}). This instrument is widely used for transient observations (i.e. ePESSTO), has the ability to switch rapidly from imaging to imaging polarimetry, and offers a field of view well suited for transient follow-up (see \citealt{ESO2016a} for full details). In addition to EFOSC2, some observations were obtained using the infrared instrument SofI (Son of ISAAC; \citealt{Moorwood1998}) which is also capable of performing polarimetric observations (see \citealt{ESO2016b} for full details). We note that both instruments exhibit large amounts of instrumental polarisation since they are both located at the Nasmyth focus (see section \ref{sec:polcalib} for full discussion). 

For the primary snapshot survey, we choose to use the \textit{V} band filter for EFOSC2 observations. It overlaps well with the {\it Gaia} pass band and that of ASAS-SN and MASTER, making it easier to extrapolate discovery magnitudes to the time of observation. Furthermore, the efficiency of EFOSC2 peaks near the \textit{V} band, and we avoid systematics from fringing by not choosing redder filters. 
The observed polarisation we measure is a combination of three contributors: the intrinsic polarisation of the target source, the polarisation induced by in-situ dust scattering and the induced polarisation from dust within the Milky Way. The \textit{V} band is close to the wavelength at which the dust induced polarisation peaks in the Milky Way (e.g. \citealt{Serkowski1975}). As such, we may not be able to separate the Galactic dust component from the intrinsic one using just a single snapshot in one filter. However, this allows us to use dust as an additional parameter of interest. For the SofI observations we used the \textit{Z} filter, to stay as close as possible to the optical bands used by the transient feed surveys.

\subsubsection{Chosen targets} \label{sec:targets}
We selected the SPLOT targets from a number of transient surveys that release rapid public notifications, generally through
the Transient Name Server\footnote{https://wis-tns.weizmann.ac.il}, on survey specific web-based lists\footnote{e.g. http://gsaweb.ast.cam.ac.uk/alerts/home} and/or via announcements in Astronomers  Telegrams\footnote{http://www.astronomerstelegram.org} and VOEvents\footnote{We use the Comet broker (\citealp{Swinbank2014}), https://github.com/jdswinbank/Comet}. The main contributors to the source list were the {\it Gaia} transient alert system, PanSTARRS, ASAS-SN, ATLAS, MASTER, CRTS, OGLE and some other, smaller, streams. We deliberately did not require prior spectroscopic classification for an object to enter our list of possible targets. The main requirement for a transient to become a target was its visibility ($\gtrsim0.5$ hours at airmass $< 2$) from La Silla observatory in our observing nights. Targets for which an alert was received within six months were entered into our target list granting us coverage of our target discovery space. Many targets received further observations since discovery and sources that had faded below magnitude $\sim21$ were culled from the target list.

During observing nights the transient feed surveys were checked continuously for new transients - we note that the {\it Gaia} transient alert system was off during our first run in 2016 resulting in the more frequent use of older transients - with earlier discovery dates. The full target list was ingested into {\it iObserve}\footnote{onekilopars.ec}, from which altitude, Moon distance and parallactic angle (PA) were obtained. We then used these observables to create an observing plan. Instrumental polarisation of these instruments is strongly dependent on PA so observations were planned near times when PA changes were small over the observation execution time. ESO Observing Blocks (OBs) were created as new alerts came in during the observing nights and a set of reserve targets (older transients) were prepared a night in advance in case of a lack of new transients or highly adverse weather. Exposure times were changed in cases where acquisition images showed a flux strongly different from expectations.

Additional criteria had to be introduced at periods of poor weather. Poor seeing and cloud coverage made observing the faintest sources very challenging.  Several nights suffered from strong winds from the North, which meant that only objects towards the South (typically with declination $\lesssim -30\deg$) could be observed. This directly impacted survey source selection with some surveys (i.e. PanSTARRS and ATLAS) unable to provide transients at low declinations. Additionally, photometric follow-up of transients is more sparse at low declinations making it harder to estimate exposure times for SPLOT observations. 

\begin{figure*}
\includegraphics[width=\linewidth]{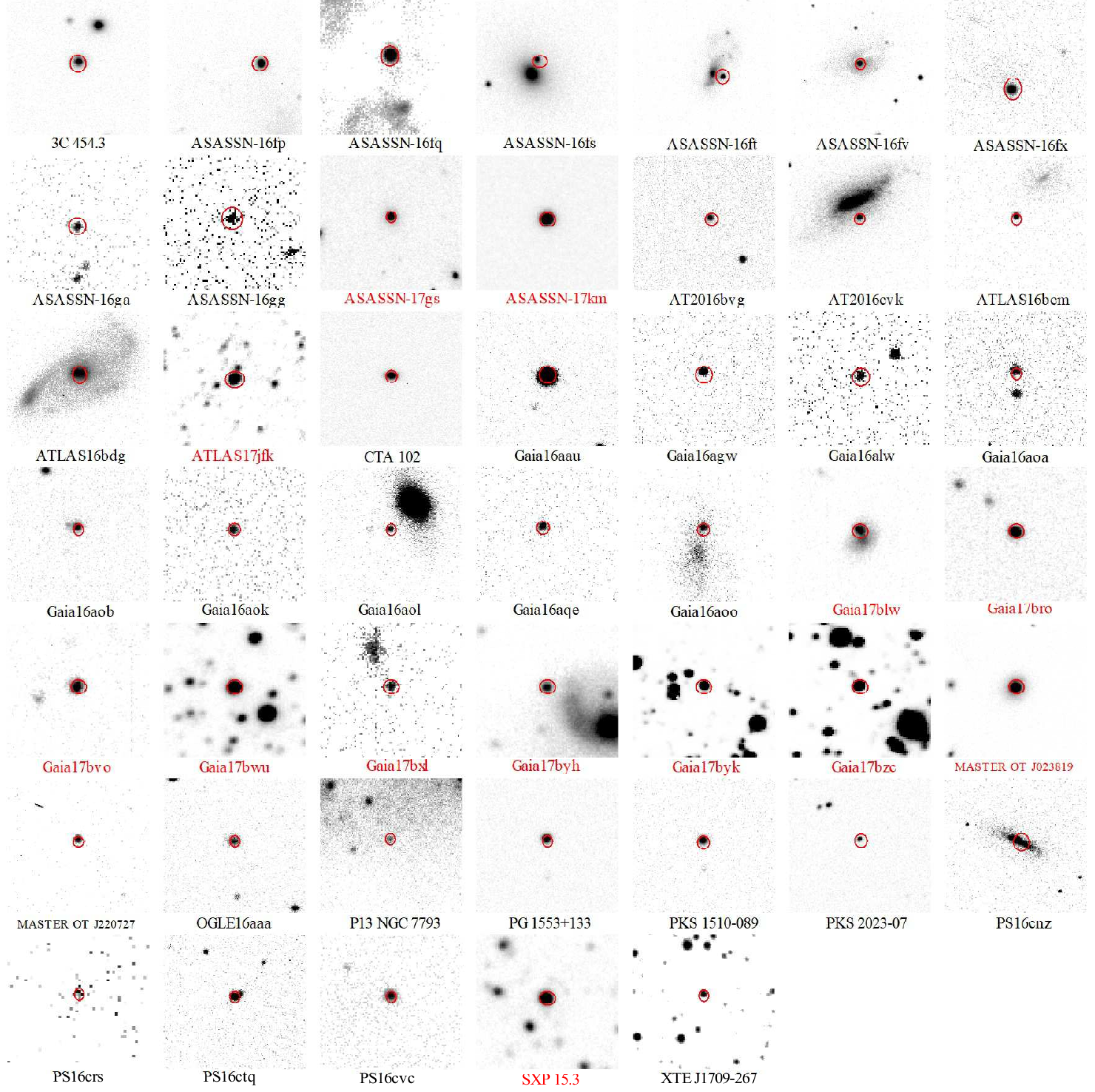}
\caption{Images of the sources that we observed as part of SPLOT. The image for GX 304-1 was saturated and therefore not included. The images are purely used as a reference for the position of each source within their hosts and/or neighbouring field stars, and are not to scale. Images were taken in \textit{V} band (EFOSC2; black text) or \textit{Z} band (SofI; red text).}
\label{fig:splotpic}
\end{figure*}

Overall, we observed 47 optical transients and an additional 8 standard stars - 3 polarised and 5 unpolarised. Images of the transients are shown in Figure \ref{fig:splotpic}.

\subsection{La Silla data acquisition} \label{sec:lasilla}
The majority of our targets were observed with EFOSC2, primarily in \textit{V} band (ESO filter \#641). We obtained our data during three observation nights (2016 June 19, 20 and 22), under poor seeing and variable thin to thick cloud conditions. Two further allocated nights (June 23, 24) were fully lost to thick cloud and high humidity. The Moon was near full throughout. Sources were additionally observed with the \textit{B} and \textit{R} filters in selected cases (ESO filters \#639 and \#642, respectively). A second block of NTT EFOSC2 observing time was awarded for SPLOT. However, the rotator encoder of the Nasmyth platform on which EFOSC2 was mounted failed, and could not be repaired on time. We therefore used SofI instead, which is mounted on the opposite Nasmyth platform, and chose to use the \textit{Z} filter. Science observations were obtained on the first two nights (2017 August 7, 8), under variable cloud and poor seeing. The third night (August 9) was lost to cloud and humidity. 

\subsubsection{EFOSC2 data} \label{sec:efosc2obs}
\begin{figure*}
\includegraphics[width=\linewidth]{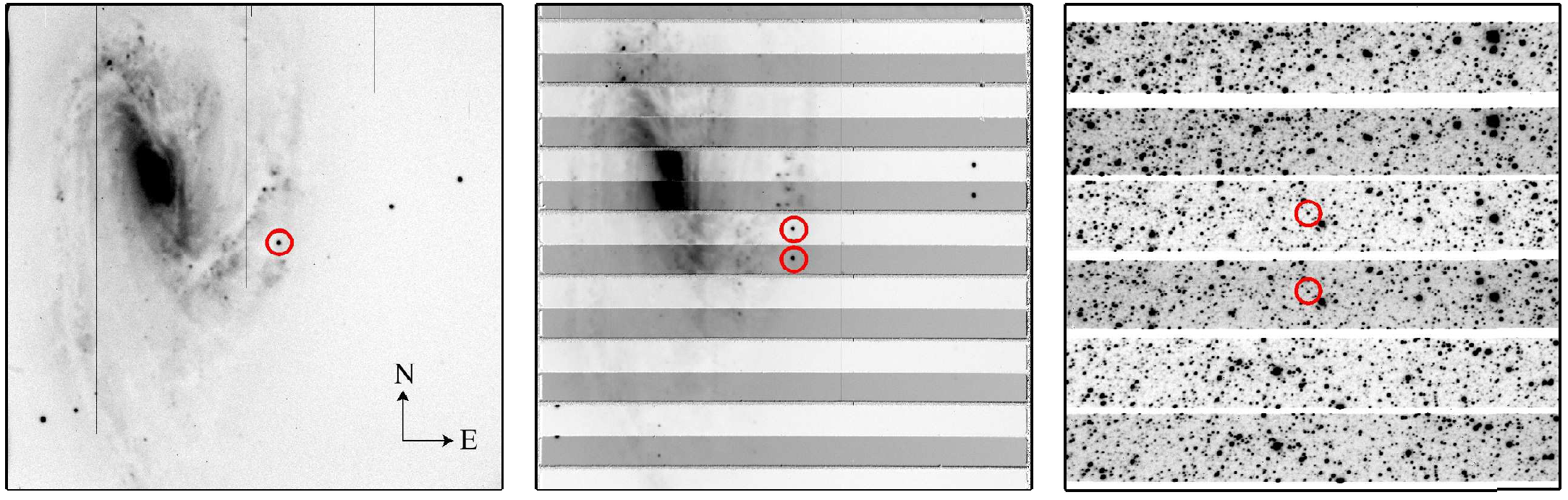}
\caption{The three panels show from left to right: a \textit{V} band EFOSC2 image of ASASSN-16fq (an example of a transient on a bright galaxy background), the same source in a single exposure of the EFOSC2 \textit{V} band polarimetric sequence and a single SofI \textit{Z} band exposure from a polarimetric sequence of Gaia17bzc (an example of a transient in a crowded field). The transient is indicated with a red circle and the orientation of EFOSC2 is indicated with a compass. The SofI orientation is different from EFOSC2 (East down, North left at zero instrument rotation), and the field is rotated within the polarimetric sequence (see Section \ref{sec:sofiobs}). 
The strips of the mask are clearly visible in the polarimetry images. }
\label{fig:panels}
\end{figure*}

Our EFOSC2 polarimetric observations used a Wollaston prism ($``$Woll\_Prism20") and a half-wave plate. The prism was used to split the incoming light into two beams, the orthogonally polarised ordinary (\textit{o}) and extraordinary (\textit{e}) beams. A mask was used to ensure the images of the two beams do not overlap (Figure \ref{fig:panels}). We used four different half-wave plate angles for our observations; $0$\,deg, $22.5$\,deg, $45$\,deg and $67.5$\,deg. The use of four angles instead of two angles allows us to obtain superior accuracy for our polarimetric measurements through beam-switching \citep{Patat2006}, as discussed in section \ref{sec:poldata}. We obtained dome screen flat field images with the polarimetric elements in place, with the half-wave plate rotating continuously, to form a flat field where polarisation response is scrambled. Bias frames were also taken, at the start of each night. The CCD readout was in `normal' mode and used $2\times 2$ binning, resulting in an image scale of 0.24 arcsecond per pixel. The gain and read noise of the CCD in this mode were 1.18 electrons per ADU and 11 electrons respectively, calculated using the method described in \cite{Janesick2001}. As EFOSC2 is mounted on a Nasmyth platform, light reflects off of a mirror set at a 45 degree angle with respect to EFOSC2 (the tertiary mirror). This leads to significant levels of instrumental polarisation \citep{Giro2003} discussed further in section \ref{sec:poldata}. To minimise part of this instrumental polarisation we always placed the transients and standard stars at (nearly) the same pixel position as part of the acquisition process (Figure \ref{fig:panels}).

Once exposures at all 4 half-wave plate positions had been completed, the images were reduced using standard {\sc IRAF}\footnote{{\sc IRAF} (Image Reduction and Analysis Facility) is distributed by the National Optical Astronomy Observatories, which are operated by the Association of Universities for Research in Astronomy, Inc., under cooperative agreement with the National Science Foundation.} tasks, and the {\sc IRAF} task {\sc appola}\footnote{Developed by E. Rol} was used to measure fluxes in the \textit{o} and \textit{e} images in each frame, using aperture photometry. A circular shaped extraction region, typically 1.5 times the full width half-maximum (FWHM) of the point spread function (PSF), was used to obtain the fluxes of the sources ($f_o$ and $f_e$). An annulus shaped sky region, with inner radius typically 3 times the FWHM, was used for the surrounding background region. Sky annuli were occasionally tweaked to avoid nearby field stars. This procedure was carried out for all point sources in the images. We ensured aperture size and annuli sizes were fixed for each of the four half-wave plate angles making up one observation. For a more detailed description, see \cite{Rol2003}. Output files were created for each half-wave plate angle observation and we created a pipeline written in Python3.5\footnote{http://www.python.org} to parse output files, calibrate the instrumental polarisation and calculate the polarisation of all sources (\citealt{Wiersema2018}; method and calibration of results discussed in section \ref{sec:polmethod}). 

Because of the aperture mask, which blocks half the field in strips (Figure \ref{fig:panels}), there are frequently not many field stars present in the polarimetry data. To perform photometry on each source, we therefore acquired a short exposure image in \textit{V} band, directly after the 4 half-wave plate rotations. We used twilight flats and bias frames to reduce these data, using standard {\sc IRAF} tasks. A further pipeline was produced to calculate the brightness of each source (discussed in section \ref{sec:photometry}.

\subsubsection{SofI data} \label{sec:sofiobs}
SofI also uses a Wollaston prism to split the light into two orthogonally polarised beams, and a mask to avoid image overlap (Figure \ref{fig:panels}). In contrast to EFOSC2, SofI has no wave plates, which means the Wollaston prism needs to be rotated with respect to the detector to acquire the Stokes parameters. We rotated the instrument through four angles: $0$\,deg, $45$\,deg, $90$\,deg and $135$\,deg - equivalent to rotating a wave plate through the four angles described in section \ref{sec:efosc2obs}. 

As SofI is an infrared instrument, its observing setup is slightly different from EFOSC2. Each observation at one angle consists of an exposure time NEXP x NDIT x DIT, where DIT is the detector integration time in seconds, NDIT the number of DIT integrations that is averaged to make a single output file, and NEXP the number of separate NDIT x DIT files.
Small dithers are applied between each (NDIT x DIT) set, typically a few arcseconds. We always used an NEXP of 5 for the polarimetric exposures. We chose the dithers and the pixel coordinate on which the source was placed (in the acquisition template) such that the transient always stayed in the central area of the mask.
The chosen NEXP x NDIT x DIT for each source is listed in Table \ref{tab:polresults}. We note that at the time of observations, no exposure time calculator for SofI \textit{Z} band imaging existed, so exposure times were estimated on the fly, using acquisition data. SofI uses a Hawaii HgCdTe array, with pixel scale of 0.288 arcseconds per pixel in its widefield mode.
To reduce the data, we acquired dark frames at the start of the nights, with a variety of DIT x NDIT to match the science and standard star observations. Flat field exposures were obtained using the `Special Flat' dome flat algorithm 
described in the Sofi manual \citep{ESO2016b}, note that these were obtained without polarisation optics (Wollaston) in the beam: unlike EFOSC2 there is no wave plate to spin continuously while taking these dome flats.
The flat fields were processed using {\sc IRAF} task {\sc flat\_special} \footnote{ https://www.eso.org/sci/facilities/lasilla/instruments/\\sofi/tools/sofi\_scripts.html}. The science and standard star exposures were reduced using {\sc IRAF} tasks, and the NEXP images (of one rotation angle for one source) were registered on a common pixel grid and combined using an average. The sky background in \textit{Z} is far lower than that in the \textit{J,H,K} bands so we do not perform sky subtraction steps used in most IR reduction. We do not perform corrections for interquadrant row cross talk in the polarimetry data: in none of the data does this effect play a role near the transient location. During SofI observations there were intermittent problems with the detector electronics, making some quadrants in some NDIT x DIT exposures highly noisy and stripy. Manual intervention in the instrument ensured only a few frames were affected, these were eliminated from the averages. The resulting average frames at four angles were analysed in the same way as the EFOSC2 data, using aperture photometry.

As with EFOSC2, we obtained SofI imaging data of the targets. These consist of fewer NDIT and NEXP (generally three successive images were taken with NDIT=NEXP=1) and therefore still show some noise residuals (e.g. from the amplifier) after reduction using dark and flat field frames. Row-by-row sky subtraction satisfactorily removed most of these. Images were further analysed in the same way as the EFOSC2 images. 

\subsection{Oadby data acquisition} \label{sec:oadby}
To get a somewhat longer timescale view of the lightcurve properties of a small number of the SPLOT transients, we observed a small subset using the University of Leicester 0.5m telescope (UL50)\footnote{Located in Oadby, Leicester, U.K.}. 
The telescope is a Planewave CDK20 \footnote{planewave.com}, a 0.5m telescope of corrected Dall-Kirkham design.  We used SBIG ST2000XM and Moravian G3-11000 CCD cameras, equipped with a broadband Johnson-Cousins \textit{B,V,R,I} filter set. Bias, dark and twilight (or dome) flat frames were obtained each observing night, data were reduced using standard recipes through a dedicated {\sc IRAF} pipeline for UL50 data.

%%%% Measuring source polarisation
\section{Measuring source polarisation} \label{sec:polmethod}

\subsection{Data analysis} \label{sec:poldata}
We represent our polarisation results as a Stokes vector, taking the form $[S]$ = $[I,Q,U,V]$ where $I$ represents the intensity, $Q$ and $U$ express the linear polarisation and \textit{V} represents the circular polarisation (described in \citealt{Chandrasekhar1960} and references therein). We do not measure circular polarisation for this investigation. When calculating our polarimetric results, we use the normalized Stokes parameters $q = Q/I$ and $u = U/I$. As mentioned in section \ref{sec:lasilla} using four rotation angles allows us to obtain smaller uncertainties on our measurements by cancelling out systematics effects caused by background subtraction and flat fielding \citep{Patat2006}.

To calculate the observed values of $q$ and $u$ we first find the normalized flux difference, $F_i$, between the ordinary and extraordinary beams for each angle, $\theta_i$, of the half-wave plate. From \citealt{Patat2006} we then use the following expressions
\begin{align}
\begin{split} \label{eq:normflux}
F_{i} = \frac{(f_{o,i} - f_{e,i})}{(f_{o,i} + f_{e,i})}
\end{split} \\
\begin{split} \label{eq:q}
q = \frac{2}{N} \sum\limits_{i=0}^{N-1}F_{i}{\rm cos}\left(\frac{i\pi}{2}\right)
\end{split} \\
\begin{split} \label{eq:u}
u = \frac{2}{N} \sum\limits_{i=0}^{N-1}F_{i}{\rm sin}\left(\frac{i\pi}{2}\right)
\end{split}
\end{align}
where $N$ is the number of rotation angles of the half wave-plate. Note that for EFOSC2, we used the (arbitrary) convention that the upper image strip is the $o$ beam and the lower the $e$ beam. For SofI, mounted at the opposite Nasmyth port, we use the opposite convention. We then calibrate and remove the instrumental polarisation from the raw measured $q,u$ values to obtain the true observed values, using a Mueller matrix fit to all standard star observations (see section \ref{sec:polcalib}).

As discussed in section \ref{sec:sofiobs} SofI
requires rotation of the Wollaston prism with respect to the detector to take the equivalent polarimetric measurements.
To convert these measured Stokes parameters into linear polarisation degree ($P$) and angle of polarisation ($\theta$), we used the following relations
\begin{align}
\begin{split} \label{eq:p}
P = \sqrt{q^{2} + u^{2}}
\end{split} \\
\begin{split} \label{eq:theta}
\theta = \frac{1}{2}{\rm arctan}\left(\frac{q}{u}\right) + \phi
\end{split} \\
\begin{split} \label{eq:phi}
\phi =
\begin{cases}
	0^{\circ},& \text{if } q > 0 \text{ and u} \geq 0\\
	180^{\circ},& \text{if } q > 0 \text{ and u} < 0\\
	90^{\circ},& \text{if } q < 0
\end{cases} 
\end{split}
\end{align}
where equation \ref{eq:phi} is for an offset angle, $\phi$, dependent on the signs of $q$ and $u$. This aligns the polarisation angle to the common definitions of position angle (where the +Q vector is North, \citealt{Wiersema2012b}; \citealt{deSerego2017}). The errors on $q$ and $u$ were calculated following the method described in \citet{Patat2006} and the errors on $P$ and $\theta$ were calculated through the propagation of the $q$ and $u$ errors (but see below for a discussion on bias).
  
The instrumentally-corrected polarisation of an optical source does not reflect the true polarisation value due to polarisation bias \citep{Serkowski1958}. This effect is a function of $P/\sigma_P$. There are a number of estimators that can correct for polarisation bias such as the Maximum Likelihood estimator \citep{Simmons1985} and Wardle-Kronberg estimator \citep{Wardle1974}. We use the modified asymptotic (MAS) estimator  defined in \citet{Plaszczynski2014} by the following expression
\begin{align} \label{eq:pmas}
P_{\rm MAS} = P - \sigma^{2} \left[\frac{1-e^{\frac{-P^2}{\sigma^{2}_{P}}}}{2P}\right]
\end{align}
where $P_{\rm MAS}$ is the modified asymptotic estimation of the true polarisation $P_{\rm 0}$ and $\sigma_{P}$ represents the standard error on the polarisation measurement.

In most cases where the SNR is high, the distribution of $P$ can be taken to be approximately Gaussian. As the SNR of a source decreases the distribution of $P$ begins to follow a Rice distribution \citep{Rice1944}. This occurs when $\eta < 2$ where $\eta=P{\rm (SNR)}$ and leads to non-symmetric and complex confidence interval calculations. For the majority of our observations, where $P_{\rm MAS}/\sigma_P \gtrsim 3$, the signal to noise is sufficiently high to quote $P_{\rm MAS}\pm \sigma_{P}$ for our results. Our quoted errors are close to the real 68\% confidence intervals but we probably underestimate the true error by a small amount \citep{Simmons1985,Sajina2011}.

For cases where the signal to noise is low (which we take as $P_{\rm MAS}/\sigma_{P} < 3$) we quote a $95\%$ upper limit on the degree of true polarisation given by
\begin{align}
P^{\alpha}_{\rm Upper} = P_{\rm MAS} + P_{\alpha}(1 - \beta e^{-\gamma P_{\rm MAS}})
\end{align}
where $\alpha = 0.95$, $P_{\alpha} = 1.95\sigma_{P}$, $\beta = 0.22$ and $\gamma = 2.54$ in the case of a $2\sigma$ upper limit \citep{Plaszczynski2014}. Given the relatively small number of sources in our survey (< 100) a full statistical treatment of the distribution of formal measurements (e.g. \citealp{Quinn2012}) would not result in changes to our conclusions.

\subsection{Calibrating instrumental polarisation} \label{sec:polcalib}
As discussed in Section \ref{sec:sample}, we require an accurate instrumental calibration to ensure that our values are not dominated by instrumental polarisation systematics and our results are meaningful. We aim for calibration accuracy $P_{\rm sys}\lesssim0.2\%$. At the time of observing there were no comprehensive investigations of SofI and EFOSC2 instrumental polarisation behaviour in the literature. Both EFOSC2 and SofI are Nasmyth mounted and should therefore exhibit high levels of polarisation, with a strong dependency on PA and wavelength. 

In the EFOSC2 run, we observed a sample of 5 unpolarised and 3 polarised standard stars over our three observing nights in \textit{V}, \textit{B}, and \textit{R} bands, for a sum total of 48 and 21 datapoints, respectively. In the SofI run, we observed 5 unpolarised and 1 polarised standard star in \textit{Z} band, for a sum total of 14 and 2 datapoints, respectively. Observation times were chosen to sample the PA dependence well. EFOSC2 is a focal reducer instrument, with somewhat similar optics to the Focal Reducer and low dispersion Spectrograph (FORS) instruments at VLT. The FORS instruments show pronounced off-axis instrumental polarisation, but low values on-axis \citep{Patat2006}. We therefore positioned each source, science and calibration object, in the centre of the CCD, near the optical axis, as part of the acquisition procedure. As such, our calibration efforts do not address off-axis instrumental polarisation patterns.

Our EFOSC2 calibration efforts are discussed in detail in a separate publication (\citealt{Wiersema2018}). We will summarise the main points below, as we use an identical approach for SofI (which is not discussed in
\citealt{Wiersema2018}).  

The SofI and EFOSC2 standard star observations are reduced and analysed in the same way as the science observations. The measured $q,u$ values for the standards are then used for the instrument modelling. As described in \cite{Wiersema2018}, we prefer a Mueller matrix approach to the instrument modelling. We use a sequence of Mueller matrices following the method described in \cite{Giro2003} and \cite{Covino2014}. The train of matrices is constructed to describe all key polarising components of the instrument and telescope.  We then fit for two unknown quantities in the resultant matrix (i.e. the wavelength dependent complex index of refraction $n_c = n - i*k$ where $n$ is the refractive index and $k$ the extinction coefficient and any angular offset between the detector and the celestial reference frame) onto the full dataset described above. For both SofI and EFOSC2, the primary cause of instrumental polarisation is found to be the tertiary mirror (M3) that feeds the light to the instrument. We use the prescription by \cite{Stenflo1994} to evaluate the matrix components of M3, and the material constants ($n,k$) from \cite{Rakic1998} at the central wavelengths of the \textit{B},\textit{V},\textit{R} filters for EFOSC2, and \textit{Z} for SofI. As demonstrated in \cite{Wiersema2018}, the resultant model fits the EFOSC2 $q,u$ values of the unpolarised and polarised standards very well, resulting in calibration accuracy to levels of $P_{\rm sys}\sim 0.1\%$. The polarisation model is expected to be dependent on time, as mirror coatings age. For SofI we follow the exact same strategy as for EFOSC2, with the only difference being the definition of $o$ and $e$ beams.

The matrix model follows the relation 
\begin{align}
[S'] = [M_{{\rm T}}]\times [S] \label{eq:invstokes}
\end{align}
where [$S$] is the Stokes vector representing the intrinsic polarisation parameters of the source, [$M_{{\rm T}}$] is the Mueller matrix representing the physical properties of the telescope (discussed above) and [$S'$] is the Stokes vector representing the measured polarisation parameters (a combination of real and instrumental polarisation). To extract the true (instrumental polarisation corrected) Stokes vector we can simply use the inverse matrix:
\begin{align}
[S] = [M_{{\rm T}}]^{-1}\times [S'] \label{eq:stokes}
\end{align}

The matrix element values depend on PA, so to correct the measured $q,u$ from section \ref{sec:polmethod}, we evaluate the matrix elements above using the PA at the middle of the polarimetric observation set. As our exposure times are relatively short, the uncertainty in PA is small. 

\begin{figure}
\includegraphics[width=\linewidth]{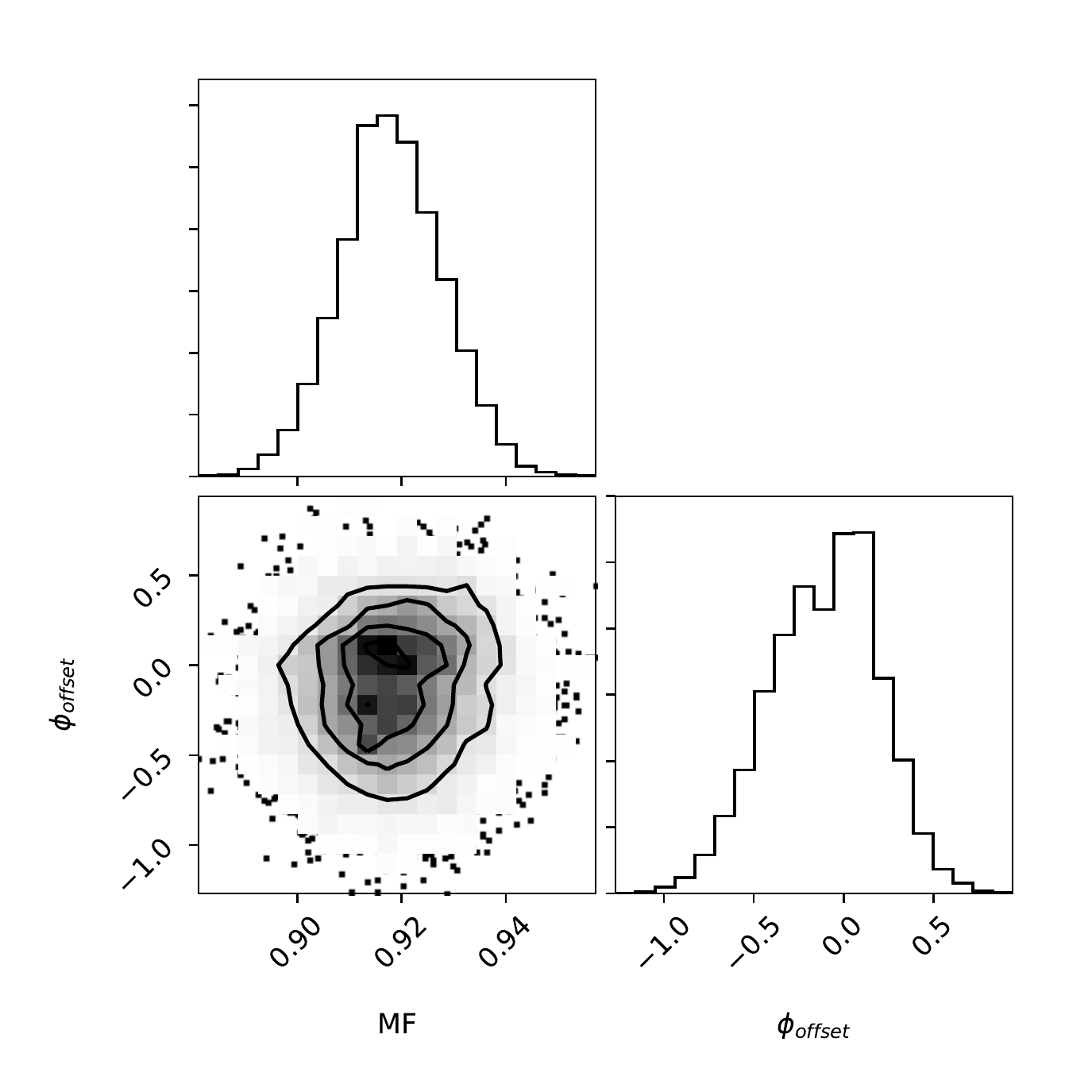}
\caption{2D projection of the probability distributions of the two fitting parameters for SofI calibration - the multiplication factor (MF) and the detector offset angle $\phi_{\rm offset}$.}
\label{fig:sofimcmc}
\end{figure}

Figure \ref{fig:sofimcmc} shows the projection of the probability distributions of the two fitting parameters, derived using a Markov Chain Monte Carlo (MCMC) code (\textsc{emcee}; \citealt{Foreman2013}). As observed with the EFOSC2 calibration \citep{Wiersema2018} we find the parameter space is non-degenerate with both parameters following a normal distribution. The median values (peaks) of the probability distributions and $1\sigma$ confidence intervals can be seen in table \ref{table:mcmcresults}.

\begin{table}
\caption{Detector angle offset and multiplication factor values derived from the MCMC analysis. Confidence intervals are $1\sigma$.}
\centering
\begin{tabular}{cc}
    \hline
\multicolumn{1}{p{2.0cm}}{\centering $\phi_{\rm offset}$ (degrees)} 
& \multicolumn{1}{p{2.0cm}}{\centering Multiplication \\ factor}\\ \hline
$-0.17\pm0.3$ & $0.92\pm0.01$ \\ \hline
\end{tabular}
\label{table:mcmcresults}
\end{table}

\begin{figure}
\includegraphics[width=\linewidth]{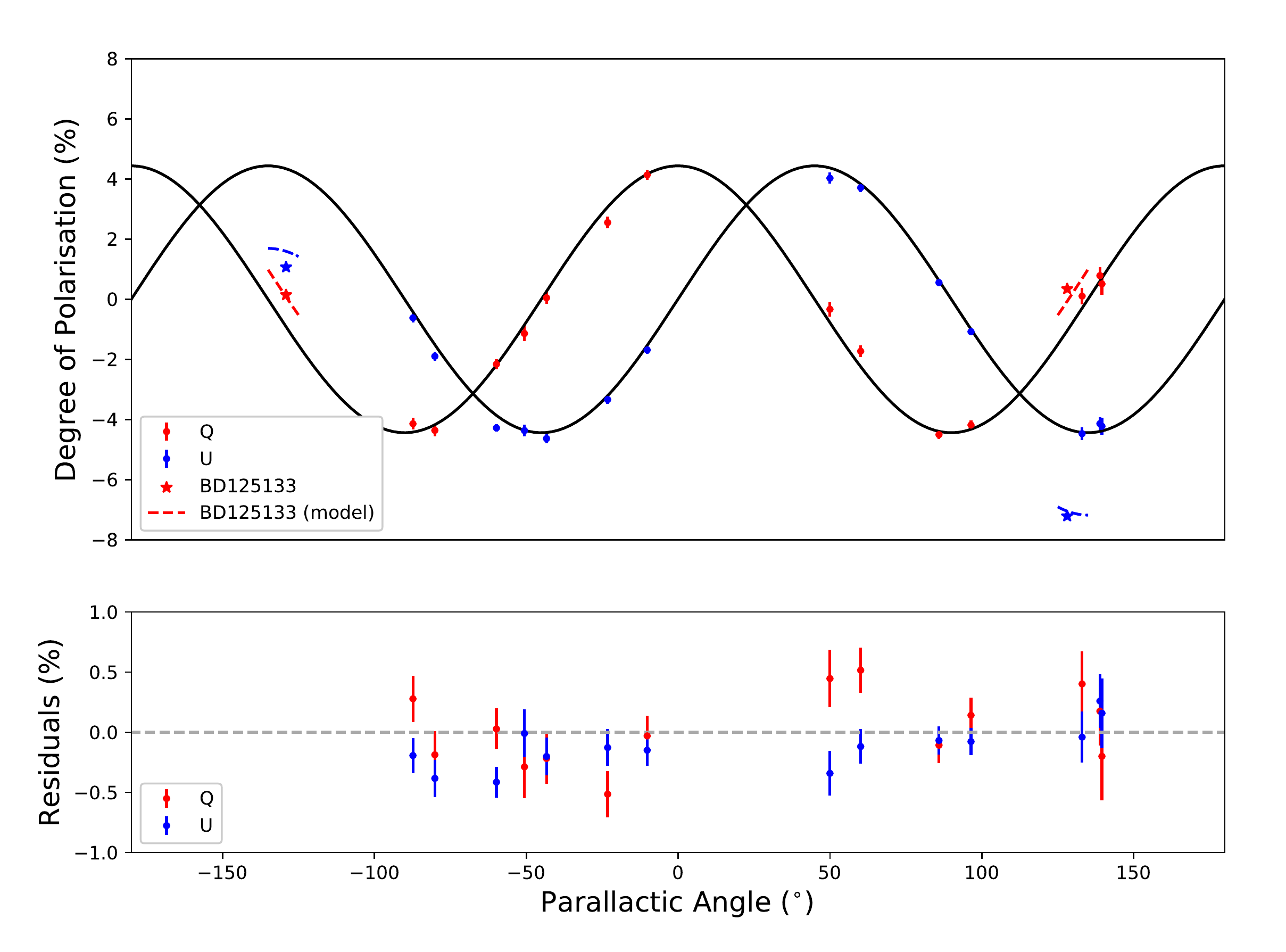}
\caption{The top panel shows the best fit model for SofI instrumental polarisation derived using the Mueller matrix method (black line) and measured $q,u$ values of the unpolarised standard stars (circles) in \textit{Z} band. We have also shown the measured $q$ and $u$ values for the polarised star BD-$12^{\circ}5133$ and the model which reproduces the measurements well. The bottom panel shows the residuals for the $q,u$, where the average residuals for the $q$ and $u$ fits are $\sim0.12\%$ and $\sim0.10\%$ respectively.}
\label{fig:sofimodel}
\end{figure}

We show the SofI calibration model using the above fitting parameter values in Figure \ref{fig:sofimodel}. We compare the models to the observed unpolarised standard star Stokes parameters $q,u$. The measured $q,u$ values from our observations of the unpolarised standard stars can be seen in table \ref{tab:sofistars}. We also observed the polarised standard star BD-$12^{\circ}5133$ and compared the observed $q,u$ values to the derived values from our model fitting. To achieve this we had to estimate the intrinsic \textit{Z} band polarisation for BD-$12^{\circ}5133$  using the empirical formula for the Serkowski parameters \citep{Serkowski1975} defined by the following relation:
\begin{align}
\begin{split} \label{eq:serkowski}
P_{\lambda} = P_{\lambda_{max}}e^{\rm -Kln^{2}(\frac{\lambda_{max}}{\lambda})}
\end{split}
\end{align}
where $P_{\lambda}$ is the linear polarisation at a given wavelength, $P_{\lambda_{max}}$ is the peak linear polarisation of a source and K is the width constant. Using values derived for BD-$12^{\circ}5133$ in \citet{Cikota2017} of $P_{\lambda_{max}} = 4.37(\pm0.01)\%$, $\lambda_{max} = 505(\pm3.5)$\,nm and K = $1.20(\pm0.04)$ we find an intrinsic polarisation of $P = 2.93(\pm0.07)\%$. From the measurements of the polarisation angle ($\sim 145$\,deg) we calculate Stokes parameters of $q = 1.00(\pm0.02)\%$ and $u = -2.75(\pm0.06)\%$ respectively. A good fit is found for both the unpolarised and polarised standard stars, with a calibration accuracy of $P_{\rm sys} \sim 0.2\%$.

There are three key points to note when comparing our SofI calibration to the EFOSC2 calibration. Firstly, we find that the SofI calibration is not as accurate as our EFOSC2 calibration (we observed fewer standard stars during the second observing run) but is still to a level required to successfully analyse our science results. Secondly, the amplitude of the instrumental $q,u$ as a function of PA is larger for SofI than it is for EFOSC2 (see figure 2 in \citealt{Wiersema2018}). This agrees with our previous calibration of EFOSC2 in the \textit{B}, \textit{V} and \textit{R} bands where a larger instrumental polarisation is observed at longer wavelengths. Thirdly, we derive a detector offset angle consistent with 0 deg. This arises from our definition of the $o$ and $e$ beams discussed in section \ref{sec:poldata}. The prescription we use for the beams is for both the SofI data analysis and calibration - the reverse of EFOSC2 and this allow us to derive this conveniently low offset. Reversing the prescription to match the beam convention used for the EFOSC2 data simply changes the derived offset angle by -90 deg. Providing the $o$ and $e$ beam ordering is kept consistent for both the calibration and data analysis, the calculated degree of polarisation and polarisation angle will be same for either prescription.

\subsection{Polarisation results} \label{sec:polresults}
\begin{figure}
\includegraphics[width=\linewidth]{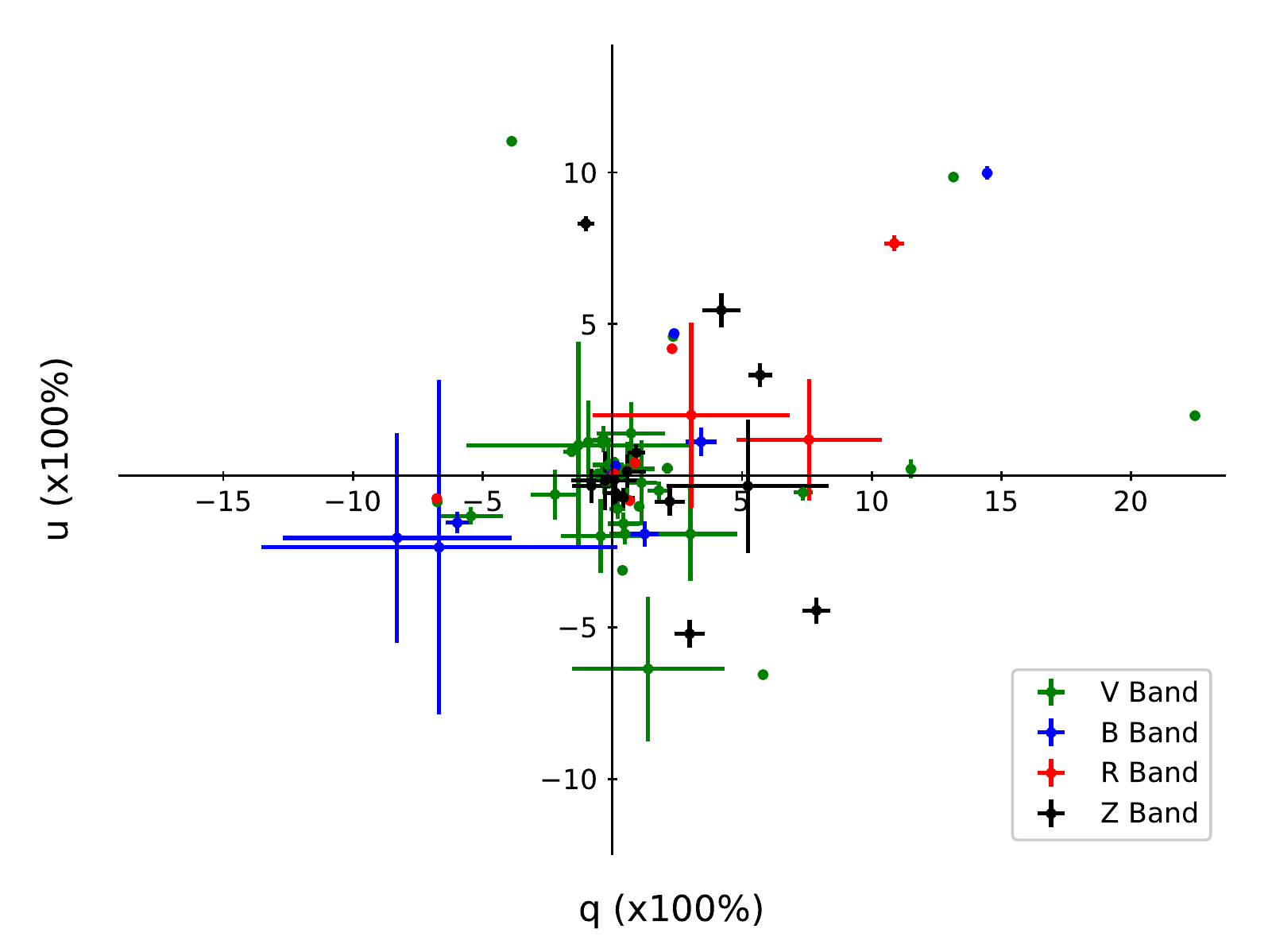}
\caption{Stokes $q$ and $u$ parameters in \textit{V}, \textit{B}, \textit{R} and \textit{Z} bands for all SPLOT sources. The plot shows that the SPLOT survey observed sources covering a large area of $q,u$ parameter space.}
\label{fig:allbandspol}
\end{figure}

\begin{figure}
\includegraphics[width=\linewidth]{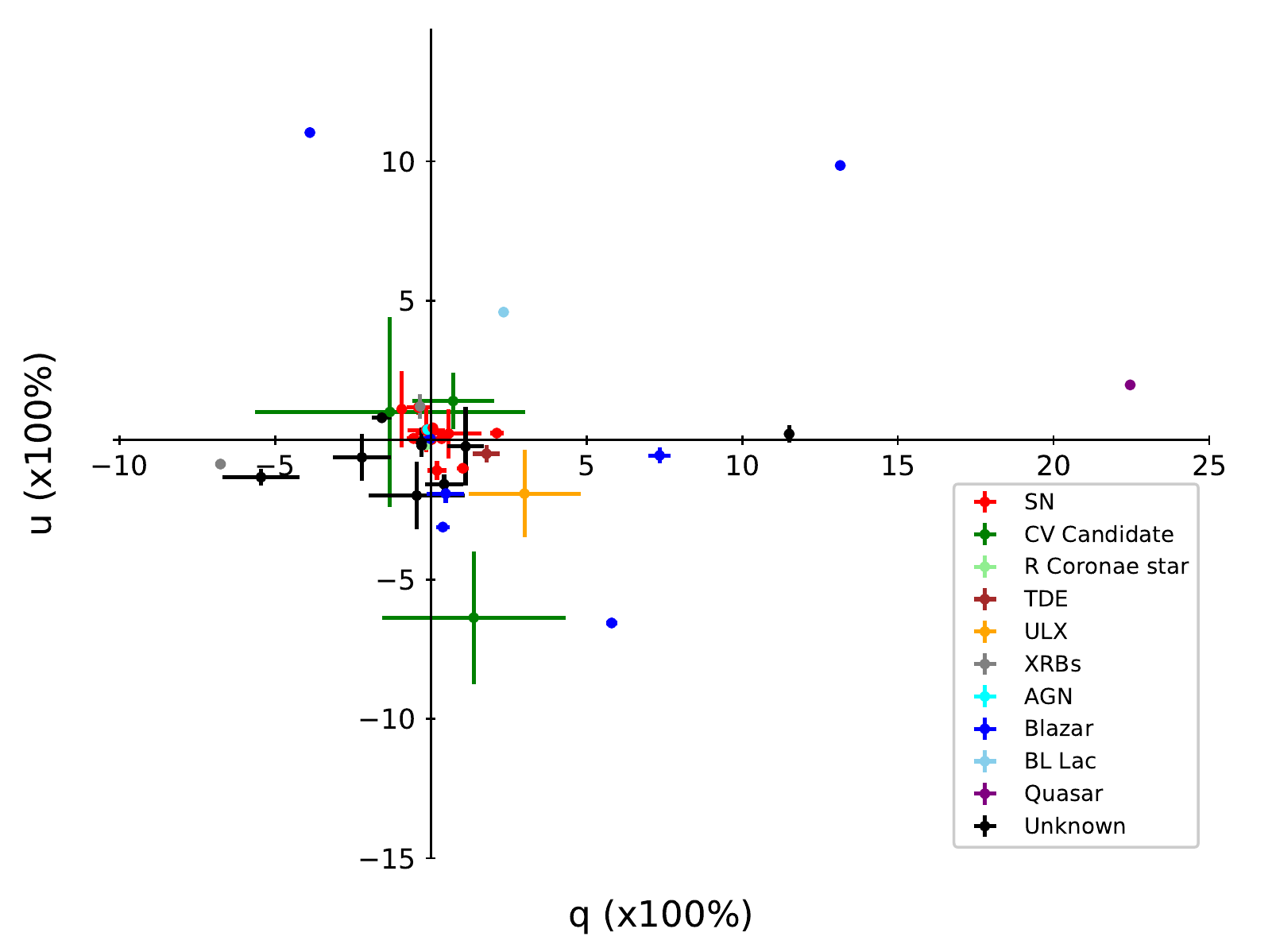}
\caption{Stokes $q$ and $u$ parameters (V band only) categorised by source type. We aimed to observe both a variety of transient sources with SPLOT and cover a large area of parameter space.}
\label{fig:polbysourcetype}
\end{figure}

\begin{figure}
\includegraphics[width=\linewidth]{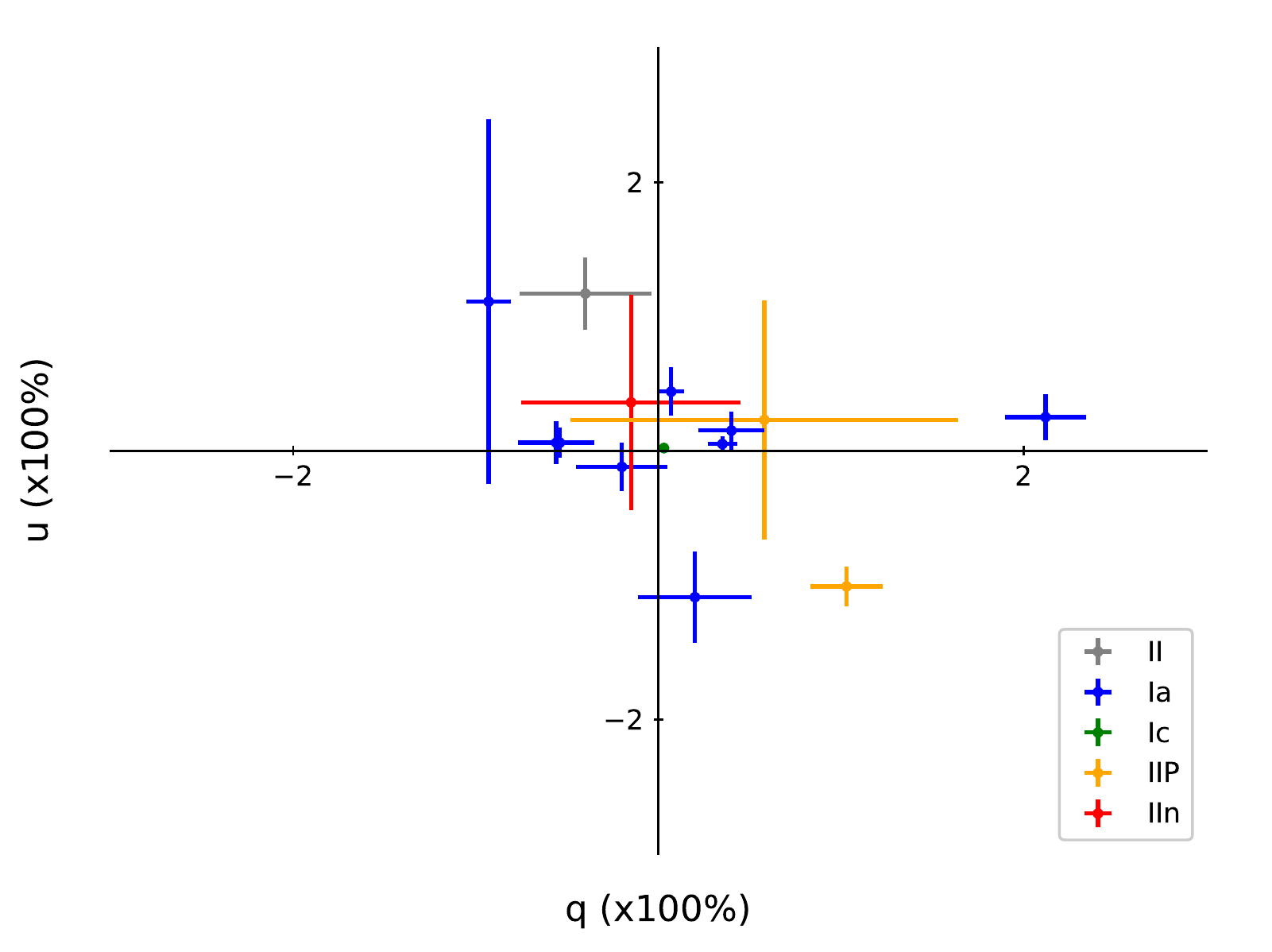}
\caption{Stokes $q$ and $u$ parameters (\textit{V} band only) separated into SN types. The $q$ and $u$ measurements shown are not corrected for line-of-sight dust. The figure highlights the significance of line-of-sight dust induced polarisation especially for type Ia SNe where we expect intrinsic $P \lesssim 0.3\%$.}
\label{fig:polbysntype}
\end{figure}

The Stokes $q$ and $u$ parameters (after correction for instrumental polarisation, section \ref{sec:polcalib}) of all observed sources can be seen in table \ref{tab:polresults} along with the bias-corrected degree of polarisation, $P$ and polarisation angle. We show the full range of Stokes $q$ and $u$ parameters for the source observations in all four filters (Figure \ref{fig:allbandspol}). This is further split into the \textit{V} band observations of each transient type (Figure \ref{fig:polbysourcetype}) and each SNe class (Figure \ref{fig:polbysntype}).

\begin{figure*}
\includegraphics[width=\textwidth]{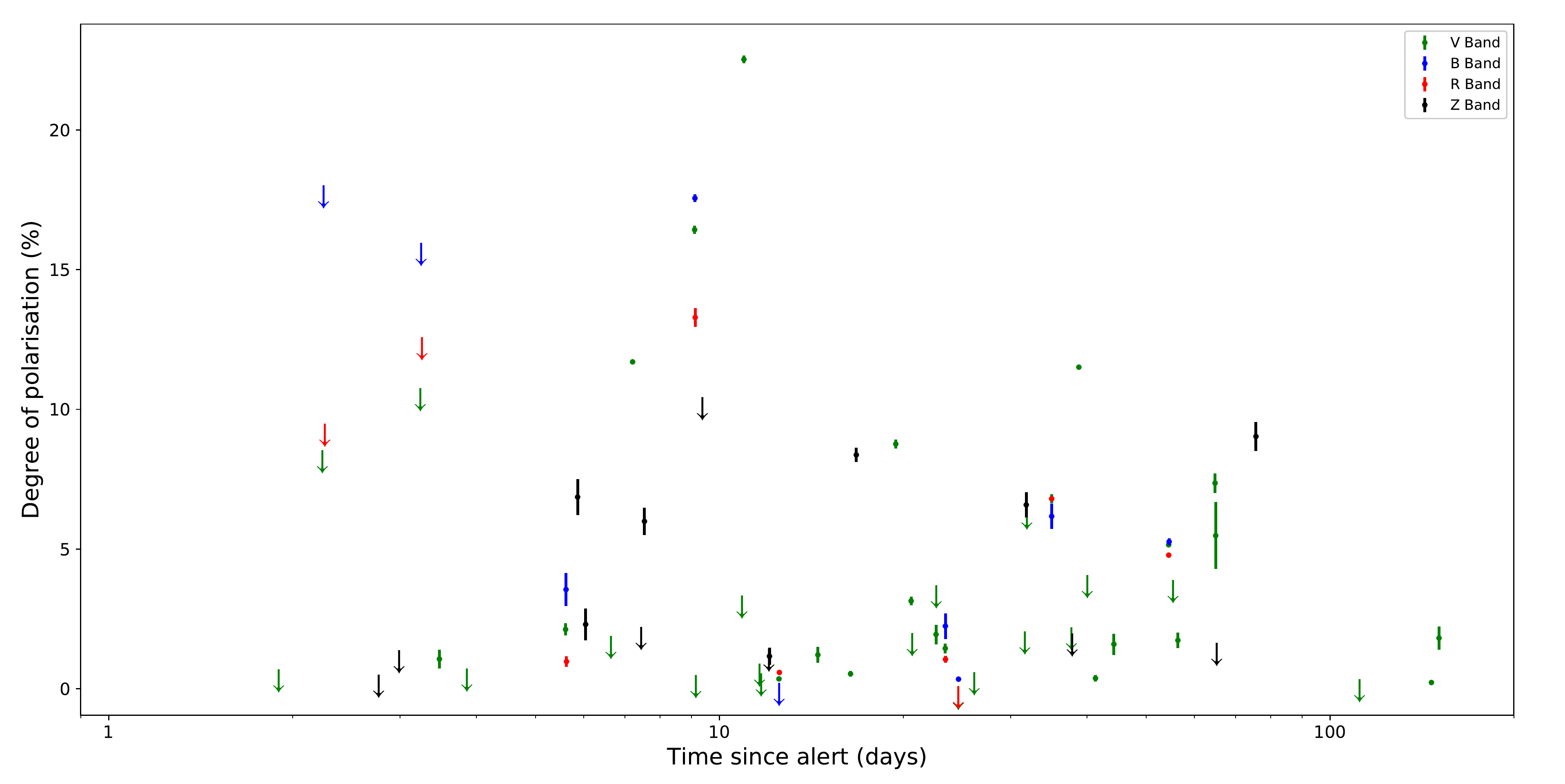}
\caption{Polarisation against time elapsed between the distributed alert and observation. Polarisation value errors are quoted to 68\% confidence and all limits are quoted to 95\% confidence. See table \ref{tab:polresults} for individual source details.}
\label{fig:pol_vs_tst}
\end{figure*}

Figure \ref{fig:pol_vs_tst} displays the polarimetric parameter space covered by SPLOT - analogous to the second panel of Figure \ref{fig:discoveryspace}. The figure shows the bias-corrected polarisation against time elapsed. The time elapsed is calculated from the time the source alert was distributed to the midpoint time when we took our observations (column three in table \ref{tab:polresults}). In the case of most new transients the alert corresponds to the discovery of the source. For sources with historic observations we use the date of a recent alert of increased activity (see appendix \ref{sec:sourceresults}), where the time elapsed is calculated from time of the recent outburst alert to the time we took our observations.

For additional information on each individual source see appendix \ref{sec:sourceresults}. We also provide light curves where possible to highlight where our observations lie with respect to the evolution of the source (e.g. are we observing before or after light curve peak for SNe and novae).

\begin{table*}
\caption{Table containing the observational time and polarisation properties of our chosen sources. The median observation date and parallactic angle values recorded are taken from the start of the third half wave-plate exposures. All errors are quoted to 68\% confidence apart from upper limits, which are quoted at 95\% confidence (see section \ref{sec:poldata}).}
	\footnotesize\setlength{\tabcolsep}{2.0pt}
    \centering
  	\begin{tabular}{|c|c|c|c|c|c|c|c|c|c|c|}
    \hline
  \multicolumn{1}{|p{1.7cm}|}{\centering Source \\ Name}
& \multicolumn{1}{|p{0.6cm}|}{\centering Filter}
& \multicolumn{1}{|p{1.3cm}|}{\centering Obs. Date \\ (mid, MJD)}
& \multicolumn{1}{|p{1.6cm}|}{\centering Exposure Time $^a$\\ (s)}
& \multicolumn{1}{|p{1.2cm}|}{\centering Parallactic Angle \\ (mid, degrees)}
& \multicolumn{1}{|p{1.1cm}|}{\centering $q$ \\ ($\times 100\%$)}
& \multicolumn{1}{|p{1.1cm}|}{\centering $u$ \\ ($\times 100\%$)}
& \multicolumn{1}{|p{1.1cm}|}{\centering $P$ \\ ($\times 100\%$)}
& \multicolumn{1}{|p{1.1cm}|}{\centering $\theta$ \\ (degrees)}
& \multicolumn{1}{|p{1.4cm}|}{\centering Time Elapsed $^b$ \\ (days)}
& \multicolumn{1}{|p{1.0cm}|}{\centering Type $^c$}
  \\ \hline
  3C 454.3 & \textit{V} & 57560.4243 & 60 & 169.9 & -3.88($\pm0.06$) & 11.03($\pm0.05$) & 11.70($\pm 0.05$) & 54.7($\pm 0.12$) & 7.21 & Blazar \\
          & \textit{V} & 57562.3199 & 2 x 60 & -149.7 & 13.15($\pm0.14$) & 9.85($\pm0.15$) & 16.43($\pm0.14$) & 18.43($\pm0.24$) & 9.11 & \\
          & \textit{B} & 57562.3289 & 2 x 60 & -152.7 & 14.45($\pm 0.05$) & 9.98($\pm0.22$) & 17.56($\pm0.14$) & 17.32($\pm0.21$) & 9.12 & \\
          & \textit{R} & 57562.3380 & 2 x 60 & -156.0 & 10.87($\pm0.38$) & 7.66($\pm0.26$) & 13.29($\pm0.34$) & 17.59($\pm0.74$) & 9.13 & \\ \hline
  ASASSN-16fp & \textit{V} & 57560.2842 & 1 x 15 + 2 x 30 & -149.4 & 0.03($\pm0.02$) & 0.02($\pm0.03$) & $\leq0.08$ & - & 24.63 & SN Ib \\
  			 & \textit{B} & 57560.2920 & 2 x 30 & -152.0 & 0.12($\pm0.14$) & 0.32($\pm0.01$) & 0.34($\pm0.05$) & 35.17($\pm4.21$) & 24.64 &  \\
  			 & \textit{R} & 57560.2986 & 2 x 30 & -154.2 & 0.07($\pm0.03$) & 0.05($\pm0.03$) & $\leq0.10$ & - & 24.65 & \\ \hline
  ASASSN-16fq & \textit{V} & 57559.9968 & 180 & 148.3 & 1.03($\pm0.20$) & -1.01($\pm0.15$) & 1.44($\pm 0.18$) & 157.89($\pm 3.54$) & 23.44 & SN IIP \\
  			 & \textit{B} & 57560.0069 & 180 & 145.0 & 1.24($\pm0.54$) & -1.92($\pm0.42$) & 2.24($\pm 0.46$) & 151.45($\pm 5.77$) & 23.45 & \\
  			 & \textit{R} & 57560.0170 & 180 & 142.0 & 0.67($\pm0.14$) & -0.82($\pm0.11$) & 1.05($\pm 0.12$) & 154.77($\pm 3.29$) & 23.46 & \\ \hline
  ASASSN-16fs & \textit{V} & 57560.0830 & 2 x 180 & 170.9 & -0.54($\pm0.10$) & 0.06($\pm0.11$) & 0.53($\pm0.10$) & 86.75($\pm5.28$) & 16.40 & SN Ia \\ \hline
  ASASSN-16ft & \textit{V} & 57559.3699 & 300 & -139.8 & -0.40($\pm0.36$) & 1.17($\pm0.27$) & 1.21($\pm 0.28$) & 54.43($\pm 6.42$) & 14.50 & SN II \\ \hline
  ASASSN-16fv & \textit{V} & 57559.1257 & 180 & -42.8 & 0.35($\pm0.08$) & 0.05($\pm0.06$) & 0.35($\pm 0.08$) & 3.86($\pm 6.41$) & 12.52 & SN Ia \\
  	         & \textit{B} & 57559.1344 & 120 & -39.1 & 0.01($\pm0.12$) & 0.06($\pm0.09$) & $\leq0.22$ & - & 12.53 & \\
  	         & \textit{R} & 57559.1418 & 120 & -35.9 & 0.57($\pm0.09$) & 0.15($\pm0.07$) & 0.58($\pm 0.09$) & 7.28($\pm 4.16$) & 12.54 & \\ \hline
  ASASSN-16fx & \textit{V} & 57559.4174 & 180 & -77.3 & -0.20($\pm0.25$) & -0.12($\pm0.18$) & $\leq0.56$ & - & 11.71 & SN Ia \\ \hline
  ASASSN-16ga & \textit{V} & 57559.2052 & 240 & 86.6 & 0.72($\pm1.32$) & 1.40($\pm1.02$) & $\leq3.35$ & - & 10.90 & CV$^{*}$ \\ \hline
  ASASSN-16gg & \textit{V} & 57559.2325 & 90 & 95.0 & -1.31($\pm4.33$) & 1.01($\pm3.41$) & $\leq8.55$ & - & 2.24 & CV$^{*}$ \\
  			 & \textit{B} & 57559.2384 & 90 & 96.4 & -6.68($\pm6.86$) & -2.36($\pm5.52$) & $\leq18.04$ & - & 2.25 & \\
  			 & \textit{R} & 57559.2437 & 60 & 97.6 & 3.04($\pm3.80$) & 2.00($\pm3.05$) & $\leq9.50$ & - & 2.26 & \\
  			 & \textit{V} & 57560.2215 & 240 & 93.0 & 1.38($\pm2.95$) & -6.37($\pm2.38$) & $\leq10.77$ & - & 3.24 & \\
  			 & \textit{B} & 57560.2408 & 240 & 97.6 & -8.30($\pm4.42$) & -2.05($\pm3.46$) & $\leq15.98$ & - & 3.25 & \\
  			 & \textit{R} & 57560.2537 & 240 & 100.5 & 7.58($\pm2.80$) & 1.19($\pm2.00$) & $\leq12.60$ & - & 3.26 & \\ \hline
  ASASSN-17gs & \textit{Z} & 57974.0350 & 5 x 3 x 60 & 138.5 & 7.87($\pm0.54$) & -4.44($\pm0.43$) & 9.03($\pm0.52$) & 165.28($\pm1.63$) & 75.64 & BL Lac \\ \hline
  ASASSN-17km & \textit{Z} & 57973.1994 & 5 x 3 x 15 & -96.8 & 0.07($\pm0.31$) & -0.14($\pm0.25$) & $\leq0.51$ & - & 2.77 & CV$^{*}$ \\
  			 & \textit{Z} & 57973.4205 & 5 x 3 x 30 & 84.6 & 0.12($\pm0.35$) & -0.57($\pm0.55$) & $\leq1.39$ & - & 2.99 & \\ \hline
  AT2016bvg & \textit{V} & 57559.1846 & 240 & 121.9 & -2.21($\pm0.93$) & -0.62($\pm0.83$) & $\leq3.91$ & - & 55.40 & Unknown \\
  			& \textit{V} & 57560.1359 & 2 x 240 & 133.8 & -1.57($\pm0.31$) & 0.80($\pm0.17$) & 1.73($\pm0.28$) & 76.58($\pm4.60$) & 56.35 & \\ \hline
  AT2016cvk & \textit{V} & 57559.2812 & 2 x 240 & -80.0 & -0.15($\pm0.60$) & 0.36($\pm0.80$) & $\leq1.90$ & - & 6.65 & SN IIn \\ \hline
  ATLAS16bcm & \textit{V} & 57560.1118 & 240 & 165.0 & -0.56($\pm0.21$) & 0.06($\pm0.16$) & $\leq0.91$ & - & 11.64 & SN Ia \\ \hline
  ATLAS16bdg & \textit{V} & 57559.0906 & 180 & 122.8 & 2.12($\pm0.22$) & 0.25($\pm0.17$) & 2.12($\pm 0.22$) & 3.33($\pm 2.96$) & 5.60 & SN Ia \\
  			 & \textit{B} & 57559.1007 & 180 & 121.4 & 3.42($\pm0.60$) & 1.12($\pm0.48$) & 3.55($\pm 0.59$) & 9.06($\pm 4.72$) & 5.61 & \\
  			 & \textit{R} & 57559.1108 & 180 & 120.2 & 0.88($\pm0.20$) & 0.43($\pm0.15$) & 0.97($\pm 0.19$) & 12.96($\pm 5.49$) & 5.62 & \\ \hline
  ATLAS17jfk & \textit{Z} & 57974.2359 & 5 x 3 x 60 & 119.9 & 2.21($\pm0.58$) & -0.85($\pm0.46$) & 2.30($\pm0.57$) & 169.47($\pm6.88$) & 6.04 & Novae \\ \hline
  CTA102 & \textit{V} & 57559.4053 & 60 & 172.3 & 22.46($\pm0.14$) & 1.98($\pm0.11$) & 22.53($\pm 0.14$) & 2.48($\pm 0.17$) & 10.97 & Quasar \\
  		 & \textit{Z} & 57973.3216 & 5 x 3 x 60 & 151.4 & 5.70($\pm0.47$) & 3.32($\pm0.40$) & 6.58($\pm0.45$) & 15.13($\pm1.97$) & 31.81 & \\ \hline
  Gaia16aau & \textit{V} & 57559.3508 & 240 & -60.0 & -0.14($\pm0.06$) & -0.17($\pm0.05$) & 0.22($\pm 0.05$) & 115.70($\pm 7.03$) & 146.58 & RCB Star \\ \hline
  Gaia16agw & \textit{V} & 57559.1566 & 240 & 101.6 & -0.01($\pm0.31$) & 0.05($\pm0.20$) & $\leq0.36$ & - & 111.86 & Blazar$^{*}$ \\ \hline
  Gaia16alw & \textit{V} & 57562.2083 & 3 x 300 & 148.7 & -5.45($\pm1.23$) & -1.33($\pm0.29$) & 5.48($\pm1.20$) & 96.84($\pm6.13$) & 64.98 & Unknown \\ \hline
  Gaia16aoa & \textit{V} & 57562.0209 & 3 x 240 & 111.0 & 0.43($\pm0.61$) & -1.58($\pm0.36$) & 1.59($\pm0.38$) & 142.59($\pm6.65$) & 44.27 & Unknown \\ \hline
  Gaia16aob & \textit{V} & 57560.0454 & 240 & 99.0 & -0.10($\pm0.17$) & 0.38($\pm0.13$) & 0.37($\pm0.12$) & 52.10($\pm9.53$) & 41.30 & AGN$^{*}$ \\ \hline
   Gaia16aok & \textit{V} & 57559.0372 & 2 x 300 & 92.8 & 11.51($\pm0.07$) & 0.22($\pm0.31$) & 11.51($\pm 0.07$) & 0.56($\pm0.18$) & 38.79 & Unknown \\ \hline
   Gaia16aol & \textit{V} & 57560.0651 & 120 & 120.7 & -0.45($\pm1.54$) & -1.99($\pm1.21$) & $\leq4.08$ & - & 40.05 & SN$^{*}$ \\ \hline
    \end{tabular}
\label{tab:polresults}
\end{table*}

\begin{table*}
	\footnotesize\setlength{\tabcolsep}{2.0pt}
    \centering
    \begin{threeparttable}
    \begin{tablenotes}
    \item{\textbf{Table \ref*{tab:polresults}} - \textit{continued...}}
    \end{tablenotes}
	\begin{tabular}{|c|c|c|c|c|c|c|c|c|c|c|}
    \hline
  \multicolumn{1}{|p{1.7cm}|}{\centering Source \\ Name}
& \multicolumn{1}{|p{0.6cm}|}{\centering Filter}
& \multicolumn{1}{|p{1.3cm}|}{\centering Obs. Date \\ (mid, MJD)}
& \multicolumn{1}{|p{1.6cm}|}{\centering Exposure Time $^a$\\ (s)}
& \multicolumn{1}{|p{1.2cm}|}{\centering Parallactic Angle \\ (mid, degrees)}
& \multicolumn{1}{|p{1.1cm}|}{\centering $q$ \\ ($\times 100\%$)}
& \multicolumn{1}{|p{1.1cm}|}{\centering $u$ \\ ($\times 100\%$)}
& \multicolumn{1}{|p{1.1cm}|}{\centering $P$ \\ ($\times 100\%$)}
& \multicolumn{1}{|p{1.1cm}|}{\centering $\theta$ \\ (degrees)}
& \multicolumn{1}{|p{1.4cm}|}{\centering Time Elapsed $^b$ \\ (days)}
& \multicolumn{1}{|p{1.0cm}|}{\centering Type $^c$}
  \\ \hline
  Gaia16aoo & \textit{V} & 57559.0088 & 240 & 137.1 & 0.58($\pm1.06$) & 0.23($\pm0.89$) & $\leq2.21$ & - & 37.74 & SN IIP \\ \hline
  Gaia16aqe & \textit{V} & 57562.4013 & 3 x 180 & -123.5 & 1.12($\pm0.59$) & -0.23($\pm1.41$) & $\leq2.07$ & -  & 31.68 & SN Ia \\ \hline 
  Gaia17blw & \textit{Z} & 57974.3484 & 5 x 3 x 60 & -70.2 & 0.57($\pm0.72$) & 0.14($\pm0.55$) & $\leq1.65$ & - & 65.32 & SN IIn \\ \hline
  Gaia17bro & \textit{Z} & 57974.3966 & 5 x 3 x 60 & -73.8 & -0.81($\pm0.74$) & -0.33($\pm0.57$) & $\leq1.99$ & - & 37.85 & SN IIn \\ \hline
  Gaia17bvo & \textit{Z} & 57974.0793 & 5 x 2 x 60 & 64.1 & -1.03($\pm0.32$) & 8.32($\pm0.25$) & 8.37($\pm0.25$) & 48.53($\pm0.86$) & 16.76 & YSO$^{*}$ \\ \hline
  Gaia17bwu & \textit{Z} & 57973.1470 & 5 x 3 x 60 & 84.1 & 0.92($\pm0.33$) & 0.76($\pm0.27$) & 1.16($\pm0.30$) & 19.81($\pm7.25$) & 12.08 & Red Star \\ \hline
  Gaia17bxl & \textit{Z} & 57973.2327 & 5 x 3 x 60 & -82.3 & 5.22($\pm3.13$) & -0.34($\pm2.20$) & $\leq10.45$ & - & 9.39 & SN \\ \hline
  Gaia17byh & \textit{Z} & 57973.0822 & 5 x 3 x 60 & -21.5 & -0.29($\pm1.29$) & -0.16($\pm0.97$) & $\leq2.22$ & - & 7.45 & SN Ic \\ \hline
  Gaia17byk & \textit{Z} & 57974.1218 & 5 x 3 x 60 & 90.7 & 2.98($\pm0.59$) & -5.21($\pm0.46$) & 5.99($\pm0.49$) & 149.88($\pm2.35$) & 7.54 & Unknown \\ \hline
  Gaia17bzc & \textit{Z} & 57974.1937 & 5 x 2 x 60 & 98.0 & 4.20($\pm0.74$) & 5.46($\pm0.57$) & 6.86($\pm0.64$) & 26.21($\pm2.65$) & 5.86 & Unknown \\ \hline
  GX 304-1 & \textit{V} & 57562.0537 & 5 & 37.9 & -6.75($\pm0.16$) & -0.86($\pm0.12$) & 6.80($\pm 0.16$) & 93.58($\pm 0.67$) & 35.01 & HMXB \\
  		  & \textit{B} & 57562.0557 & 5 & 37.9 & -5.98($\pm0.45)$ & -1.55($\pm0.35$) & 6.17($\pm 0.45$) & 97.29($\pm 2.07$) & 35.01 & \\
  		  & \textit{R} & 57562.0578 & 5 & 37.9 & -6.77($0.08$) & -0.75($\pm0.06$) & 6.80($\pm 0.08$) & 93.08($\pm 0.34$) & 35.01 & \\ \hline
  \scriptsize{MASTER J023819} & \textit{Z} & 57974.2921 & 5 x 2 x 60 & -75.4 & 0.20($\pm0.25$) & -0.66($\pm0.20$) & 0.66($\pm0.20$) & 143.25($\pm8.36$) & 0.34 & AGN$^{*}$ \\ \hline
  \scriptsize{MASTER J220727} & \textit{V} & 57559.3162 & 2 x 240 & -143.9 & 0.20($\pm0.31$) & -1.09($\pm0.34$) & 1.06($\pm0.34$) & 140.24($\pm8.76$) & 3.48 & SN Ia \\ \hline
  OGLE16aaa & \textit{V} & 57560.3271 & 3 x 240 & -76.6 & 1.79($\pm0.43$) & -0.49($\pm0.31$) & 1.81($\pm0.42$) & 172.33($\pm6.44$) & 150.82 & TDE \\ \hline  
  P13 NGC 7793 & \textit{V} & 57560.3716 & 3 x 240 & -88.8 & 3.01($\pm1.80$) & -2.06($\pm1.62$) & $\leq6.54$ & - & 31.92 & ULX \\ \hline
  PG 1553+113 & \textit{V} & 57560.2030 & 30 & 142.7 & 2.34($\pm0.10$) & 4.59($\pm0.08$) & 5.15($\pm 0.09$) & 31.50($\pm 0.49$) & 54.42 & BL Lac \\
  			 & \textit{B} & 57560.2062 & 30 & 141.8 & 2.38($\pm0.16$) & 4.69($\pm0.13$) & 5.26($\pm 0.13$) & 31.55($\pm 0.73$) & 54.42 & \\
  			 & \textit{R} & 57560.2094 & 30 & 140.8 & 2.30($\pm0.08$) & 4.19($\pm0.07$)& 4.78($\pm 0.07$) & 30.62($\pm 0.43$) & 54.43 & \\ \hline
  PKS 1510-089 & \textit{V} & 57558.9952 & 45 & -128.1 & 5.81($\pm0.18$) & -6.56($\pm0.15$) & 8.76($\pm 0.16$) & 155.77($\pm 0.54$) & 19.45 & Blazar \\
  			  & \textit{V} & 57560.1548 & 45 & 132.8 & 0.39($\pm0.21$) & -3.12($\pm0.16$)& 3.14($\pm 0.16$) & 138.55($\pm 1.49$) & 20.61 & \\
  			  & \textit{V} & 57562.1839 & 60 & 124.4 & 0.48($\pm0.60$) & -1.92($\pm0.33$) & 1.94($\pm 0.35$) & 142.03($\pm 5.08$) & 22.64 & \\ \hline
  PKS 2023-07 & \textit{V} & 57559.2568 & 240 & -147.2 & 7.35($\pm0.36$) & -0.55($\pm0.28$) & 7.36($\pm 0.35$) & 177.84($\pm 1.38$) & 64.83 & Blazar \\ \hline
  PS16cnz & \textit{V} & 57559.0751 & 240 & 160.3 & -0.29($\pm0.18$) & -0.19($\pm0.13$) & $\leq0.60$ & - & 26.16 & Unknown \\ \hline
  PS16crs & \textit{V} & 57562.1494 & 2 x 300 & 158.6 & -0.99($\pm0.13$) & 1.11($\pm1.36$) & $\leq3.72$ & - & 22.68 & SN Ia \\ \hline
  PS16ctq & \textit{V} & 57560.1844 & 2 x 240 & 102.4 & -0.20($\pm0.23$) & -0.04($\pm0.44$) & $\leq0.50$ & - & 9.16 & Unknown \\ \hline
  PS16cvc & \textit{V} & 57560.4040 & 240 & 150.0 & 0.40($\pm0.18$) & 0.15($\pm0.14$) & $\leq0.71$ & - & 1.90 & SN Ia\\
  		  & \textit{V} & 57562.3638 & 3 x 180 & 163.7 & 0.03($\pm0.07$) & 0.39($\pm0.18$) & $\leq0.74$ & - & 3.86 & \\ \hline
  SXP 15.3 & \textit{Z} & 57973.2826 & 5 x 3 x 30 & -31.2 & 0.43($\pm0.42$) & -0.72($\pm0.33$) & $\leq1.45$ & - & 12.07 & XRB \\ \hline
  XTE J1709-267 & \textit{V} & 57562.0869 & 3 x 240 & -102.6 & -0.34($\pm0.11$) & 1.21($\pm0.44$) & $\leq2.00$ & - & 20.71 & LMXB\\ \hline
    \end{tabular}
\begin{tablenotes} \footnotesize
\item{$^a$ Exposure times are given per angle. In the case of SofI \textit{Z} band data, the exposure time is shown as NEXP x NDIT x DIT, where DIT is the detector integration time in seconds, NDIT the number of DIT integrations that is averaged to make a single output file, and NEXP the number of separate NDIT x DIT files. In the case of EFOSC2, the exposure time is shown as NSET x EXPT, where EXPT is the integration time per angle, and NSET is the number of consecutive 4-angle cycles within the observation.}
\item{$^b$ The reader is reminded that time elapsed refers to the time between the distribution of the alert and our polarimetric observations.}
\item{$^c$ For additional information about the classification of the tabulated sources see appendix \ref{sec:sourceresults}.}
\item{$^{\bf *}$ Classification not spectroscopically confirmed.}
\end{tablenotes}
\end{threeparttable}
\end{table*}

%%%% Source Photometry
\section{Source photometry} \label{sec:photometry}
Each target in the first observing run (EFOSC2) was imaged in the \textit{V} band, directly following the polarimetric sequence. In the second observing run (SofI) the same method was followed in the \textit{Z} band. A small subset were also observed in the \textit{V} and/or \textit{B} bands using the UL50 at Oadby as part of an ongoing transient programme.  Photometry was performed in the same manner for all observations.

Due to the weather conditions at both La Silla and in Oadby, our observing nights were not photometric and field stars were used for calibration, wherever possible. We cross-match field stars within the telescopes respective field-of-views (FOV) with the AAVSO Photometric All-Sky Survey (APASS), the Sloan Digital Sky Survey - DR13 (SDSS), the PanSTARRs DR1 \citep{Chambers2016i} and the Skymapper Southern Sky Survey \citep{Keller2007} catalogues for the \textit{V} and \textit{B} band images. SDSS, Pan-STARRS and Skymapper do not have direct photometric observations in the \textit{V} and \textit{B} bands, so both the \textit{V} and \textit{B} magnitudes and associated errors were calculated from the following expressions
\begin{align}
\begin{split} \label{eq:vbandmag}
{\rm M_{\textit{V}}} = {\rm M}_{g} - 0.5784({\rm M}_{g} - {\rm M}_{r}) - 0.0038
\end{split} \\
\begin{split} \label{eq:bbandmag}
{\rm M_{\textit{B}}} = {\rm M}_{g} + 0.3130({\rm M}_{g} - {\rm M}_{r}) + 0.2271
\end{split} \\
\begin{split} \label{eq:vbanderr}
\sigma_{{\rm M_{\textit{V}}}} = \sqrt{(0.4216\sigma_{{\rm M}_{g}})^{2} + (0.5784\sigma_{{\rm M}_{r}})^{2}}
\end{split} \\
\begin{split} \label{eq:bbanderr}
\sigma_{{\rm M_{\textit{B}}}} = \sqrt{(1.3130\sigma_{{\rm M}_{g}})^{2} + (0.3130\sigma_{{\rm M}_{r}})^{2}}
\end{split}
\end{align}
where ${\rm M}_{g}$ and ${\rm M}_{r}$ are the catalogue field star magnitudes in the SDSS $r$ and $g$ bands and ${\rm M_V}$ and ${\rm M_B}$ are the calculated equivalent star magnitudes in the \textit{V} and \textit{B} bands (the expressions are taken from Lupton 2005\footnote{\label{magconvert} http://www.sdss3.org/dr8/algorithms/sdssUBVRITransform.php}). Additionally, $\sigma_{{\rm M}_{g}}$ and $\sigma_{{\rm M}_{r}}$ are the $1\sigma$ errors on in the $g$ and $r$ bands with $\sigma_{{\rm M}_{V}}$ and $\sigma_{{\rm M}_{B}}$ the derived errors for the \textit{V} and \textit{B} bands.

Source Extractor (SExtractor; \citealt{Bertin1996} was used to calculate the magnitudes of all sources using apertures matched to the seeing FWHM. The SExtractor catalogue output was then cross-matched with the catalogues listed above. Any APASS, SDSS or Pan-STARRS objects that were coincident with a detected source to within $\leq 1$ arcsec were matched up. Objects that we suspected were not stars but other astrophysical objects (i.e. galaxies) were filtered out. The
relation between the SExtractor instrumental magnitudes and catalogue magnitudes was fit with a first degree polynomial to calculate zero points (we ignore colour terms and atmospheric extinction); outliers that were $> 3\sigma$ away from the best-fit line were clipped during the fitting process. 

We note that although the SDSS, Pan-STARRs and Skymapper $r$ and $g$ filters are very similar, they are not identical in properties. The effect on measured magnitudes is small but not negligible when we apply the filter transformations described above; there is a small uncertainty associated with this effect. The SDSS filter transformations were calculated using measurements from a large sample of stars. Therefore, there is a small additional uncertainty on the resulting magnitudes (typically 0.01 mag). In light of these issues, the errors on our calculated magnitudes may be underestimated by up to $\sim 0.1$ mag. 

We incorporated a similar method for the sources for the \textit{Z} band images during the second observing run. All sources were at low declinations due to high wind observing constraints, with a large number residing at declination $<-30$ deg and therefore most targets only appeared in the Skymapper catalogue. The SofI \textit{Z} filter is not identical to either the SDSS, Pan-STARRS or Skymapper $z$ filters and transformations between the bands is not well known. We therefore only provide a rough estimate for the magnitudes.

Some SPLOT target fields had very few field stars that could be used: the EFOSC2 and SofI field of views are $4.1\times4.1$ and $4.9\times4.9$ arcminutes respectively. These cases could not be calibrated using this method. As the weather on our observing run was highly variable we could not accurately interpolate between images to estimate the magnitude zero points, and so we do not calculate a magnitude for these sources. See table \ref{tab:photresults} for full set of results. 

\begin{table*}
\caption{Table containing the calculated brightness of each source and the observation date for images where a magnitude could be obtained. All errors on the magnitudes are quoted to $1\sigma$. Approximate magnitudes are given for SofI photometry (see section \ref{sec:photometry}).}
    \centering
	\begin{tabular}{|c|c|c|c|c|}
    \hline
  \multicolumn{1}{|p{3.0cm}|}{\centering Source \\ Name}
& \multicolumn{1}{|p{1.2cm}|}{\centering Filter}
& \multicolumn{1}{|p{1.8cm}|}{\centering Exposure Time \\ (s)}
& \multicolumn{1}{|p{1.4cm}|}{\centering Obs. Date \\ (mid, MJD)}
& \multicolumn{1}{|p{2.0cm}|}{\centering Magnitude \\ (AB)}
  \\ \hline
  3C 454.3 & \textit{V} & 30 & 57560.4269 & 14.26($\pm0.02$) \\
  		  & \textit{V} & 30 & 57605.0446 & 15.11($\pm0.01$) \\
  		  & \textit{V} & 30 & 57645.0503 & 15.99($\pm0.03$) \\
  		  & \textit{B} & 30 & 57663.9217 & 16.50($\pm0.04$) \\
  		  & \textit{B} & 30 & 57696.8793 & 16.04($\pm0.01$) \\
  		  & \textit{B} & 30 & 57710.8737 & 16.55($\pm0.02$) \\
  		  & \textit{V} & 30 & 57721.8178 & 15.82($\pm0.10$) \\ \hline
  ASASSN-16fp & \textit{V} & 20 & 57560.3017 & 14.10($\pm0.02$) \\
  			 & \textit{V} & 30 & 57605.0196 & 15.85($\pm0.02$) \\ \hline
  ASASSN-16fs & \textit{V} & 30 & 57560.0934 & 17.21($\pm0.04$) \\ \hline
  ASASSN-16ft & \textit{V} & 60 & 57559.3780 & 17.15($\pm0.02$) \\ \hline
  ASASSN-16fv & \textit{V} & 30 & 57559.1458 & 15.04($\pm0.01$) \\ \hline
  ASASSN-16fx & \textit{V} & 30 & 57559.4228 & 17.06($\pm0.03$) \\ \hline
  ASASSN-16ga & \textit{V} & 30 & 57559.2120 & 19.04($\pm0.04$) \\ \hline
  ASASSN-16gg & \textit{V} & 30 & 57559.2463 & 19.44($\pm0.08$) \\
  	         & \textit{V} & 60 & 57560.2604 & 19.78($\pm0.09$) \\ \hline
  ASASSN-17gs & \textit{Z} & 60 & 57974.0054 & $\sim 16.5$ \\ \hline
  ASASSN-17km & \textit{Z} & 5 & 57973.1920 & $\sim 13.7$ \\
  			 & \textit{Z} & 5 & 57973.4067 & $\sim 13.7$ \\ \hline
  AT2016bvg & \textit{V} & 30 & 57559.1913 & 18.10($\pm0.07$) \\
  			& \textit{V} & 60 & 57560.1491 & 18.26($\pm0.02$) \\ \hline
  AT2016cvk & \textit{V} & 60 & 57559.2907 & 17.77($\pm0.05$) \\ \hline
  ATLAS16bcm & \textit{V} & 60 & 57560.1186 & 17.61($\pm0.02$) \\ \hline
  ATLAS16bdg & \textit{V} & 30 & 57559.1162 & 16.70($\pm0.02$) \\ \hline
  ATLAS17jfk & \textit{Z} & 60 & 57974.2101 & $\sim18.6$ \\ \hline
  CTA 102 & \textit{V} & 20 & 57559.4079 & 15.48($\pm0.02$) \\
  		 & \textit{V} & 30 & 57605.0320 & 16.58($\pm0.02$) \\
  		 & \textit{B} & 30 & 57663.9612 & 16.48($\pm0.02$) \\
  		 & \textit{B} & 30 & 57696.8702 & 15.05($\pm0.01$) \\
  		 & \textit{B} & 30 & 57710.8650 & 14.65($\pm0.02$) \\
  		 & \textit{V} & 30 & 57721.8110 & 13.12($\pm0.01$) \\
  		 & \textit{B} & 30 & 57721.8600 & 13.89($\pm0.02$) \\
  		 & \textit{V} & 30 & 57721.8647 & 13.19($\pm0.01$) \\
         & \textit{Z} & 60 & 57973.2963 & $\sim 15.7$ \\
         & \textit{B} & 30 & 58062.8560 & 17.04($\pm0.03$) \\ \hline
  Gaia16aau & \textit{V} & 60 & 57559.3576 & 14.74($\pm0.18$) \\ \hline
  Gaia16agw & \textit{V} & 30 & 57559.1634 & 17.58($\pm0.01$) \\ \hline
  Gaia16alw & \textit{V} & 60 & 57562.2399 & 19.26($\pm0.06$) \\ \hline
  Gaia16aoa & \textit{V} & 60 & 57562.0405 & 19.16($\pm0.03$) \\ \hline
  Gaia16aob & \textit{V} & 60 & 57560.0522 & 17.27($\pm0.01$) \\ \hline
  Gaia16aok & \textit{V} & 60 & 57559.0532 & 19.83($\pm0.11$) \\ \hline
  Gaia16aoo & \textit{V} & 30 & 57559.0156 & 18.37($\pm0.04$) \\ \hline
  Gaia17blw & \textit{Z} & 60 & 57974.3190 & $\sim 17.6$ \\ \hline
  Gaia17bro & \textit{Z} & 60 & 57974.3714 & $\sim 16.8$ \\ \hline
  \end{tabular}
\label{tab:photresults}
\end{table*}

\begin{table*}
  \centering
  \begin{tablenotes}
  \item{\textbf{Table \ref*{tab:photresults}} - \textit{continued...}}
  \end{tablenotes}
  \begin{tabular}{|c|c|c|c|c|}
  \hline
  \multicolumn{1}{|p{3.0cm}|}{\centering Source \\ Name}
& \multicolumn{1}{|p{1.2cm}|}{\centering Filter}
& \multicolumn{1}{|p{1.8cm}|}{\centering Exposure Time \\ (s)}
& \multicolumn{1}{|p{1.4cm}|}{\centering Obs. Date \\ (mid, MJD)}
& \multicolumn{1}{|p{2.0cm}|}{\centering Magnitude \\ (AB)}
  \\ \hline
  Gaia17bxl & \textit{Z} & 60 & 57973.2071 & $\sim 19.4$ \\ \hline 
  Gaia17byh & \textit{Z} & 60 & 57973.0562 & $\sim 17.3$ \\ \hline
  MASTER OT J023819 & \textit{Z} & 60 & 57974.2733 & $\sim 14.8$ \\ \hline
  MASTER OT J220727 & \textit{V} & 60 & 57559.3293 & 18.29($\pm 0.02$) \\ \hline
  PG 1553+113 & \textit{V} & 30 & 57508.0162 & 16.18($\pm 0.02$) \\ \hline
  PKS 1510-089 & \textit{V} & 20 & 57558.9975 & 16.02($\pm 0.01$) \\
  			  & \textit{V} & 20 & 57560.1571 & 16.12($\pm 0.02$) \\ \hline
  PKS 2023-07 & \textit{V} & 90 & 57559.2635 & 18.13($\pm 0.01$) \\ \hline
  PS16cnz & \textit{V} & 60 & 57559.0819 & 17.28($\pm0.02$) \\ \hline
  PS16ctq & \textit{V} & 60 & 57560.1976 & 18.64($\pm0.02$) \\ \hline
  PS16cvc & \textit{V} & 30 & 57560.4108 & 16.74($\pm0.01$) \\
  		  & \textit{V} & 30 & 57605.0560 & 16.55($\pm0.02$) \\ \hline
  SXP 15.3 & \textit{Z} & 10 & 57973.2692 & $\sim 15.0$ \\ \hline
  XTE J1709-267 & \textit{V} & 90 & 57562.1066 & 17.87($\pm0.01$) \\ \hline 
  \end{tabular}
\end{table*}

% Discussion
\section{Discussion: survey results} \label{sec:surveyresults}
The SPLOT survey was conducted as a pilot investigation to determine the feasibility of an optical polarimetric survey of transient astrophysical sources. To do this we set ourselves a number of goals for SPLOT, outlined in section \ref{sec:intro}. Below, we discuss how well the results of SPLOT fit in our initial aims.

\subsection{Transient selection and sample breadth}
Our first and arguably most important goal was to observe a fairly large number of sources during our observing runs. This was important for a number of reasons: we wanted to sufficiently sample a host of different transient phenomena, sample a representative fraction of the contents of real-time alert streams produced by current facilities and cover a large volume of the multi-dimensional parameter space of properties where transient events exist.

\begin{figure}
\includegraphics[width=\linewidth]{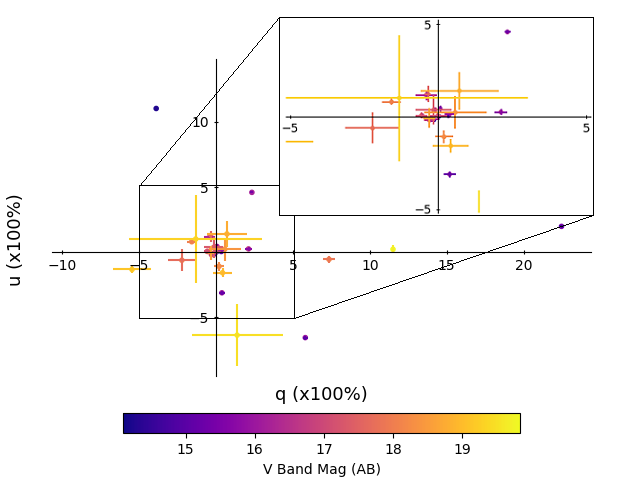}
\caption{Stoke's $q$ and $u$ parameters (V band only) over plotted with the accompanying AB magnitude. This plot demonstrates the photometric-polarimetric parameter space covered by SPLOT. The insert represents a zoomed in view of the central sources in the figure ($q,u$ values within $\pm5\%$).}
\label{fig:magplot}
\end{figure}

To maximise the number of sources we could observe, and reduce the uncertainties on the calibration, we aimed for short exposure times - with the longest observation blocks requiring execution times of no more than one hour. Our shortest execution times were $\sim 15$\,mins where we were limited by the overheads (i.e. source acquisition, read-out times). As discussed in detail in section \ref{sec:sample} we had variable weather conditions throughout our two observing runs. We lost over two and a half nights out of a scheduled eight to bad weather with the addition of the {\it Gaia} alert system becoming unavailable for the duration of our first observing run. For weather conditions where the seeing FWHM was above $\sim1.5$ arcseconds we struggled to observe the very faintest sources whilst also keeping our exposures relatively short. These limitations restricted the total parameter space we could fully explore - experience from the Palomar Transient Facility has shown that studying transients found at magnitudes $\gtrsim20$\,mag greatly expands this transient parameter space yield. As seen from table \ref{tab:polresults}, we observed 48 optical transients excluding calibration sources utilising a whole host of transient survey streams. If we further break down our sample into classifications, we observed the following:
\begin{itemize}
{\item 19 Supernovae/Supernovae candidates}
{\item 8 sources with no follow-up classification observations}
{\item 9 Active Galactic Nuclei (including BL Lacs, Blazars, strong candidate variables, etc)}
{\item 3 X-ray binaries}
{\item 3 Cataclysmic variable candidates}
{\item 1 Ultraluminous X-ray source}
{\item 1 Tidal disruption event}
{\item 1 Extragalactic Novae}
{\item 1 R Coronae Borealis star}
{\item 1 Young stellar object}
{\item 1 Brightening red star}
\end{itemize}
The above list shows that we observed a fairly diverse range of transient sources and by design a large number of sources with no prior classification. Additionally we covered a reasonable volume of the multi-dimensional parameter space, partially described by the two windows in Figure \ref{fig:discoveryspace}. Our sample representation of the polarimetric - time domain (Figure \ref{fig:pol_vs_tst}) highlights the depth of the survey. Our photometry further supports this - we covered sources whose apparent magnitudes lay between $14-20$\,mag. However, the variable weather limited how faint we could observe during times of poor conditions. Figure \ref{fig:magplot} represents a visualisation of the explored parameter space of the SPLOT survey.

\subsection{Galactic dust induced polarisation}
The polarisation measurements we made have not been corrected for line-of-sight dust and therefore contain the effects of dust scattering from both the Milky Way and host. The magnitude of this effect cannot be diagnosed directly from SPLOT \textit{V} data alone. Some effects of dust can be seen in our sample results of extragalactic sources, such as the small number of type Ia SNe that exhibit significant polarisation measurements - suggesting a large contribution from column dust (see Figure \ref{fig:polbysntype}). SPLOT also contains several sources that are at low Galactic latitude and several sources that were additionally observed in \textit{B} and \textit{R} bands. 3C454.3 was observed in multiple bands and showed significant wavelength variations; a decrease in $\sim 4\%$ between \textit{B} and \textit{R} band polarisation. Likewise, ASASSN-16fq, ATLAS16bdg and GX 304-1 exhibited similar behaviour but to a smaller extent. Therefore, to fully characterise wavelength dependent behaviour, multi-band snapshots would be required. 

A future survey can therefore estimate the Galactic dust contribution to polarisation measurements in several ways. By using field stars measurements in each set of polarimetry data, an average field star polarisation value could be derived. This could be used as a proxy for the Milky Way dust contribution to polarisation at those coordinates and with a high number of sources could slowly build up a Galactic map - with the {\it Gaia} DR2 release providing accurate astrometry and distances to a vast number of sources \citep{Lindegren2018} this could be achieved, however, it must be noted that relatively nearby field stars do not probe the full Galactic line-of-sight. The field of views of both EFOSC2 and SofI are too small to obtain a sufficient number of field stars with most sources so we were unable to attempt this during SPLOT. A value could also potentially be estimated via polarimetric sky surveys (e.g. SOUTH POL, \citealt{Magalhaes2012}) or via high resolution reddening and distance maps of field stars to name a few methods. Dust could then become a crucial parameter in many of these surveys. Many explosive transients show interplay between local dust, gas and photon emission. A large snapshot sample would be able to couple the retrieved polarisation values to models and spectroscopic observations (e.g. \citealt{Zelaya2017}). Similarly, in recent years unexpected Galactic filamentary structures have been found in long wavelength radio polarisation observations, some of which have also been seen in Planck dust polarisation maps (e.g. \citealt{Zaroubi2015}). Intrinsically unpolarised transients can play a useful role in tests of the dust induced polarisation in transmission in these fields, as they are bright and can probe the full Galactic dust column. 

\subsection{The effect of practical constraints}
We also highlighted our goal to investigate the impact of practical constraints on the success of a SPLOT-like survey. We discuss the effects these constraints had on our survey below.

\subsubsection{Weather conditions}
The varied weather conditions had a significant impact on the survey. In total we lost two and a half out of eight observing nights completely ($\sim31\%$ of our allocated time) restricting our total sample size. In periods of poor conditions (thick cloud, very poor seeing, wind), we favoured some bright sources and/or sources with long lasting outburst durations. This resulted in a sample made of some sources brighter than we had initially aimed for.

During both of our runs, we were in Bright time where the Moon was near full and up most of the night. This creates an additional sky background which is highly polarised, and therefore affects $q$ and $u$ in different ways for a source near to the Moon (see Figure \ref{fig:panels}: in the middle panel the background is very different for the $o$ and $e$ image). In nights of thin cloud, prominent Moon haloes created an additional annulus zone with strongly enhanced polarised sky background, resulting in additional pointing restrictions. The Moons influence on the sample is largely limited to an increased $\sigma_P$ for a small subset of sources, and similarly it limited exposure times for a subsample.

If we had obtained eight nights of decent weather conditions - our sample size would have been closer to $80-100$ sources and perhaps we could have sampled a larger number of sources fainter than $\sim19.5$\,mag.

\subsubsection{Instrumental calibration}
As discussed in section \ref{sec:lasilla} and \ref{sec:polcalib} the Nasymth mounted EFOSC2 and SofI both induce a high level of polarisation which must be corrected for to retrieve accurate science measurements. We used a Mueller matrix approach to model the physical telescopic system. The tertiary mirror (M3) that reflects the light towards the detector at a 45\,deg angle was found to induce the vast majority of the instrumental polarisation. We successfully calibrated both EFOSC2 and SofI with calibration accuracies of $P_{\rm sys} \lesssim 0.1\%$ and $P_{\rm sys} \lesssim 0.2\%$, respectively. The success of this calibration not only is sufficient to achieve our initial aims but has the potential to be expanded to other similar instruments and to various other optical filters. For a full discussion on our calibration method see \citealt{Wiersema2018}. Future calibration pipelines could also include correcting for instrumental polarisation away from the optical axis - something not covered by our efforts.

\subsubsection{Extrapolating light curves}
Many of the transients targeted with SPLOT had fairly large delays between discovery/alert and SPLOT observation (see Figure \ref{fig:pol_vs_tst}). We had to extrapolate the discovery magnitude to the epoch of SPLOT observation. This was often uncertain, especially for sources with no additional follow-up. Fast decaying transients (e.g. some CV outbursts and some unknown transients) were occasionally much fainter than expected, and therefore have larger polarimetric uncertainties than the uncertainty limits we aimed to achieve (see Table \ref{tab:polresults}). The periods of poor weather conditions often added to this problem meaning exposure times had to be adjusted. If, during the four angle polarimetric sequence, the weather deteriorated quickly it was harder to avoid increases in polarimetric error. We aborted observations for two sources which had faded too much to provide a reasonable polarimetric uncertainty within a reasonable execution time.

\subsection{Science results precision}
Our final aim was to achieve results with enough precision to deliver scientific conclusions for individual sources. We had aimed for polarimetric uncertainties of $\sigma_P \sim 0.2\%$ for bright sources and $\sigma_P \sim 0.5\%$ for the faintest sources. The calibration discussed above achieved our required target for constraining the induced instrument polarisation. However, the majority of the measured uncertainties were dependent on the weather and during periods of poor weather our polarimetric uncertainties were greater than we had aimed for (see table \ref{tab:polresults}).  

\subsection{Overall feasibility of a SPLOT-like survey}
\subsubsection{SPLOT results}
For SPLOT we aimed at a sample size of $\sim50-60$ sources in \textit{V} band, with a magnitude cap described in section \ref{sec:obs}. Though we did not achieve this number, we can conclude the following about the survey. The SPLOT polarisation results (see Table \ref{tab:polresults}) showed a mixed success rate. For observing periods with clear weather and little cloud, the required flux sensitivity and result precision could be achieved in the short execution times we set ourselves (below $\sim$1 hour), in particular in the EFOSC2 run. The effects of rapidly deteriorating seeing and cloud coverage resulted in some measurements failing to reach our aims. The SofI measurements have larger uncertainties than the EFOSC2 ones, and the SofI sample is brighter than the EFOSC2 one due to several factors. The instrument sensitivity considerations (a typical blue or flat-spectrum transient would require longer execution time with SofI \textit{Z} band observations than with EFOSC2 \textit{V} band observations), the weather and seeing conditions and the polarimetric accuracy achievable from the calibration of SofI. Our polarimetric results do highlight that a SPLOT-like imaging polarimetry survey of transients is not more expensive than a run-of-the-mill spectroscopic transient classification program, for the snapshot single-band strategy targets.

\subsubsection{Single or multi-band measurements}
There is no doubt that spectropolarimetry would provide scientifically superior datasets than broadband imaging polarimetry. This is especially true for sources that exhibit intrinsic wavelength dependent continuum polarisation, strong emission lines exhibiting polarisation structure and sources with high levels of foreground dust. However, as also stated in section \ref{sec:survey}, the execution time will limit such a survey to only the very brightest subsample. This would result in similar spectropolarimetry surveys being unable to sample the fainter transient events, cutting out volumes of parameter space containing transients of high interest.

To make comparisons between single and multi-band measurements, observing time was set aside to observe a small fraction of SPLOT sources in \textit{B},\textit{V} and \textit{R} rather than only in \textit{V}. These were mainly bright sources, but were not otherwise pre-selected on source type. Bad weather meant this sample is small, but some show wavelength dependent polarisation that is consistent with dust scattering dominating the signal. As discussed above, separating this dust components would require repeat visit observations with multiple broad bands, or observations deeper into the infrared. The SofI \textit{Z} band data should show lower dust polarisation effects, but the sample is smaller and we cannot make any general conclusions on the dust contributions.

\subsubsection{Snapshot or multi-epoch measurements}
The arguments for snapshots as opposed to multi-epoch polarimetry is similar to that of single band or multi-band polarimetry. In the first run with EFOSC2, half a night was set aside for repeat visits of a small subset of transients, to get variability timescales from hours to several days. The weather conditions meant that only a small subset could be done, and as such the sample with repeat visits is small (Table \ref{tab:polresults}). We see the benefit of multiple epoch observations from measurements of PKS1510-089 where the polarisation significantly decreases over a period of four days, highlighting important science such as how the internal structure of a source can vary over small timescales. As interstellar dust polarisation is not time dependent you can be confident that short scale polarisation variability between observing epochs (as discussed above) is, at least in part, intrinsic to the target source. Multi-epoch observations also have the added benefit of probing the wavelength-dependent contribution of the host galaxy dust contribution, which can be significant and vary from the Galactic Serkowski like model. The downside to this multi-epoch type of survey is that uncovering the temporal behaviour of these sources comes at the cost of survey sample size. This trade-off between sample size and depth of follow-up must always be addressed for polarimetric surveys such as SPLOT.

Our results have shown that even short exposure, single epoch photometry can provide scientific value for a number of sources. Future surveys may opt to run multi-epoch observations to increase the scientific value obtained per source on smaller samples and to explore any short time variability. However, for the majority of SPLOT we opted for a single snapshots to fit our initial aims of exploring as large a volume of the polarimetric parameter space, discussed in section \ref{sec:survey}, as possible.

\subsubsection{Highlighting sources of astrophysical interest}
The sample contains some sources that belong to rare subclasses, and as such even a single polarisation data point is of astrophysical interest, and helps to fill out blanks in the parameter space sketched in Figure \ref{fig:discoveryspace}. We highlight a few interesting sources below but for a full discussion on all individual sources see appendix \ref{sec:sourceresults}.

\paragraph{Gaia16aok}
Gaia16aok discovered as an outburst from a previously quiescent source with observed radio emission, exhibited very high levels of polarisation - $P = 11.51(\pm0.07)\%$ in \textit{V} band. A source with these properties coupled with an unknown progenitor warrants further follow-up observations to uncover the underlying physical mechanism.

\paragraph{Gaia17bvo}
Gaia17bvo a galactic variable with no previous classification also exhibited significant polarisation. We measured a polarisation of $P = 8.37(\pm0.37)\%$ in \textit{Z} band. As in the case of Gaia16aok, the single snapshot polarimetric observation highlights the potential interest in this source.

\paragraph{OGLE16aaa}
We observed OGLE16aaa, a TDE with a \textit{V} band polarisation of $P = 1.81(\pm0.42)\%$ - lower than previous measurements of relativistic TDEs and one of only a handful of TDE polarimetric observations (\citealt{Wiersema2012a}; Wiersema et al., in prep).

\paragraph{P13 NGC 7793}
We measured a polarisation of $P < 6.54\%$ from our \textit{V} band observation of P13 NGC 7793, a pulsating ULX with a period of $\sim 0.42$\,s comprising of a black hole and a donor star. This is the first polarisation measurement of a ULX but ideally under better weather conditions this limit would have been more constraining. A strongly beamed jet could lead to strongly polarised optical light in some ULXs.

\subsection{Looking to the future}
The real test looking forward is if a survey like SPLOT can detect sources of astrophysical interest within the stream of alerts through its polarimetry alone, even for sources without prior spectroscopic classification. This ability will be greatly increased by targeting a more homogeneous set of transients (e.g. coming from one, well defined, stream like ZTF) on nights less affected by weather. For a future, more mature, imaging polarimetry survey, an algorithmic target selection process could be implemented using one of these transient streams and would likely result in a higher science return for the sample as a whole, by allowing proper statistics. Limits could be placed on the age of the transient to get a higher scientific return for transients where the timescale of polarimetric change is similar to the time since first source detection, though case should be taken to scan the full polarimetric parameter space, especially for sources with ambiguous or unknown classification. 

The SPLOT survey was conducted during Visitor nights, with a visiting observer (KW+AH for EFOSC2, KW for SofI) at the observatory as the NTT is run almost entirely in Visitor mode. A service mode operated programme or robotic telescope would give a larger yield of transients for future surveys, a better ability to deal with changing conditions and a better ability to target rarer classes of transient. However, future larger volume transient feeds may negate some of the above points. During the SPLOT runs we always had available transients to observe, even in periods of strict pointing and poor weather, and the ePESSTO project has shown that transient programmes can be run well in Visitor mode. In a future survey, our SPLOT-like survey results can all be disseminated via ATels (i.e: \citealt{Higgins2016,Wiersema2016}) or using rapid automated channels (e.g. VOEvent), so that they can be linked to alerts via a broker like ANTARES\footnote{https://www.noao.edu/ANTARES} (\citealt{Saha2016}), which annotates alerts with radio to X-ray catalogue information, as well as time-domain information, on short timescales.

% Conclusion
\section{Conclusions} \label{sec:conclusion}
We undertook our SPLOT survey to test the feasibility of using linear optical polarimetry as a tool to both add value to large transient data streams and to highlight objects of potential scientific interest, in near real-time. We obtained polarimetric measurements of $\sim50$ optical transients including OGLE16aaa, a TDE and P13 NGC 7793, a pulsating ULX - where the number of previous polarimetric observations of these transient classes is very limited. We also observed a number of previously unclassified transients, some of which exhibited high levels of polarisation and significant variability in brightness (i.e. Gaia16alw, Gaia16aok). In addition, we have produced a calibration method that successfully removes instrumental polarisation effects for both EFOSC2 and SOFI. This resulted in the creation of software that allows semi-automated reduction, analysis and calibration of incoming imaging polarimetry data fast enough that dissemination of results can be done within hours of data taking.

With the advent of much larger transient missions mapping out huge volumes of transient parameter space, SPLOT has demonstrated that similar polarimetric surveys would be a welcome addition in highlighting sources for further follow-up. In combination with rapid radio and X-ray data, polarisation can provide a fast way to aid in selection of transients for studying of astrophysical sources non-thermal emission processes and increase the exploration of this vast multi-dimensional parameter space.

%%%% Acknowledgements and Bibliograhy
\section{Acknowledgments}
We thank the anonymous referee for their constructive report that improved this paper.
Based on observations collected at the European Organisation for Astronomical Research in the Southern Hemisphere under ESO programme 097.D-0891(A) and 099.D-0262(A).
This research made use of data from the Steward Observatory spectropolarimetric monitoring project which is supported by Fermi Guest Investigator grants NNX08AW56G, NNX09AU10G, NNX12AO93G and NNX15AU81G.
This work has also made use of data from the European Space Agency (ESA)
mission {\it Gaia} (\url{https://www.cosmos.esa.int/gaia}), processed by
the {\it Gaia} Data Processing and Analysis Consortium (DPAC,
\url{https://www.cosmos.esa.int/web/gaia/dpac/consortium}) and the {\it Gaia}, DPAC and the Photometric Science Alerts Team (http://gsaweb.ast.cam.ac.uk/alerts). 
Funding for the DPAC has been provided by national institutions, in particular the institutions participating in the {\it Gaia} Multilateral Agreement.
ABH is supported by an STFC studentship granted by the University of Leicester. KW, RLCS and NRT acknowledge support from STFC. RLCS acknowledges support from Royal Society Research Grant RG170230. HFS is supported by a PhD studentship granted by the University of Sheffield. 
\L W acknowledges the Polish NCN grant OPUS 2015/17/B/ST9/03167. We are grateful to all ESO support staff at La Silla, for their assistance and encouragement; and we particularly thank Ivo Saviane for allowing us to use SofI when EFOSC2 was unusable.
We thank Cristina Baglio for her kind assistance with SofI observing blocks. KW thanks Dipali Thanki and Ray McErlean for their excellent support of science operations at the University of Leicester observatory (UL50).  
This research made extensive use of the app {\em iObserve}, written by C. Foellmi (https://onekilopars.ec). We are grateful to Dr Foellmi for his commitment to this app.
This research made use of Astropy, a community-developed core Python package for Astronomy \citep{Astropy2018}.
We acknowledge ESA {\it Gaia}, DPAC and the Photometric Science Alerts Team (http://gsaweb.ast.cam.ac.uk/alerts).
We thank S. Zane and E. Rol for their contributions to the paper.
We thank the cooks at La Silla observatory for their crucial work.
\bibliographystyle{mnras}
\bibliography{splot}

% Appendix
\appendix
\section{Observations of unpolarised standard stars with SofI}
Table \ref{tab:sofistars} gives the unpolarised standard star measurements used to calibrate the instrumental polarisation of SofI.
\begin{table*}
\caption{Observations of unpolarised standard stars using SofI, with the \textit{Z} band filter. The $q,u$ values are used to calibrate the instrumental polarisation behaviour. For a full discussion see section \ref{sec:polcalib}.}
	\centering
	\begin{tabular}{|c|c|c|c|c|c|c|}
  \hline
  \multicolumn{1}{|p{1.8cm}|}{\centering Source \\ Name}
& \multicolumn{1}{|p{2.0cm}|}{\centering Obs. Date \\ (mid, MJD)}
& \multicolumn{1}{|p{2.0cm}|}{\centering Parallactic Angle \\ (mid, degrees)}
& \multicolumn{1}{|p{1.8cm}|}{\centering $q$ \\ ($\times 100\%$)}
& \multicolumn{1}{|p{1.8cm}|}{\centering $q$ error \\ ($\times 100\%$)}
& \multicolumn{1}{|p{1.8cm}|}{\centering $u$ \\ ($\times 100\%$)}
& \multicolumn{1}{|p{1.8cm}|}{\centering $u$ error \\ ($\times 100\%$)}   \\ \hline
WD 0310-688 & 57973.2626 & -80.1 & -4.36 & 0.20 & -1.90 & 0.15 \\
		   & 57973.3567 & -43.3 & 0.05 & 0.21 & -4.63 & 0.16 \\
           & 57973.4007 & -23.3 & 2.55 & 0.19 & -3.34 & 0.15 \\
           & 57974.3143 & -59.8 & -2.16 & 0.17 & -4.28 & 0.13 \\
           & 57974.4247 & -10.2 & 4.13 & 0.17 & -1.69 & 0.13 \\
WD 1344+106 & 57972.9930 & 139.5 & 0.51 & 0.36 & -4.22 & 0.29 \\
		   & 57973.9927 & 138.9 & 0.79 & 0.29 & -4.14 & 0.22 \\
WD 1615-154 & 57973.0455 & 133.0 & 0.11 & 0.27 & -4.47 & 0.21 \\
WD 1620-391 & 57973.1131 & 85.9 & -4.50 & 0.15 & 0.56 & 0.12 \\ 
		   & 57974.1577 & 96.4 & -4.18 & 0.15 & -1.08 & 0.11 \\
           & 57975.0302 & 50.0 & -0.33 & 0.24 & 4.03 & 0.19 \\
WD 2359-434 & 57973.1813 & -87.3 & -4.14 & 0.19 & -0.62 & 0.15 \\
		   & 57973.3807 & 60.1 & -1.73 & 0.19 & 3.71 & 0.14 \\
           & 57974.2676 & -50.6 & -1.14 & 0.26 & -4.37 & 0.20 \\ \hline     
     \end{tabular}
\label{tab:sofistars}
\end{table*}

{\section{Individual source results} \label{sec:sourceresults}
This section outlines the polarisation and photometry results of each source individually with details on the classification and, for long-lived sources, any historic observations. We also discuss the scientific value our snapshot polarimetry has given to each source. Where available we complement our photometry and polarisation results with additional observations using data from the Steward Observatory spectropolarimetric monitoring project \citep{Smith2009}, the {\it Gaia} transient alert system \citep{Wyrzykowski2012,Hodgkin2013} and the ASAS-SN Sky Patrol \citep{Shappee2014,Kochanek2017b}. To test for the significance of the total dust contribution from the Milky Way to measurements of extragalactic sources with detected polarisation we use $P_{\rm Gal, dust} \leq 9\times E(B-V)\%$ \citep{Serkowski1975}. Galactic colour excess, $E(B-V)$ was calculated using the method described in \citet{Schlafly2011}. We do not estimate the host dust contribution to the polarisation signal but acknowledge that in some sources this may be significant. The reader is reminded that the SofI magnitudes displayed in the light curves are only estimates.

\subsection{3C 454.3}
3C 454.3 is a well known Blazar that periodically enters outburst phases (i.e. \citealt{Hunstead1972,Pauliny-Toth1987,Raiteri2007}) and is one of the brightest gamma-ray sources in the sky (see \citealt{Ackermann2010} and \citealt{Britto2016} for record \textit{Fermi} observations). The source entered a recent outburst period on 2016 June 11 and was detected by several observatories \citep{Jorstad2016,Lucarelli2016,Ojha2016}. We observed the source twice, on 2016 June 20 and 22 with polarisation measurements of $P = 11.70(\pm0.05)\%$ in the \textit{V} band on the first night and $P = 16.43(\pm0.14)\%$, $P = 17.56(\pm0.14)\%$ and $P = 13.29(\pm0.34)\%$ on the second night in the \textit{V}, \textit{B} and \textit{R} bands respectively. We also measured a change in the polarisation angle - it varied in the \textit{V} band from $\theta = 54.7(\pm0.12)$ deg $\theta = 18.43(\pm0.24)$ deg. We calculate $E(B-V) = 0.09$ at the source position corresponding to $P_{\rm Gal, dust} \leq 0.81\%$. Figure \ref{fig:3c454_plot} highlights that our observations took place during a period of very high activity. We also saw a change in \textit{V} band polarisation of $\sim6\%$ over a 24 hour period. The high levels of polarisation and short duration variability are in agreement with previous observations of 3C 454.3 (i.e. \citealt{Smith2009,Sasada2012}). The data is also consistent with a non-thermal emission mechanism - expected from Blazars.

\begin{figure}
\includegraphics[width=8.5cm]{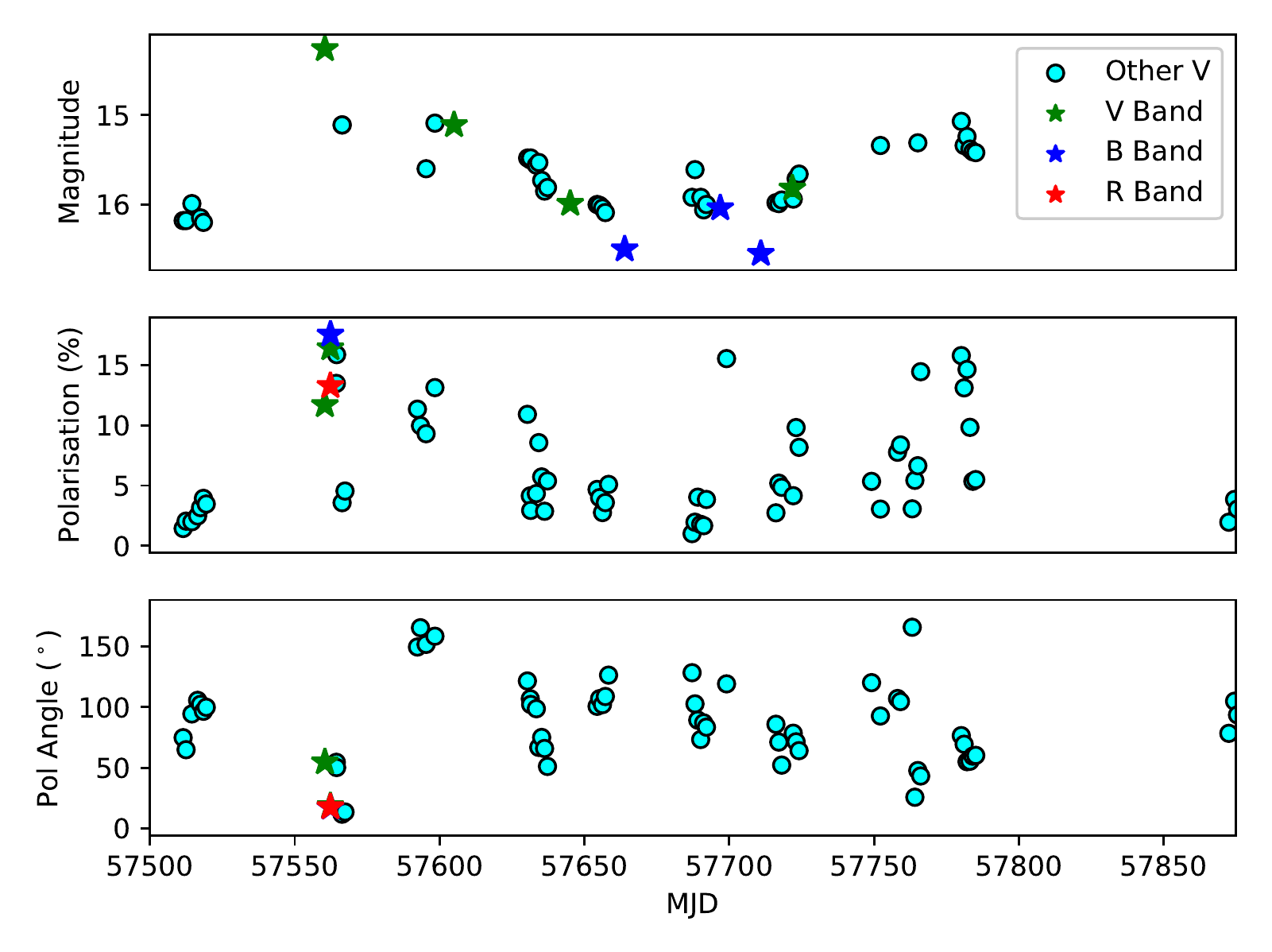}
\caption{Plot of 3C 454.3 showing the temporal evolution of source brightness, degree of polarisation and polarisation angle. The 'Other V' legend label refers to data taken from the Steward Observatory spectropolarimetric monitoring project.}
\label{fig:3c454_plot}
\end{figure}

\subsection{ASASSN-16fp}
ASASSN-16fp (also known as AT2016coi, Gaia16arp, PS16cvj and SN2016coi) was discovered on 2016 May 27 in the galaxy UGC 11868 \citep{holoien2016}. There has been some disagreement in the literature as to the classification of ASASSN-16fp. The source was initially classified as a type Ic (broad line) SNe via spectroscopic observations \citep{EliasRosa2016} but presence of Helium I absorption lines in the early-time spectrum ($< 12$ days post-alert) suggested that the source may be a new type Ib broad line SNe \citep{Yamanaka2017}. We observed the source approximately three weeks post-alert with a polarisation of $P < 0.08\%, P = 0.34(\pm0.05)\%$ and $P < 0.10\%$ for the \textit{V}, \textit{B} and \textit{R} bands respectively. We additionally obtained photometry of the source twice. We measured a brightness of 14.03($\pm0.01$)\,mag on the same night we obtained the polarisation observations and a brightness of 15.80($\pm0.02$)\,mag approximately six weeks later (see Figure \ref{fig:ASASSN16fp_lc}). We calculate $E(B-V) = 0.07$ at the source position relating to $P_{\rm Gal, dust} \lesssim 0.6\%$. This result is consistent with an intrinsically unpolarised source, where the polarisation signal is almost entirely due to Galactic dust. 

\begin{figure}
\includegraphics[width=8.5cm]{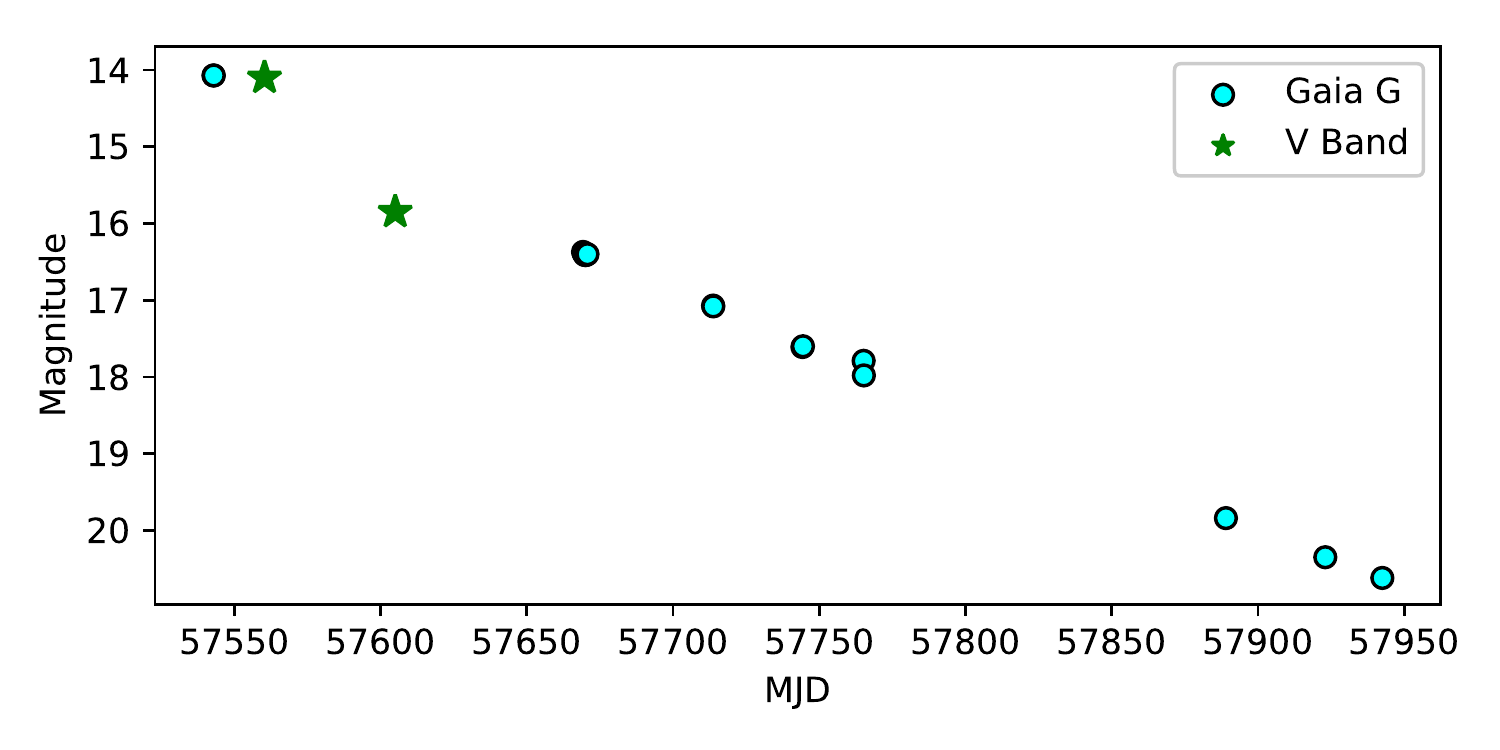}
\caption{Lightcurve of ASASSN-16fp. {\it Gaia} observed this source numerous times and has well sampled the decline in brightness We observed the source near peak and obtained photometry again, six weeks later during the decline.}
\label{fig:ASASSN16fp_lc}
\end{figure}

\subsection{ASASSN-16fq}
ASASSN-16fq (also known as AT2016cok and SN2016cok) was discovered in M66 on 2016 May 28 \citep{Bock2016}. The source was initially classified as a type IIP SN via spectroscopic observations \citep{Zhang2016}. A pre-explosion counterpart to the source is discussed in detail in \citealt{Kochanek2017a}. We observed the source approximately three weeks post-alert and measured a polarisation of $P = 1.44(\pm0.18)\%, P = 2.24(\pm0.46)\%$ and $P = 1.05(\pm0.12)\%$ for the \textit{V}, \textit{B} and \textit{R} bands respectively. We calculate $E(B-V) = 0.03$ at the source position corresponding to $P_{\rm Gal, dust} \leq 0.27\%$. We can estimate the maximum level of asphericity (where $P_{\rm host, dust} = 0\%$) of the source. Using figure 4 in \citealt{Hoflich1991} and assuming $\tau_{\rm max} = 1$, $N(r)\propto r^{-2}$, $\epsilon = 0.05$ and $i = 90$\,deg where $\tau$ is the optical depth, $N$ is the density profile, $\epsilon$ is the extension of the inner boundary and $i$ is the angle of inclination we find that the axis ratio, $E\gtrsim 0.65$. Therefore the maximum ellipticity of the source, $1 - E \lesssim 0.35$. 

\subsection{ASASSN-16fs}
ASASSN-16fs (also known as AT2016cpy and SN2016cpy) was discovered in UGC09523 on 2016 June 4 \citep{Masi2016}. The source was classified as a Type Ia SNe via spectroscopic observations \citep{Pan2016}. We observed the source approximately two weeks post-alert and measured a polarisation of $P = 0.53(\pm0.10)\%$ in \textit{V} band. We calculate $E(B-V) = 0.03$ at the source position corresponding to $P_{\rm Gal, dust} \leq 0.27\%$. If we remove the Galactic dust contribution from the measured polarisation we get $P \lesssim 0.3\%$, consistent with previous broadband type Ia SNe observations.

\subsection{ASASSN-16ft}
ASASSN-16ft (also known as AT2016cqj and SN2016cqj) was discovered in CGCG 382-005 on 2016 June 5 \citep{Brimacombe2016i}. The source was classified as a type II SN via spectroscopic follow-up observations \citep{Hosseinzadeh2016}. We observed the source two weeks post-alert and measured a polarisation of $P = 1.21(\pm0.28)\%$ in \textit{V} band. We calculate $E(B-V) = 0.03$ at the source position corresponding to $P_{\rm Gal, dust} \leq 0.27\%$. We again estimate the maximum level of asphericity of the source following the same method used for ASASSN-16ft. We find that the axis ratio, $E\gtrsim 0.65$ and the maximum ellipticity of the source, $1 - E\lesssim 0.15$. 

\subsection{ASASSN-16fv}
ASASSN-16fv (also known as AT2016cgz and SN2016cgz) was discovered in IC4705 on 2016 June 7 \citep{Brimacombe2016ii}. The source was classified as a type Ia SN via spectroscopic observations \citep{Prieto2016}. We observed the source approximately 13 days post-alert and measured a polarisation of $P = 0.35(\pm0.08)\%, P < 0.22\%$ and $P = 0.58(\pm0.09)\%$ for the \textit{V}, \textit{B} and \textit{R} bands respectively.  We calculate $E(B-V) = 0.09$ at the source position corresponding to $P_{\rm Gal, dust} \leq 0.81\%$. The Galactic dust contribution probably dominates the measured source polarisation.

\subsection{ASASSN-16fx}
ASASSN-16fx (also known as AT2016csd, Gaia16avj and SN2016csd) was discovered in GALEXASC J020044.56-461644.0 on 2016 June 8 \citep{Brown2016i}. The source was classified as a type Ia supernovae via spectroscopic observations \citep{Morrell2016}. We observed the source 12 days post-alert and measured a polarisation of $P \leq0.56\%$ and a brightness of $17.02(\pm0.03)$\,mag in \textit{V} band (see Figure \ref{fig:ASASSN16fx_lc}).

\begin{figure}
\includegraphics[width=8.5cm]{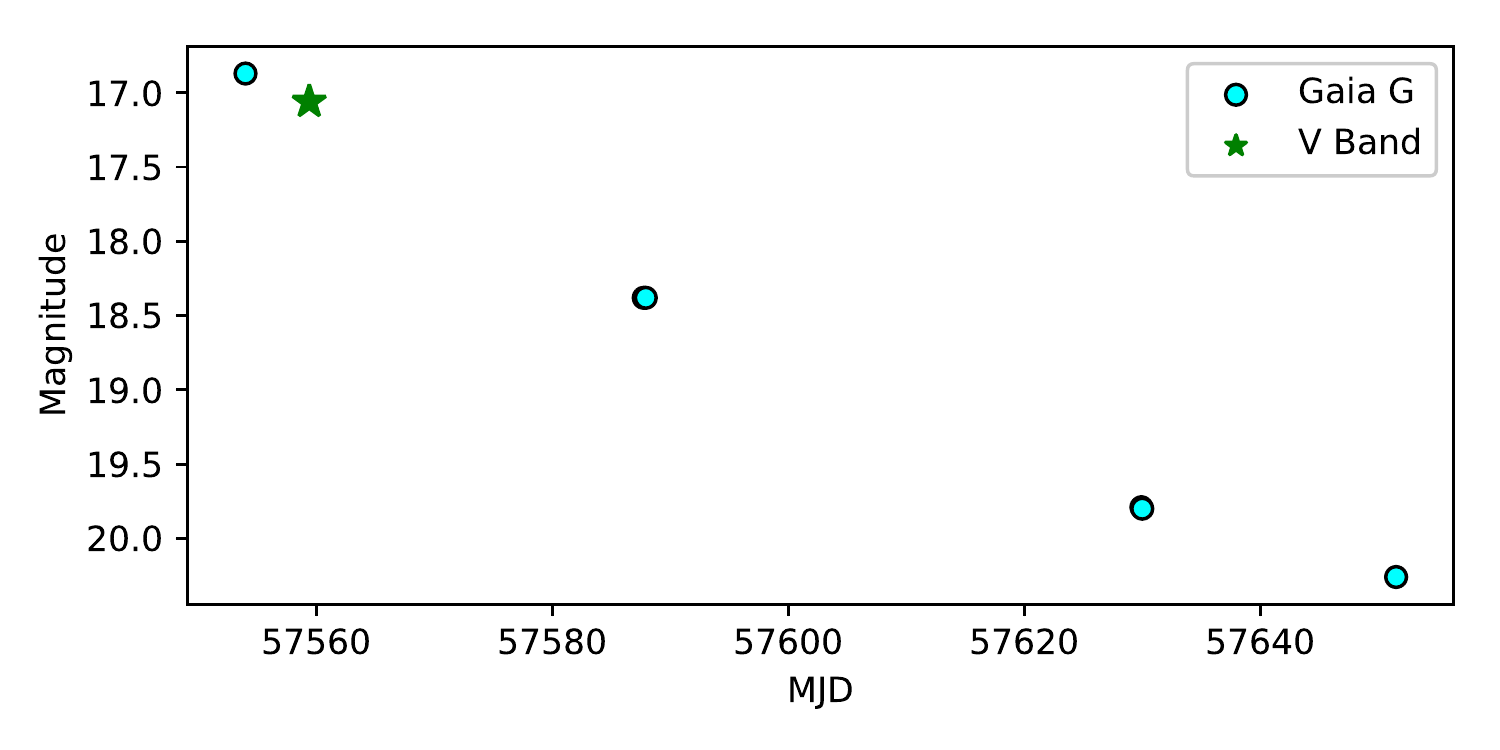}
\caption{Lightcurve of ASASSN-16fx. {\it Gaia} observed this source numerous times and has well sampled the decline in brightness We observed the source post-peak as it was declining in brightness.}
\label{fig:ASASSN16fx_lc}
\end{figure}

\subsection{ASASSN-16ga}
ASASSN-16ga was detected on 2016 June 9 and tentatively classified as a CV candidate due to a short time increase in brightness - $V < 18$\,mag non-detection on 2016 May 14 and $V\sim15$\,mag on 2016 June 9 \citep{Shappee2014}. We observed the source on 2016 June 19, 11 days post-alert and measured a polarisation of $P\leq3.35\%$.

\subsection{ASASSN-16gg}
ASASSN-16gg was detected on 2016 June 17 and was tentatively classified as a CV candidate due to a short time increase in brightness - $V >18.0$\,mag non-detection on 2016 June 9, $V =15.0$\,mag on 2016 June 17 \citep{Shappee2014}. We observed the source twice (2016 June 19 and 20). On the first night we obtained polarisation limits of $P\leq8.55\%$, $P\leq18.04\%$ and $P\leq9.50\%$ in the \textit{V}, \textit{B} and \textit{R} bands respectively and a magnitude of $V =19.52(\pm0.08)$\,mag. On the second night we obtained polarisation limits of $P < 10.77\%$, $P < 15.98\%$ and $P < 12.60\%$ in the \textit{V}, \textit{B} and \textit{R} bands respectively and a magnitude of $V =19.78(\pm0.09)$\,mag.

\subsection{ASASSN-17gs}
ASASSN-17gs (also known as AT2017egv) was an optical transient detected on 2017 May 25 \citep{Stanek2017b} coincident with a new gamma-ray source, Fermi J1544-0649, initially detected by \textit{Fermi}/LAT on 2017 May 15 \citep{Ciprini2017}. Initially the source was seen as a candidate relativistic tidal disruption event (TDEs with a relativistic jet, similar to \textit{Swift} J1644 and \textit{Swift} J2058). The source was later classified as a BL Lac object via follow-up multi-band radio observations \citep{Bruni2018}. We observed the source approximately ten weeks post-alert and measured a polarisation of $P = 9.03(\pm0.52)\%$ in \textit{Z} band. We calculate $E(B-V) = 0.14$ at the source position corresponding to $P_{\rm Gal, dust} \leq 1.26\%$. The result agrees with previous observations of high levels of polarisation in BL Lac objects (i.e. \citealt{Smith2007}) and is consistent with a non-thermal emission mechanism. Interestingly, the two relativistic tidal disruption events with optical polarimetry (\textit{Swift} J1644 and \textit{Swift} J2058) seem to show polarisation values of similar amplitude (\citealt{Wiersema2012a}; Wiersema et al. in prep.).

\begin{figure}
\includegraphics[width=8.5cm]{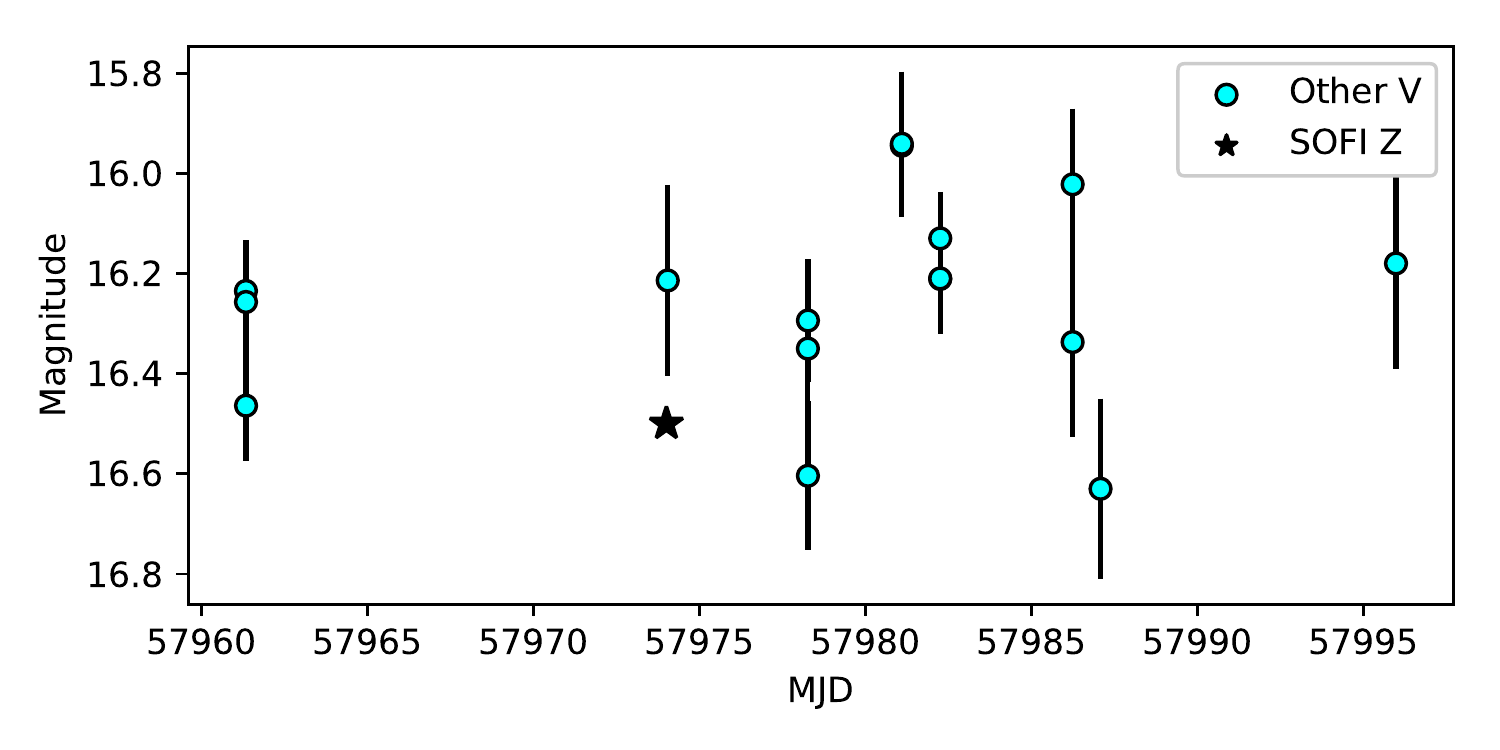}
\caption{Lightcurve of ASASSN-17gs. The V band data points (cyan) correspond to ASAS-SN observations of the source.}
\label{fig:ASASSN17gs_lc}
\end{figure}

\subsection{ASASSN-17km}
ASASSN-17km was detected on 2017 August 5 and tentatively classified as a CV candidate due to a short time increase in brightness - $V < 17.2$\,mag non-detection on 2017 August 4, $V \sim14.7$ on 2017 August 5 and a further increase in brightness to $V\sim12.7$\, mag on 2017 August 7 \citep{Shappee2014}. We observed the source twice on the 2017 August 8, three days post-alert and several hours apart and measured polarisations of $P\leq0.51\%$ and $P\leq1.39\%$ in \textit{Z} band. We also took two photometric measurements of the source at the described times above and measured \textit{Z} band magnitudes of $\sim13.7$\,mag and $\sim13.7$\,mag respectively. Figure \ref{fig:ASASSN17km_lc} shows that we observed the source very close to peak brightness in V band.

\begin{figure}
\includegraphics[width=8.5cm]{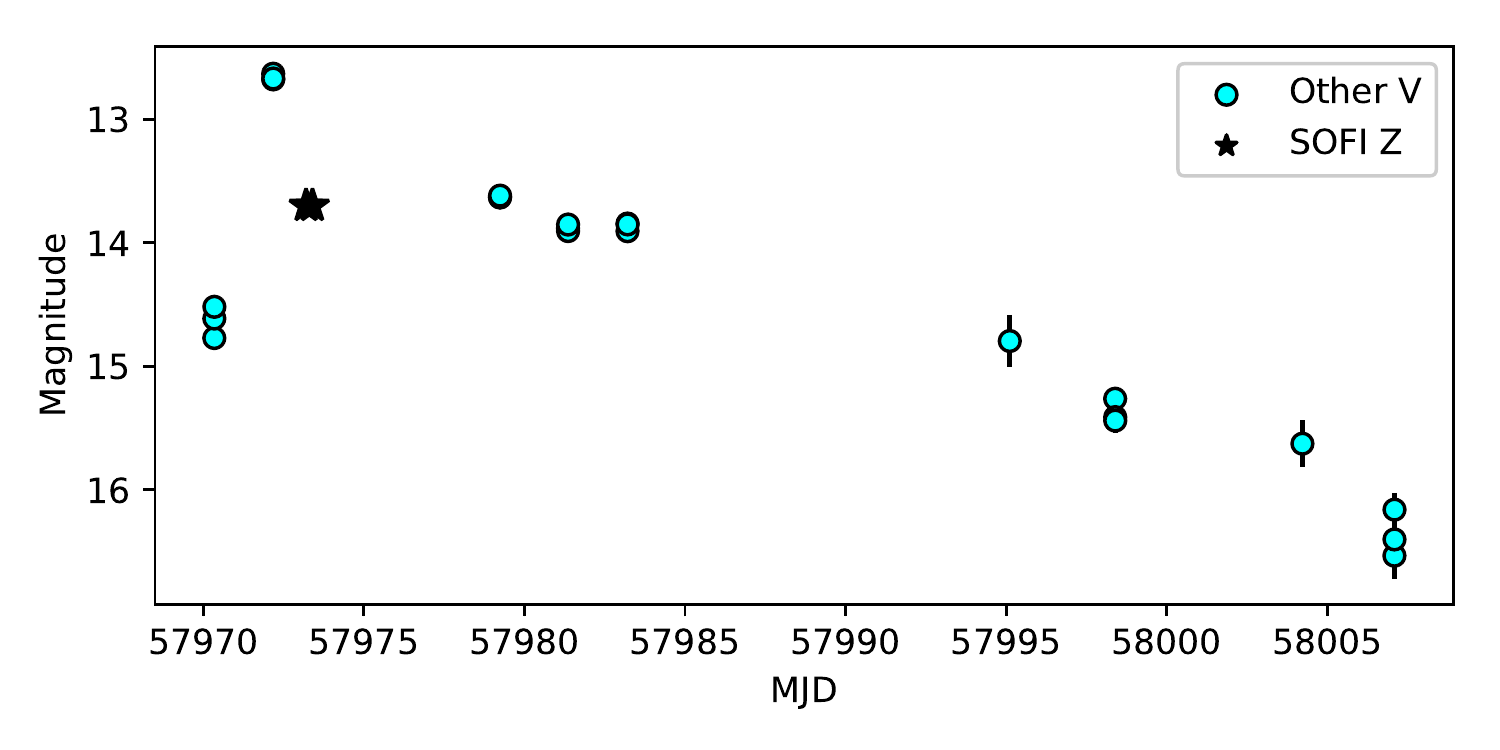}
\caption{Lightcurve of ASASSN-17km. The V band data points (cyan) correspond to ASAS-SN observations of the source. We appear to have observed the source within $\sim 1-2$\,days of the peak.}
\label{fig:ASASSN17km_lc}
\end{figure}

\subsection{AT2016bvg}
AT2016bvg (also known as CTRS 160505-150133 and PS16bux) was discovered on 2016 April 16 \citep{Chambers2016iv}. We observed the source twice on 2016 June 19 and 20 - just over two months post-alert. We measured a polarisation limit of $P < 3.91\%$ and a value $P = 1.73(\pm0.28)\%$ in \textit{V} band on the first and second nights respectively.

\subsection{AT2016cvk}
AT2016cvk (also known as ASASSN-16jt and SN2016cvk) was discovered on 2016 June 12 \citep{Brimacombe2016iii}. AT2016cvk and ASASSN-16jt were initially detected several days apart but further observations concluded that the source was most probably a type IIn SN and that these two source positions were coincident \citep{Brown2016ii}. We observed the source approximately one week post-alert and measured a polarisation of $P \leq 1.90\%$ in \textit{V} band. \cite{Brown2016ii} show that this object resembles the rare unusual transient SN2009ip, a SN which has shown prior violent outbursts. Our limit on the polarisation is consistent with those obtained for SN 2009ip (\citealt{Mauerhan2014}).

\subsection{ATLAS16bcm}
ATLAS16bcm (also known as AT2016csr and SN2016crs) was discovered in SDSS J151431.52+064123.9 on 2016 June 3 \citep{Tonry2016i}. The source was classified as a type Ia SN via follow-up spectroscopic observations \citep{Hangard2016}. We observed the source 12 days post-alert and measured a polarisation of $P < 0.91\%$ in \textit{V} band.

\subsection{ATLAS16bdg}
ATLAS16bdg (also known as AT2016cvn and SN2016cvn) was discovered in NGC4708 on 2016 June 5 in NGC4708 on June $5^{th}$ 2016 \citep{Tonry2016ii}. It was classified as a type Ia SN via spectroscopic observations \citep{Mundell2016}. We observed the source two weeks post-alert and measured a polarisation of $P = 2.12(\pm0.22)\%$, $P = 3.55(\pm0.59)\%$ and $P = 0.97(\pm0.19)\%$ for the \textit{V}, \textit{B} and \textit{R} bands respectively. We calculate $E(B-V) = 0.04$ at the source position corresponding to $P_{\rm Gal, dust} \leq 0.36\%$.

\subsection{ATLAS17jfk}
ATLAS17jfk (also known as AT2017fvz and kait-17bm) was discovered on 2017 August 2 \citep{Hestenes2017} and later classified as a extragalactic Novae in NGC 6822 via follow-up spectroscopic observations \citep{Williams2017}. We observed the source approximately six days post-alert and measured a polarisation of $P=2.30(\pm0.57)\%$ in \textit{Z} band. We calculate $E(B-V) \sim 0.20$ at the source position corresponding to $P_{\rm Gal, dust} \leq 1.80\%$. 

\subsection{CTA 102}
CTA 102 was first discovered at radio wavelengths in 1960 \citep{Harris1960} and was subsequently classified as a Quasar via follow-up optical observations \citep{Sandage1965}. We observed the source twice during our observing runs. The first was in response to increased optical activity reported on 2016 June 9 \citep{Larionov2016} and we observed the source 11 days later and measured a polarisation of $P = 22.53(\pm0.14)\%$ in \textit{V} band. The second was in response to report of an increase of gamma-ray emission detected on 2017 July 7 \citep{Bulgarelli2017}. We observed the source approximately one month after the report and measured a polarisation of $P = 6.58(\pm0.45)\%$ in \textit{Z} band. We calculate $E(B-V) = 0.06$ at the source position corresponding to $P_{\rm Gal, dust} \leq 0.54\%$. The polarisation measurements are both consistent with non-thermal emission. Figure \ref{fig:cta102_plot} shows that this source exhibits significant variability - both in brightness and polarisation over very short timescales.

\begin{figure}
\includegraphics[width=8.5cm]{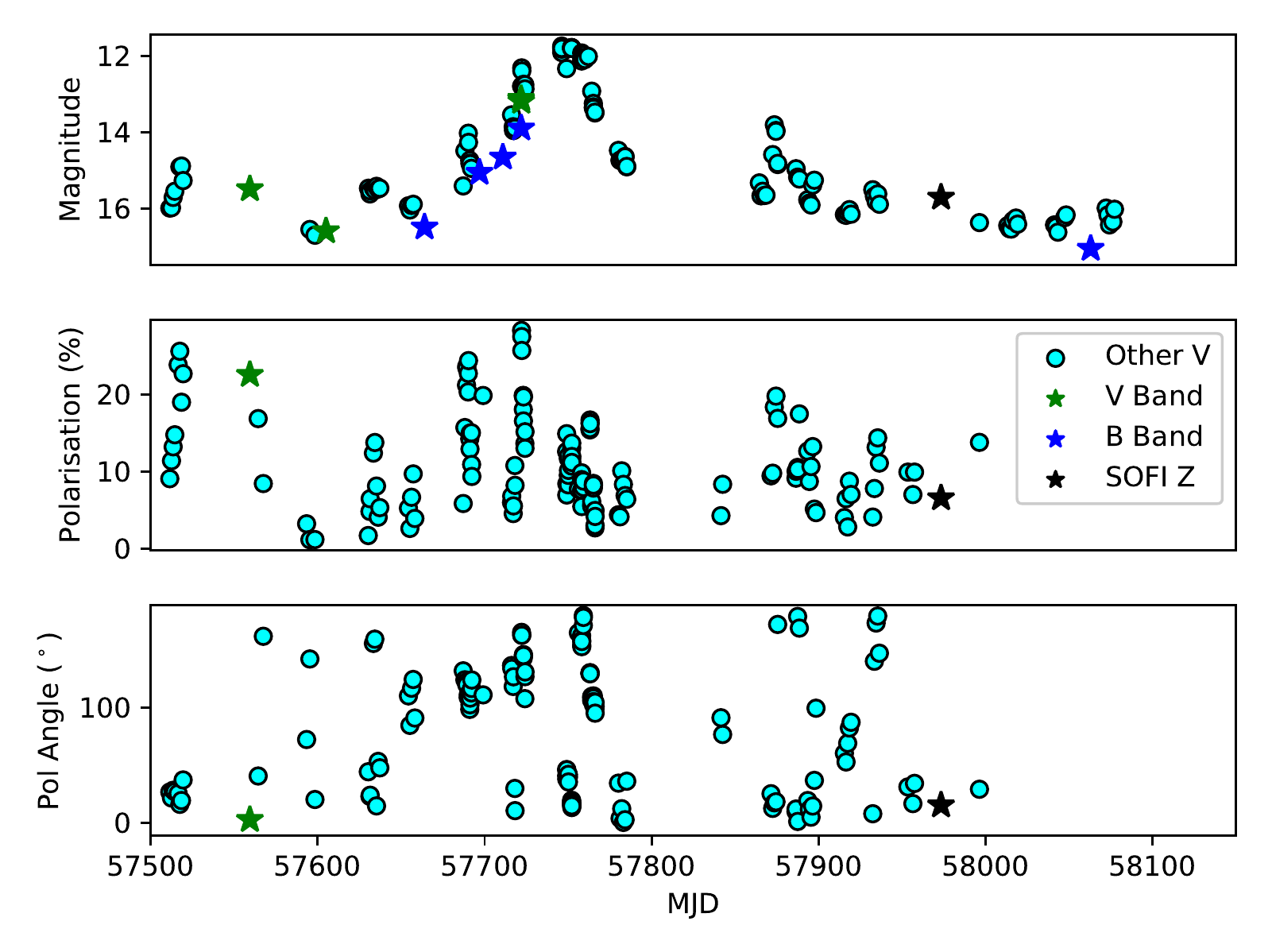}
\caption{Plot of CTA 102 showing the temporal evolution of source brightness, degree of polarisation and polarisation angle. The 'Other V' legend label refers to data taken from the Steward Observatory spectropolarimetric monitoring project.}
\label{fig:cta102_plot}
\end{figure}

\subsection{Gaia16aau}
Gaia16aau (also known as AT2016dbu and OGLE-SMC 710.08.1) was detected on 2016 January 25 \citep{Delgado2016ii} and classified as a RCB star within the SMC \citep{Tisserand2016}. We observed the source approximately five months post detection and measured a polarisation of $P=0.22(\pm0.05)\%$ in \textit{V} band. Since the first {\it Gaia} detection in 2014 November, the source has increased by over 5\,mags in brightness (see Figure \ref{fig:Gaia16aau_lc}). {\it Gaia} reports the source has a proper motion of $0.47(\pm0.09)$ and $-1.33\pm(0.07)$\,mas yr$^{-1}$ in RA and Dec, respectively.

\begin{figure}
\includegraphics[width=8.5cm]{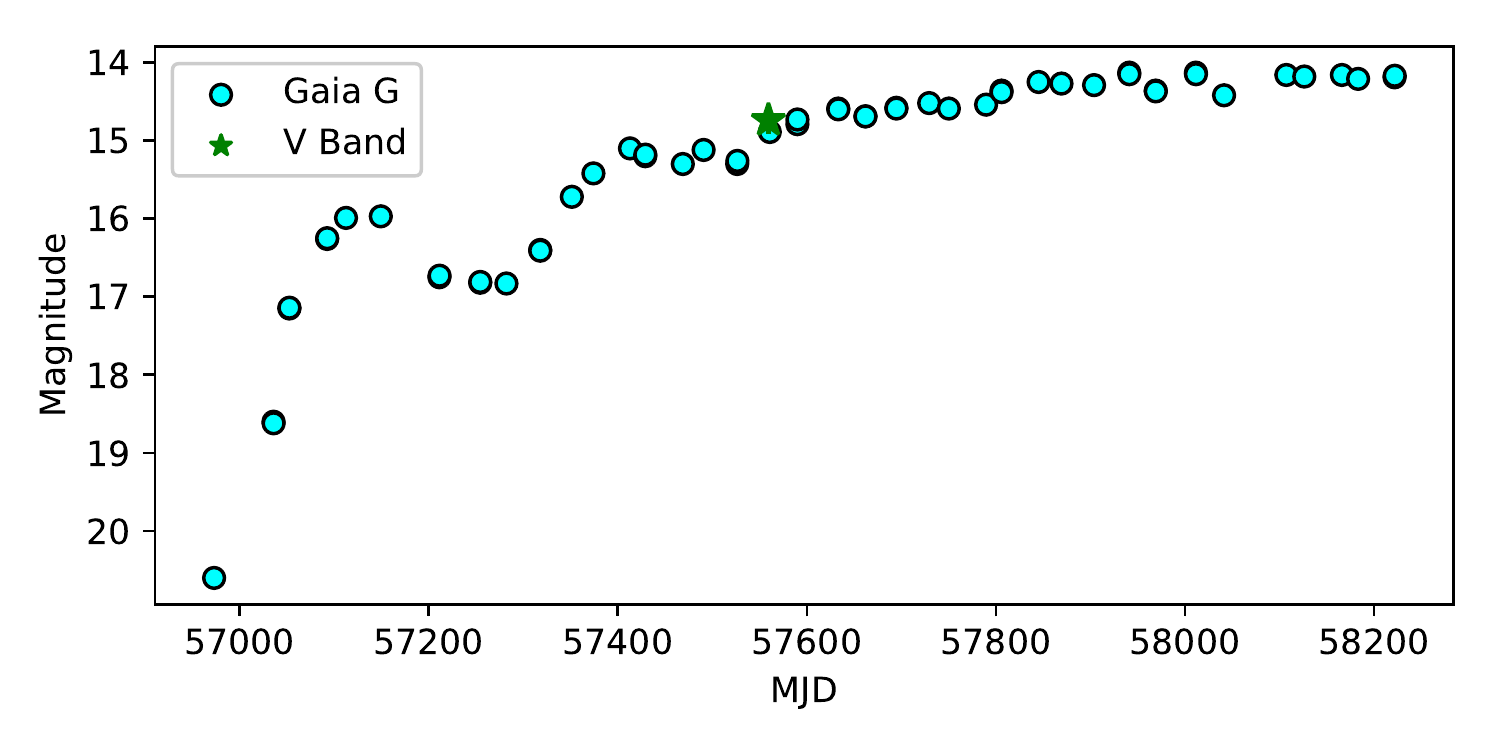}
\caption{Lightcurve of Gaia16aau. Apart from a brief period in 2015 the source has been steadily rising in brightness since the first detection back in 2014.}
\label{fig:Gaia16aau_lc}
\end{figure}

\subsection{Gaia16agw}
Gaia16agw (also known as AT2016dth) was detected on 2016 February 29 \citep{Delgado2016}. The source was coincident with a previous detection from ASASSN (ASASSN15mw; \citealt{Shappee2014}). Figure \ref{fig:Gaia16agw_lc} shows the source exhibits periods of high variability variability since its most historic {\it Gaia} detection in 2015 - changes on timescales of months by $1-2$\,mag. It therefore was classified as a Blazar candidate. We observed the source approximately 16 weeks post-{\it Gaia} detection, as the source appeared to enter a period of higher activity and measured a polarisation of $P \leq 0.36\%$ and a brightness of $17.58(\pm0.01)$ mag in \textit{V} band.

\begin{figure}
\includegraphics[width=8.5cm]{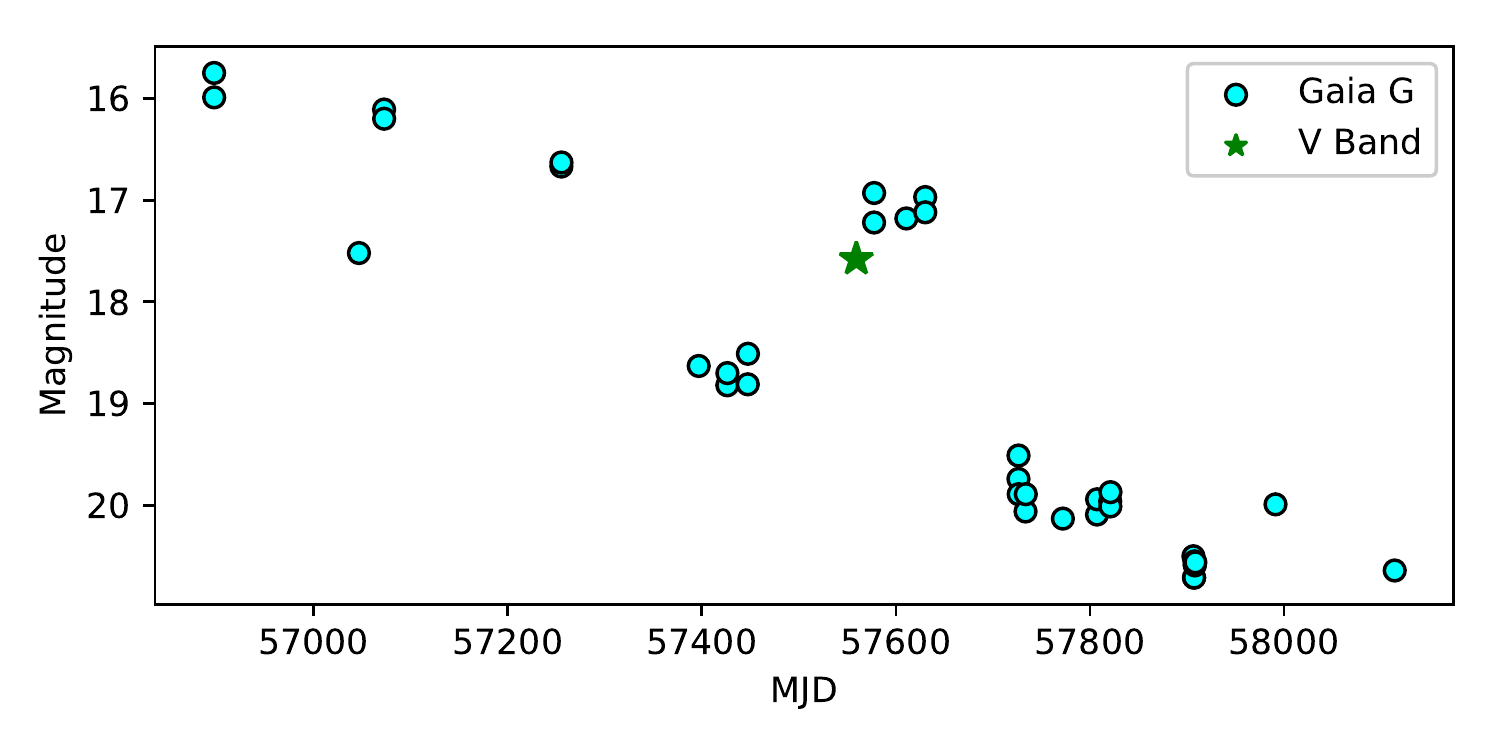}
\caption{Lightcurve of Gaia16agw. From the lightcurve, we appear to have observed the source as it was entering a period of higher activity.}
\label{fig:Gaia16agw_lc}
\end{figure}

\subsection{Gaia16alw}
Gaia16alw (also known as AT2016dxp) was detected on 2016 April 19 \citep{Delgado2016}. We observed the source approximately two months post detection and measured a polarisation of $P = 5.48(\pm1.20)\%$ in \textit{V} band. We measured a brightness of $19.65(\pm0.06)\%$ in \textit{V} band. See Figure \ref{fig:Gaia16alw_lc} for light curve.

\begin{figure}
\includegraphics[width=8.5cm]{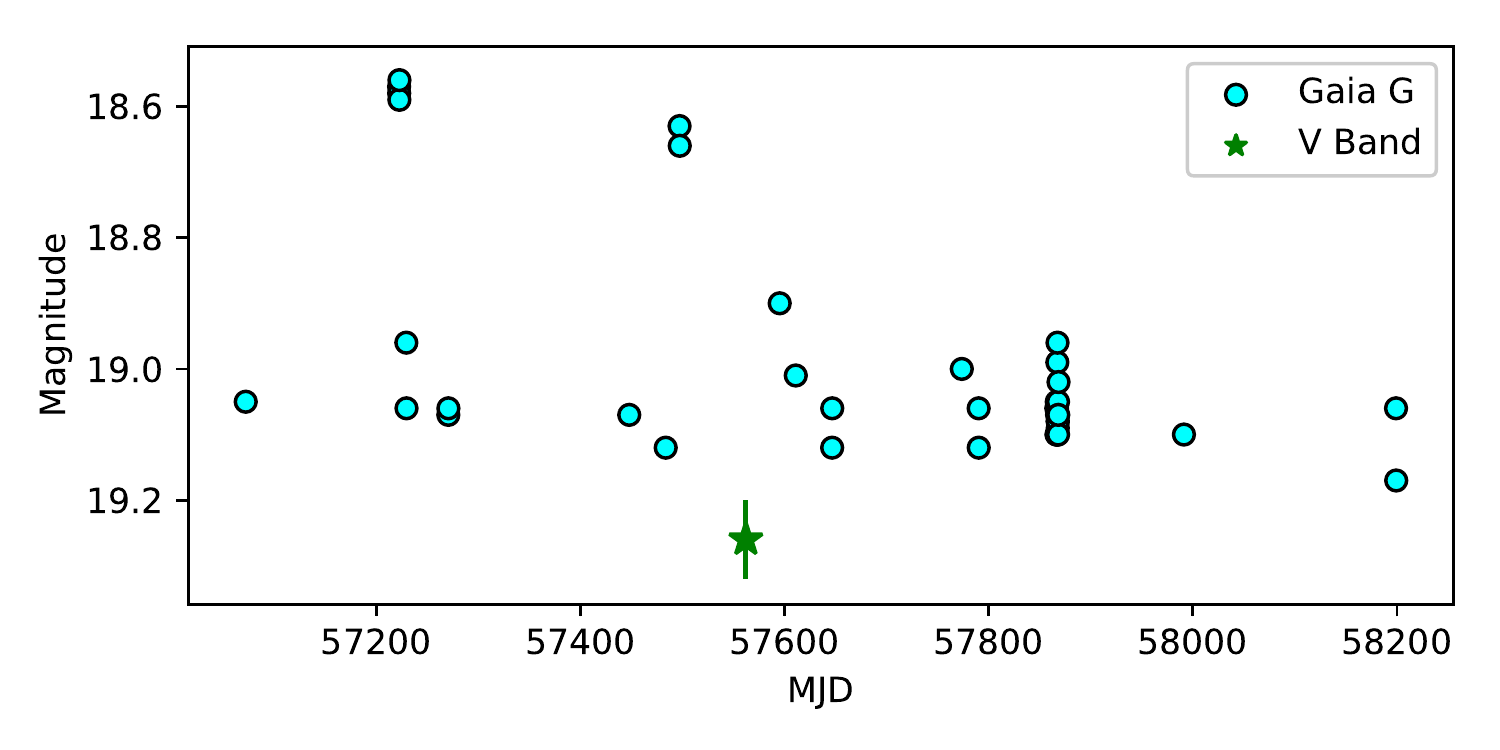}
\caption{Lightcurve of Gaia16alw. We observed the source during a period of low activity.}
\label{fig:Gaia16alw_lc}
\end{figure}

\subsection{Gaia16aoa}
Gaia16aoa (also known as AT2016eab) was detected on 2016 May 9 \citep{Delgado2016}. The alert was in response to an increase in brightness from a faint Digitized Sky Survey (DSS2) source. We observed the source approximately six weeks post-alert, coinciding with a period of high activity (see Figure \ref{fig:Gaia16aoa_lc}) and measured a polarisation of $P = 1.59(\pm0.38)\%$ in \textit{V} band. We measured a brightness of $19.17(\pm0.03)$ in \textit{V} band.

\begin{figure}
\includegraphics[width=8.5cm]{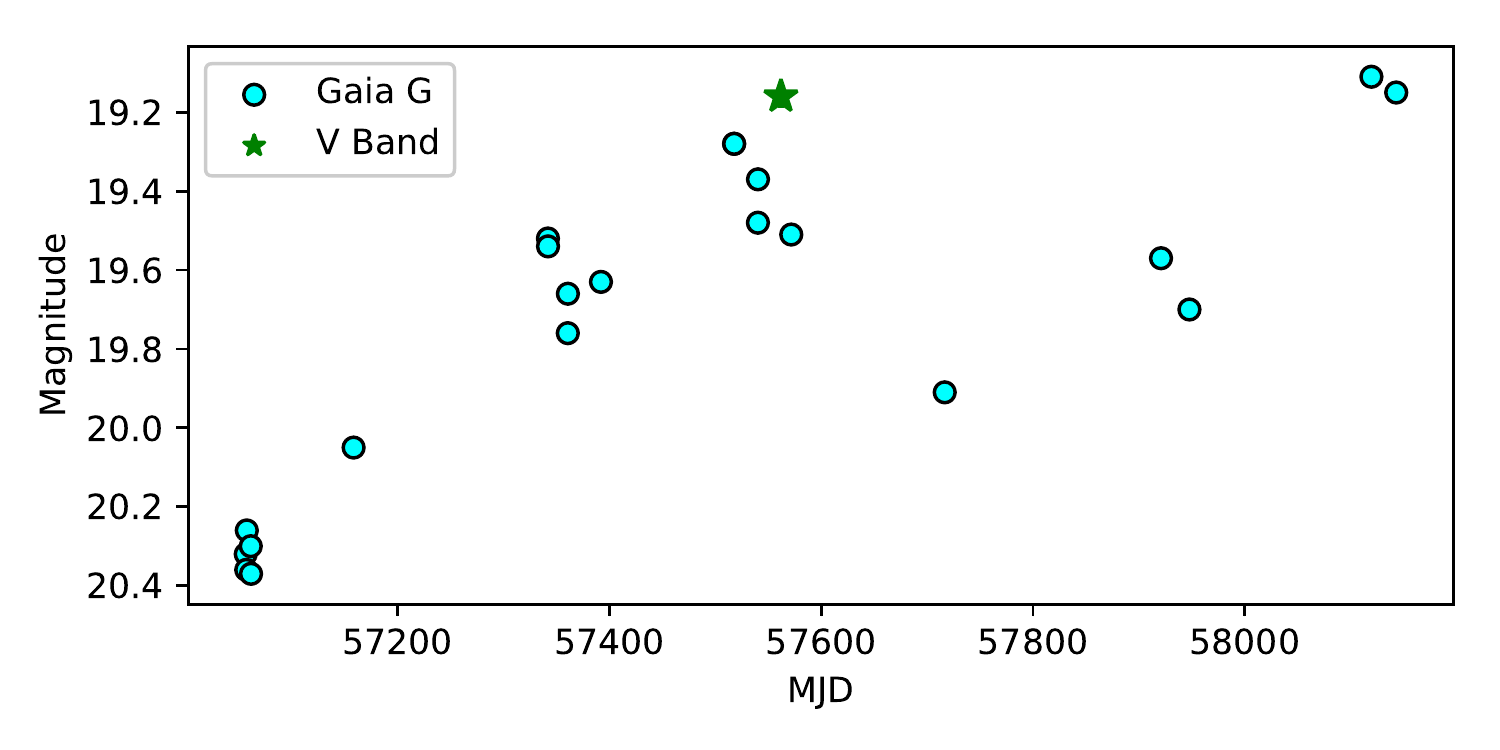}
\caption{Lightcurve of Gaia16aoa. We observed the source during a period of high activity.}
\label{fig:Gaia16aoa_lc}
\end{figure}

\subsection{Gaia16aob}
Gaia16aob (also known as AT2016eaa) was detected on 2016 May 10 as a brightening of candidate AGN 2MASX J11431053-2946384 \citep{Delgado2016}. We observed the source approximately six weeks post-alert, during a period of high variability (see Figure \ref{fig:Gaia16aob_lc}) and measured a polarisation of $P = 0.37(\pm0.12)\%$ and a brightness of $17.13(\pm0.01)$ mag in \textit{V} band. We calculate $E(B-V) = 0.06$ at the source position corresponding to $P_{\rm Gal, dust} \leq 0.54\%$. The polarisation measurement is therefore consistent with an intrinsically unpolarised source. 

\begin{figure}
\includegraphics[width=8.5cm]{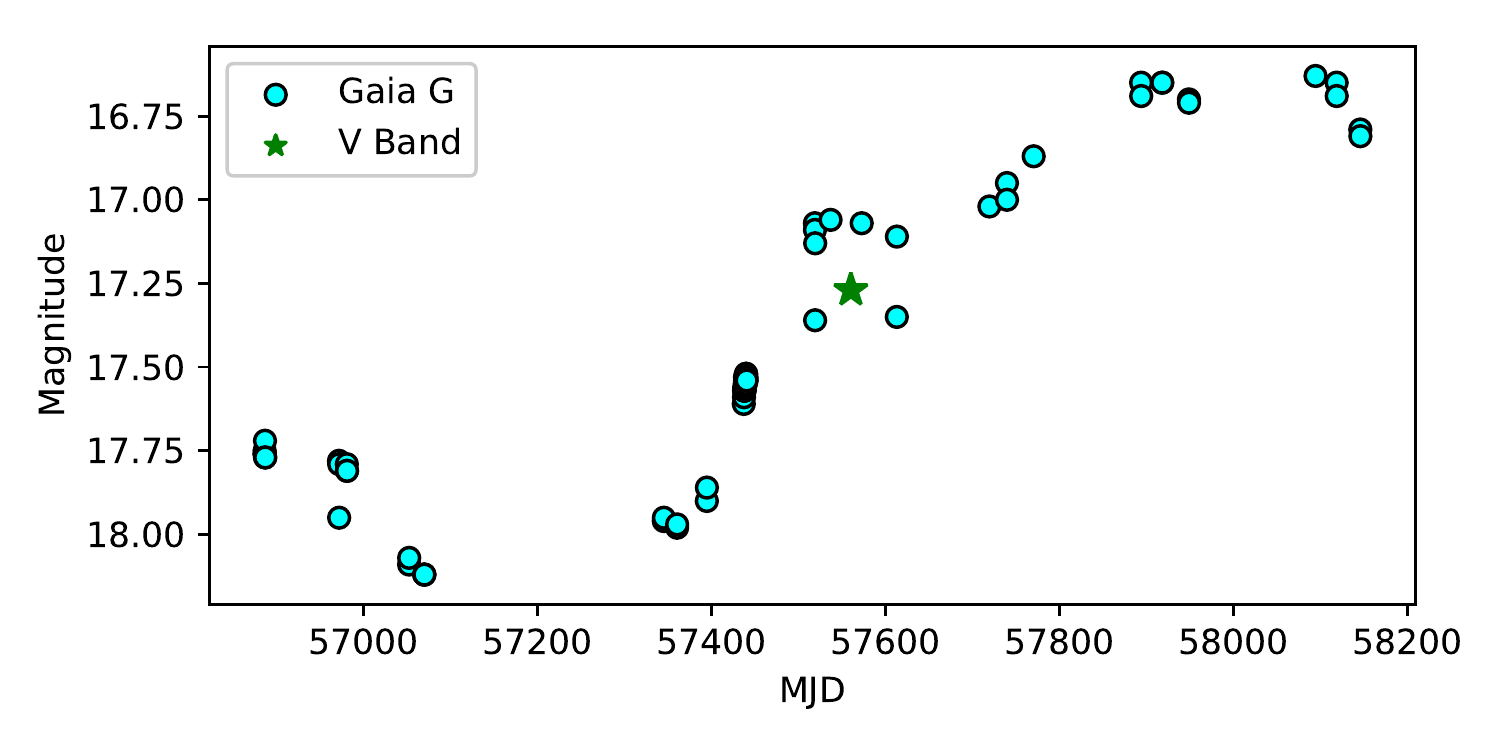}
\caption{Lightcurve of Gaia16aob. From the lightcurve, we appear to have observed the source during a period of high activity.}
\label{fig:Gaia16aob_lc}
\end{figure}

\subsection{Gaia16aok}
Gaia16aok (also known as AT2016eap) was detected on 2016 May 12 \citep{Delgado2016}. The alert was in response to an increase in brightness of radio source NVSS J115815-314702. We observed the source approximately five weeks post-alert, during a period of short-time, high level variability (see Figure \ref{fig:Gaia16aok_lc} and measured a polarisation of $P = 11.51(\pm0.07)\%$ and a brightness of 19.83($\pm0.11$)\,mag in \textit{V} band.

\begin{figure}
\includegraphics[width=8.5cm]{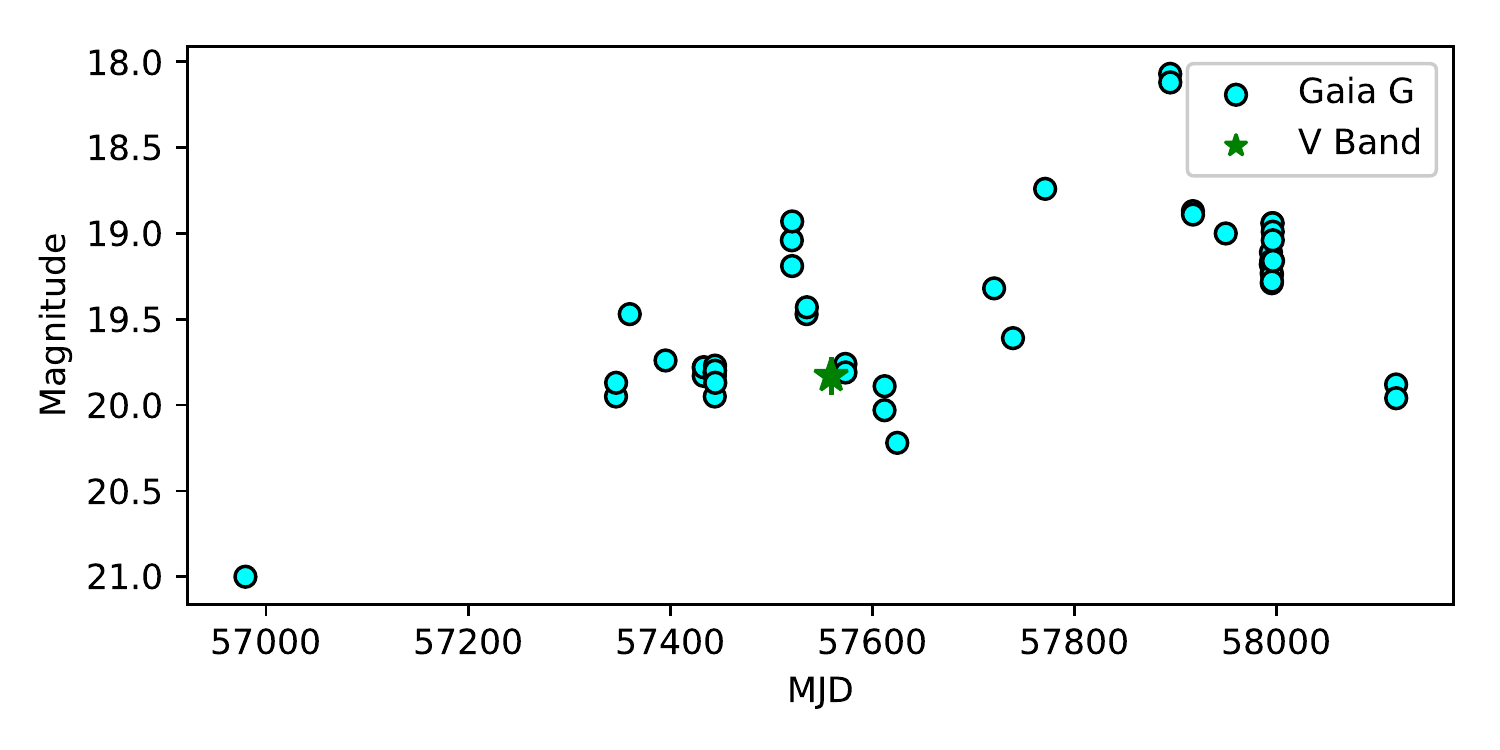}
\caption{Lightcurve of Gaia16aok. We observed the source during a period of high variability.}
\label{fig:Gaia16aok_lc}
\end{figure}

\subsection{Gaia16aol}
Gaia16aol (also known as AT2016eaq and PS16cni) was detected on 2016 May 12 \citep{Delgado2016}. The source was located in the galaxy IC 690 and tentatively classified as a SN candidate. We observed the source approximately five weeks post-alert and measured a polarisation of $P \leq 4.08\%$ in \textit{V} band. Figure \ref{fig:Gaia16aol_lc} suggests we may have observed within a week or so of peak brightness. Unfortunately we could not obtain a magnitude for our observation.

\begin{figure}
\includegraphics[width=8.5cm]{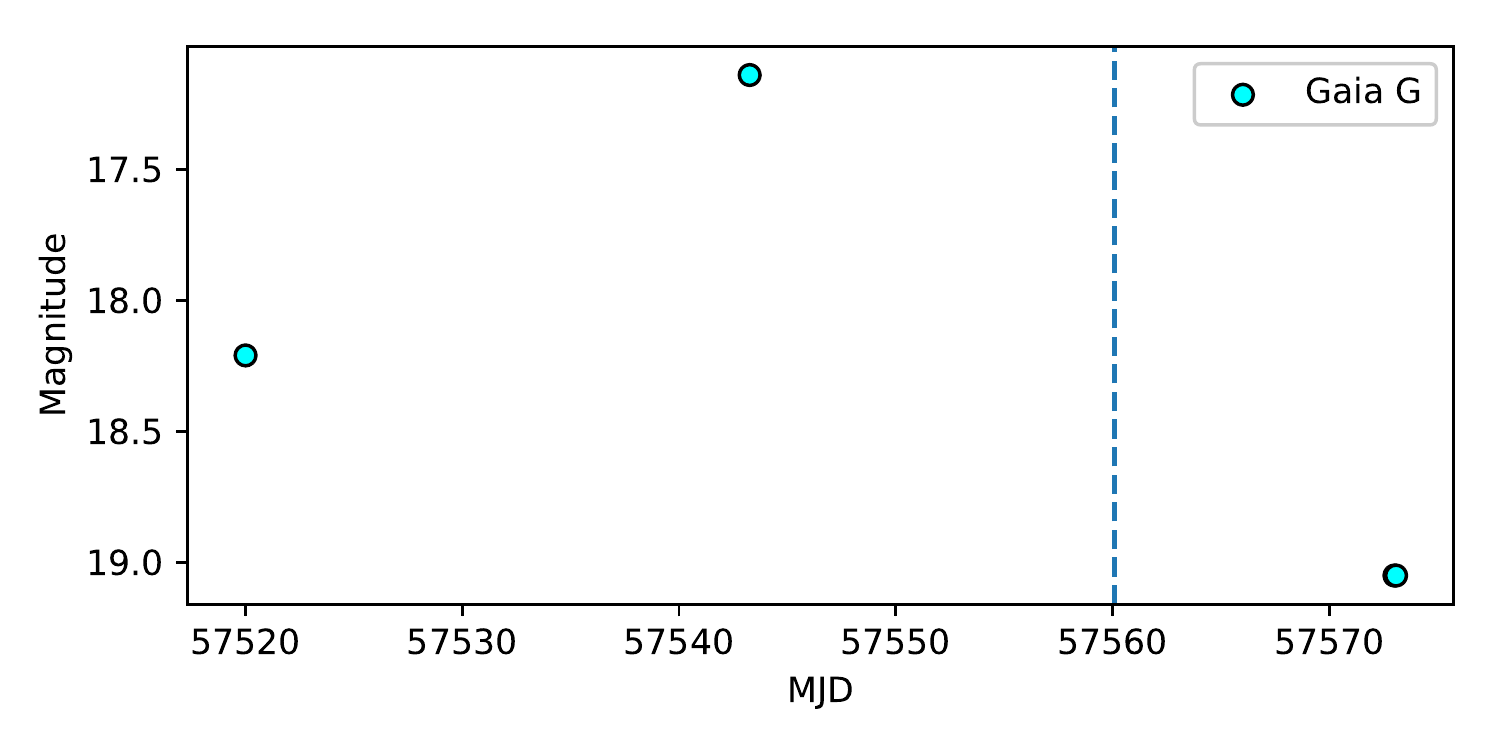}
\caption{Lightcurve of Gaia16aol. The dashed line indicates the date when we observed the source.}
\label{fig:Gaia16aol_lc}
\end{figure}

\subsection{Gaia16aoo}
Gaia16aoo (also known as ASASSN-16dm, AT2016blb, PS16bop and SN2016blb) was discovered in 2MASX J11372059-0454450 on 2016 March 30 \citep{kiyota2016}. The source was classified as a type IIP SNe via spectroscopic follow-up observations \citep{Falco2016}. We observed the source five weeks post-alert, as it was declining in brightness (see Figure \ref{fig:Gaia16aoo_lc}) and measured a polarisation of $P\leq 2.21\%$ in \textit{V} band.

\begin{figure}
\includegraphics[width=8.5cm]{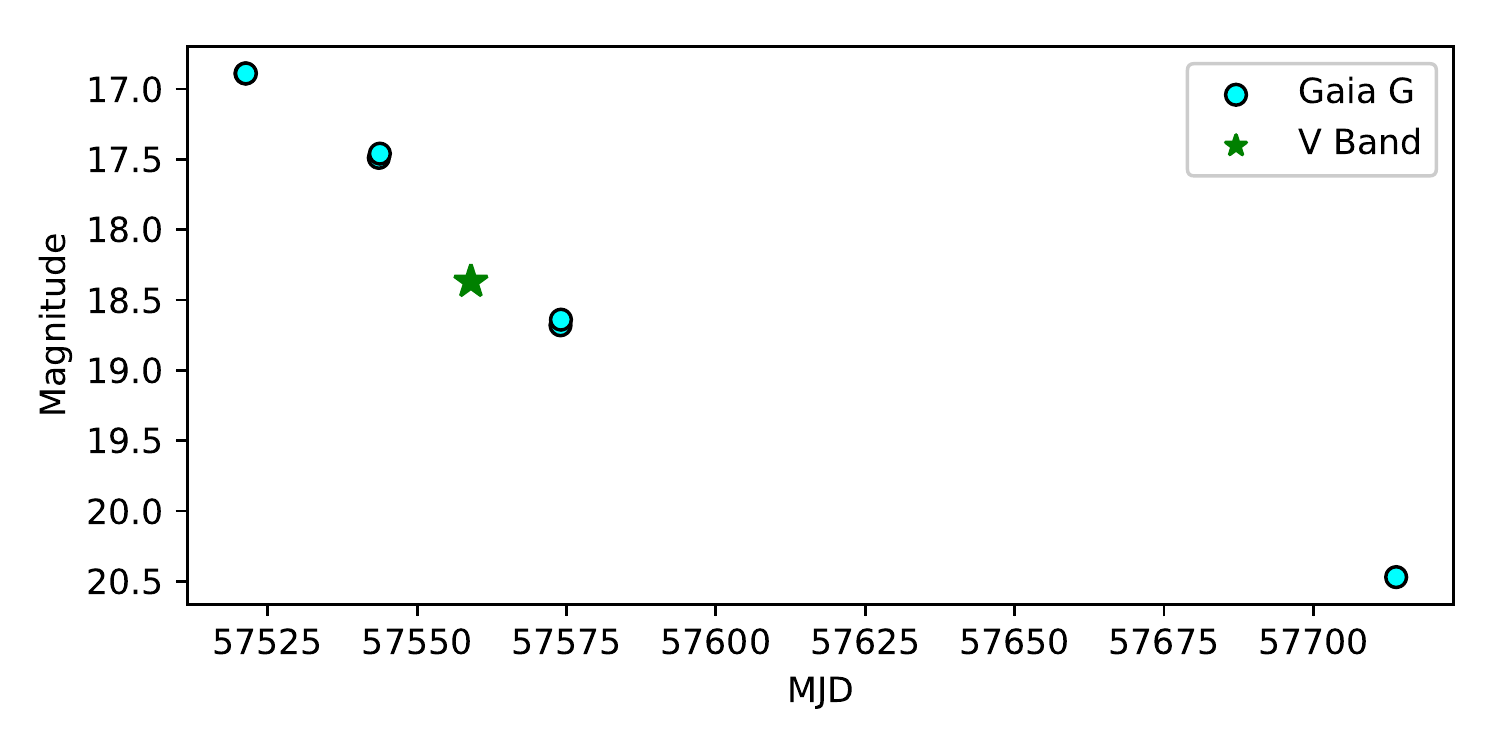}
\caption{Lightcurve of Gaia16aoo. We observed the source post-peak as it was declining brightness.}
\label{fig:Gaia16aoo_lc}
\end{figure}

\subsection{Gaia16aqe}
Gaia16aqe (also known as AT2017fqg) was detected on 2016 May 22 \citep{Delgado2016}. It was tentatively classified as a type Ia SNe. We observed the source approximately four weeks post-alert and measured a polarisation of $P\leq 2.07\%$ in \textit{V} band. Figure \ref{fig:Gaia16aqe_lc} suggests we observed the source as it was fading. Unfortunately we could not obtain a magnitude for our observation.

\begin{figure}
\includegraphics[width=8.5cm]{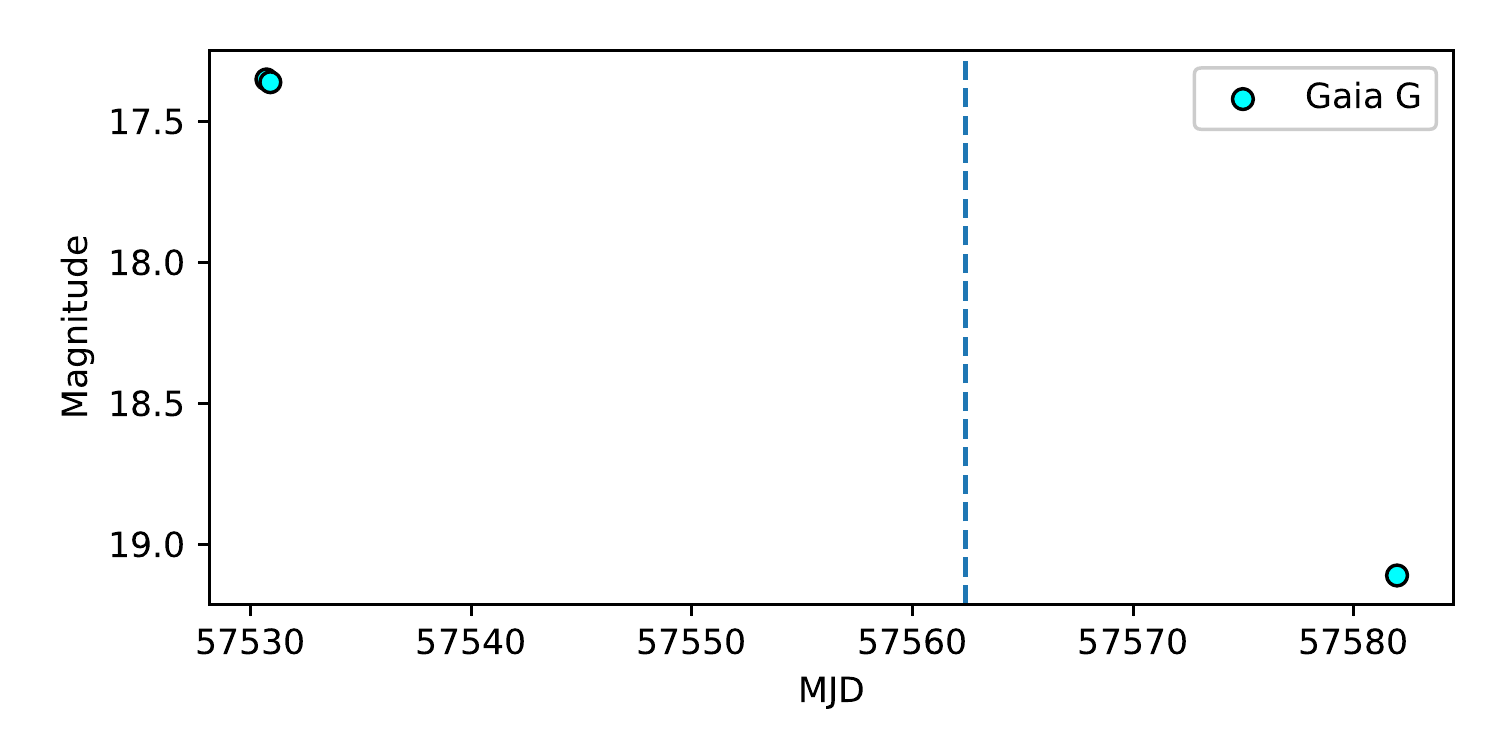}
\caption{Lightcurve of Gaia16aqe. The dashed line indicates the date when we observed the source.}
\label{fig:Gaia16aqe_lc}
\end{figure}

\subsection{Gaia17blw}
Gaia17blw (also known as AT2017eni and SN2017eni) was discovered on 2017 June 5 \citep{Delgado2017iv}. It was classified as a type IIn SN via follow-up spectroscopic observations \citep{Strader2017}. We observed the source approximately two months post-alert and measured a polarisation of $P \leq 1.65\%$ and a brightness of $\sim16$\,mag in \textit{Z} band. See Figure \ref{fig:Gaia17blw_lc} for light curve.

\begin{figure}
\includegraphics[width=8.5cm]{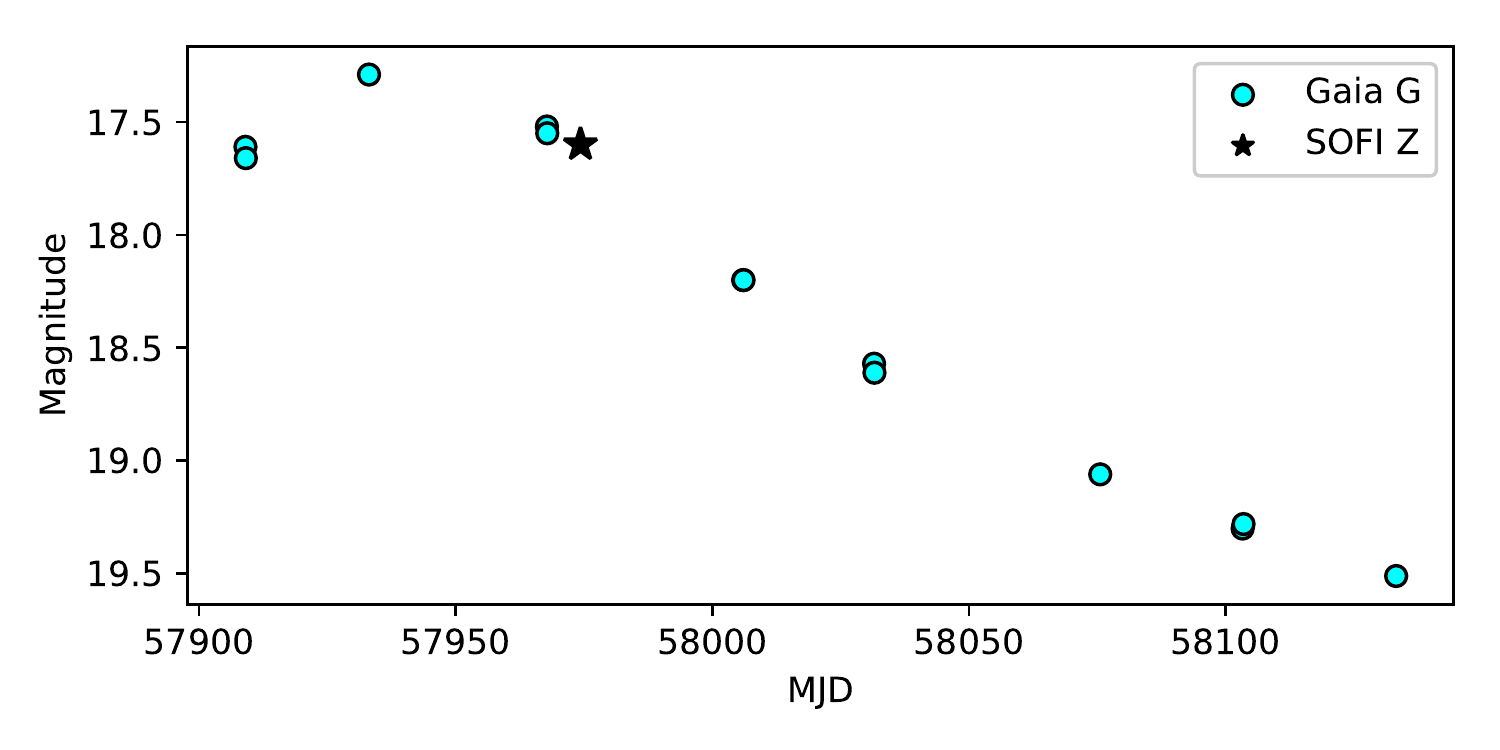}
\caption{Lightcurve of Gaia17blw. We appear to observed the source post maximum.}
\label{fig:Gaia17blw_lc}
\end{figure}

\subsection{Gaia17bro}
Gaia17bro (also known as AT2017fck and SN2017fck) was discovered on 2017 July 2 \citep{Delgado2017i}. The source was classified as a type IIn SN via follow-up spectroscopic observations \citep{Strader2017}. We observed the source five weeks post-alert, as it was declining in brightness (see Figure \ref{fig:Gaia17bro_lc}) and measured a polarisation of $P\leq 1.99\%$ and a brightness of $\sim16.2$\,mag in \textit{Z} band.

\begin{figure}
\includegraphics[width=8.5cm]{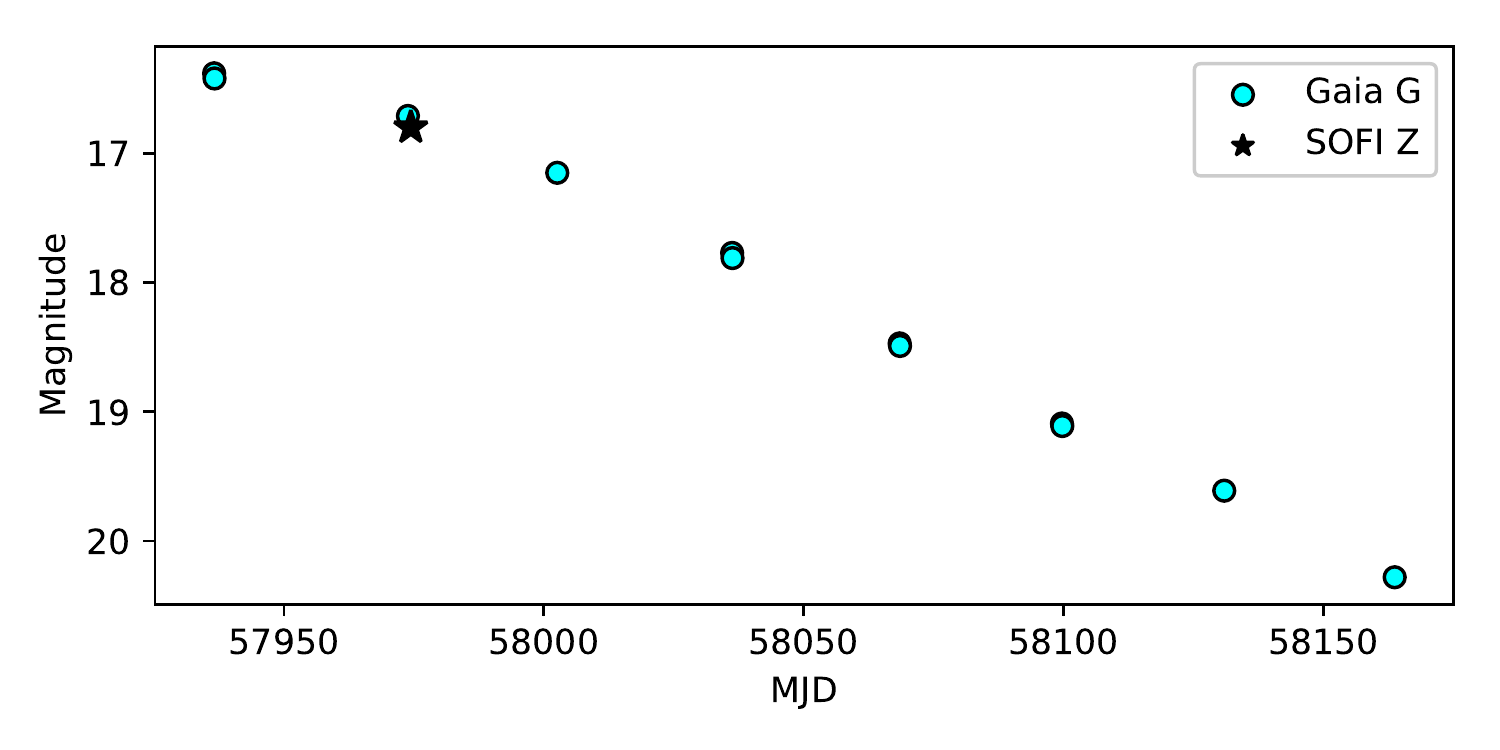}
\caption{Lightcurve of Gaia17bro. We appear to have observed the source a couple of weeks post-peak brightness.}
\label{fig:Gaia17bro_lc}
\end{figure}

\subsection{Gaia17bvo}
Gaia17bvo (also known as AT2017fqg) was detected on 2017 July 23 \citep{Delgado2017ii}. The alert was in response to an increase in brightness ($\sim0.5$\,mag) from a previously known variable source residing in the Galactic plane, tentatively classifying the source as a candidate YSO. We observed the source approximately 17 days post-alert and measured a polarisation of $P = 8.37(\pm0.25)\%$ in \textit{Z} band. Figure \ref{fig:Gaia17bvo_lc} suggests that we observed the source during an outburst phase.

\begin{figure}
\includegraphics[width=8.5cm]{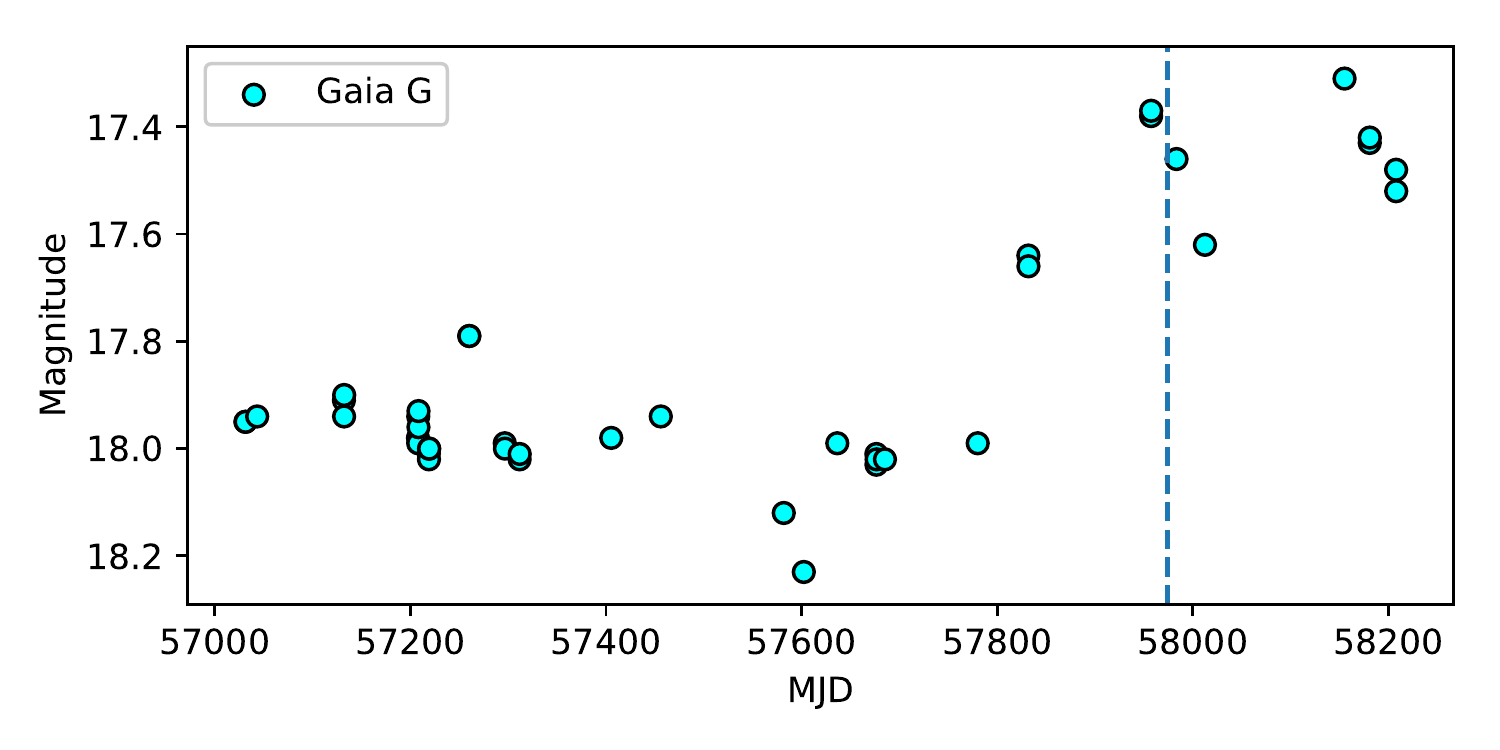}
\caption{Lightcurve of Gaia17bvo. The blue dashed line indicates the date when we observed the source.}
\label{fig:Gaia17bvo_lc}
\end{figure}

\subsection{Gaia17bwu}
Gaia17bwu (also known as AT2017fum) was detected on 2017 July 27 \citep{Delgado2017v}. The alert was in response to an increase in brightness ($\sim 1$ mag) from a previously observed red star and exhibited strong emission lines. We observed the source 12 days post-alert and measured a polarisation of $P = 1.16(\pm0.30)\%$ in \textit{Z} band. Figure \ref{fig:Gaia17bwu_lc} suggests we observed the source during a period of very high activity. Unfortunately we were unable to obtain a magnitude for our observation.

\begin{figure}
\includegraphics[width=8.5cm]{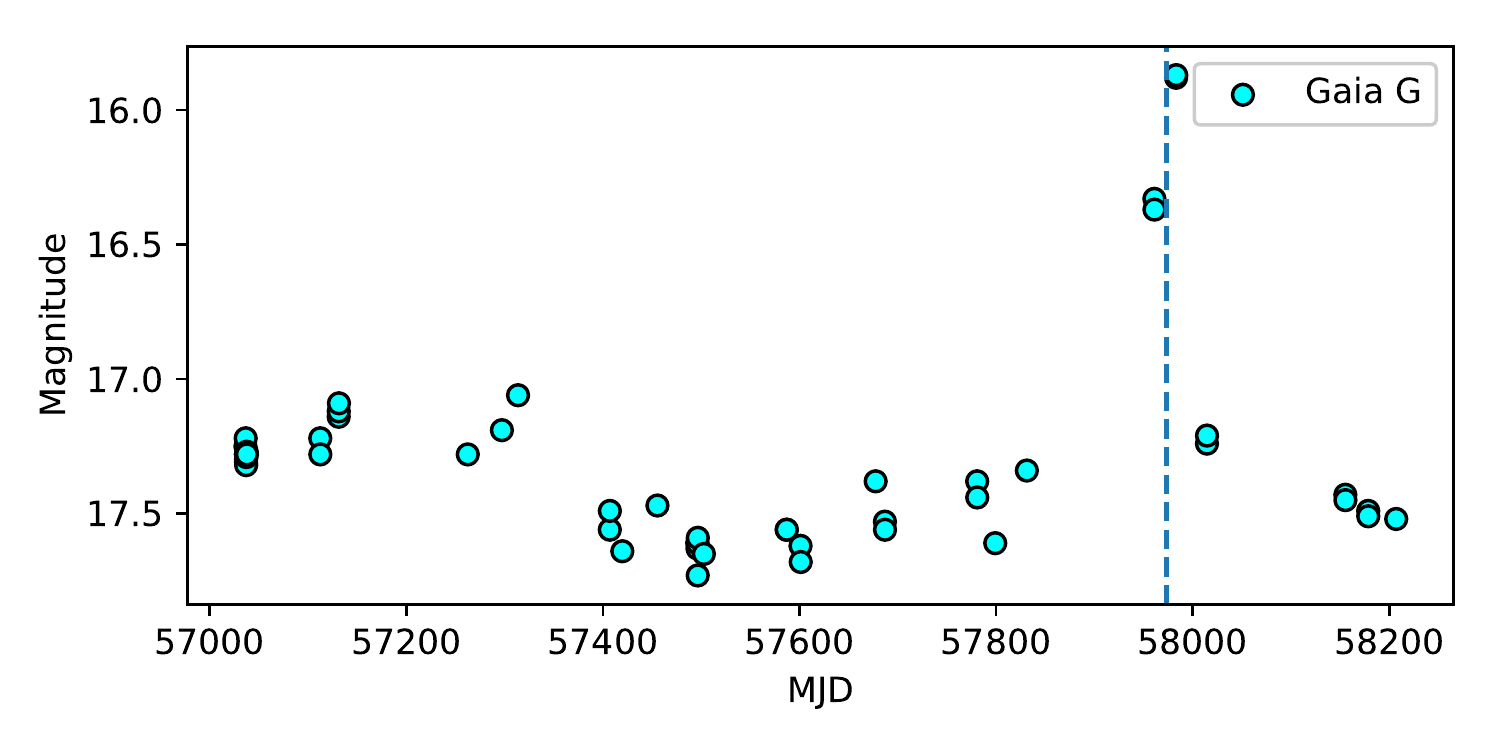}
\caption{Lightcurve of Gaia17bwu. The blue dashed line indicates the date when we observed the source.}
\label{fig:Gaia17bwu_lc}
\end{figure}

\subsection{Gaia17bxl}
Gaia17bxl (also known as AT2017fve) was detected on 2017 July 29 \citep{Delgado2017vi}. The source was tentatively classified as a SNe located close to the galaxy GALEXASC J012208.86-484752.8. We observed the source nine days post-alert and measured a polarisation of $P \leq 10.45\%$ in \textit{Z} band. We measured a brightness of $\sim19.9$\,mag in \textit{Z} band. Figure \ref{fig:Gaia17bxl_lc} clearly shows we observed the source as it had faded from peak brightness.

\begin{figure}
\includegraphics[width=8.5cm]{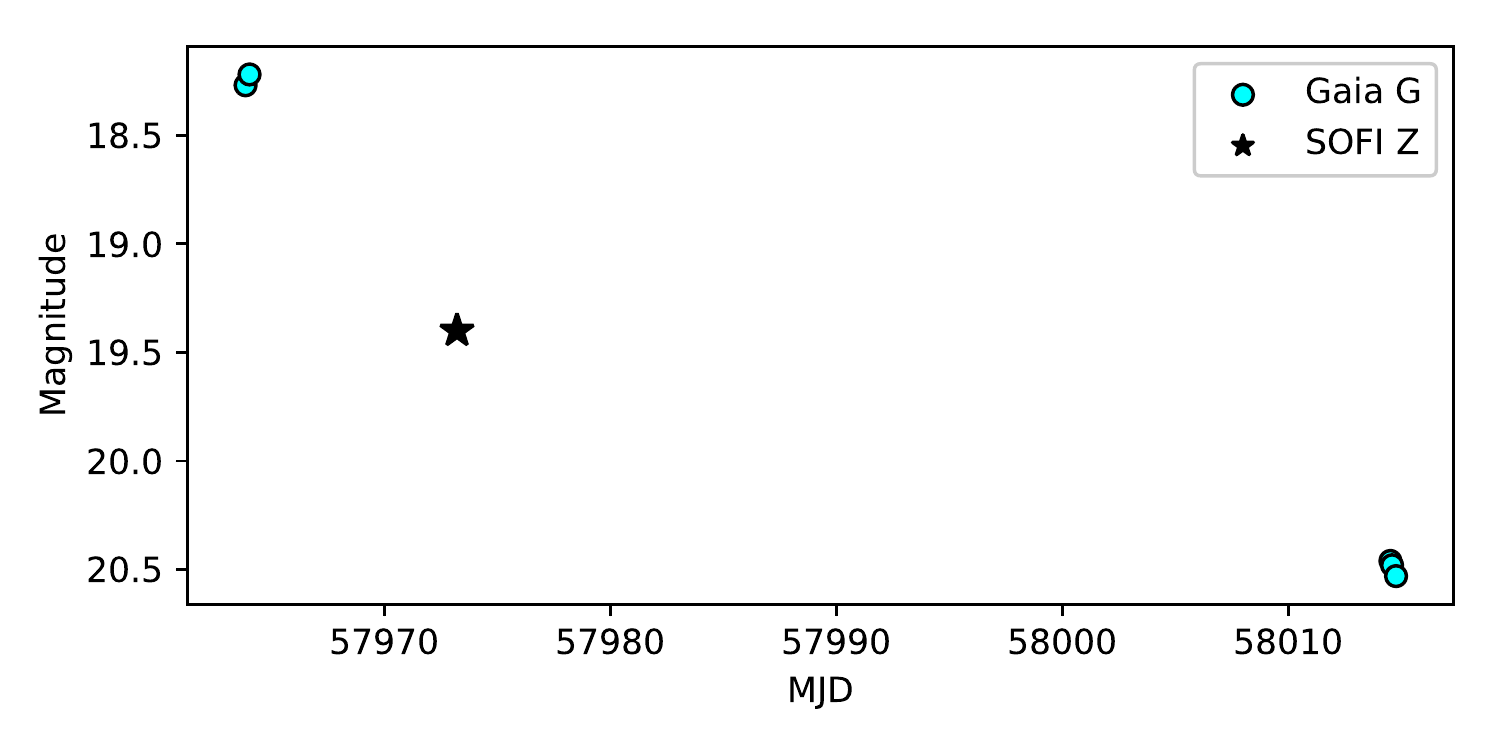}
\caption{Lightcurve of Gaia17bxl. We observed the source as it was fading.}
\label{fig:Gaia17bxl_lc}
\end{figure}

\subsection{Gaia17byh}
Gaia17byh (also known as AT2017fwm and SN2017fwm) was discovered on 2017 July 31 \citep{Delgado2017iii}, and shows a well resolved rise to maximum in the {\it Gaia} data. The source was classified as a type Ic SN via follow-up spectroscopic observations \citep{Lyman2017}. We observed the source one week post-alert and measured a polarisation of $P \leq 2.22\%$ and a brightness of $\sim17.2$\,mag in \textit{Z} band. The {\it Gaia} lightcurve presented in Figure \ref{fig:Gaia17byh_lc} suggests we observed the source at peak brightness.

\begin{figure}
\includegraphics[width=8.5cm]{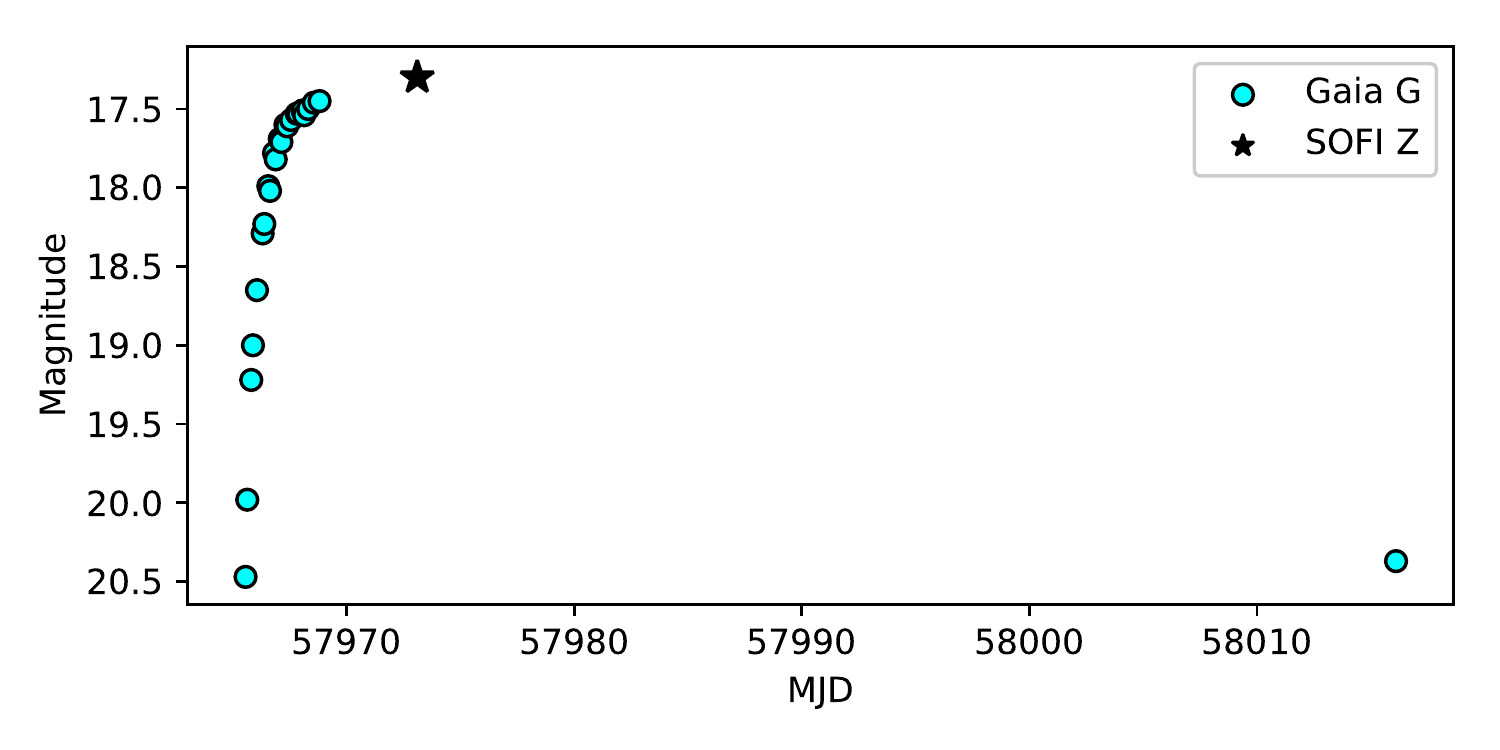}
\caption{Lightcurve of Gaia17byh, this source has a very well sampled rise by {\it Gaia}. We observed the source at peak brightness.}
\label{fig:Gaia17byh_lc}
\end{figure}

\subsection{Gaia17byk}
Gaia17byk (also known as AT2017fwt) was detected on 2017 August 1 \citep{Delgado2017iii}. The alert was in response to an increase in brightness ($\sim0.5$\,mag) of a source residing in the Galactic plane. We observed the source eight days post-alert and measured a polarisation of $P = 5.99(\pm0.49)\%$ in \textit{Z} band. Unfortunately we were unable to obtain a magnitude measurement for our observation but Figure \ref{fig:Gaia17byk_lc} highlights our time of observation.

\begin{figure}
\includegraphics[width=8.5cm]{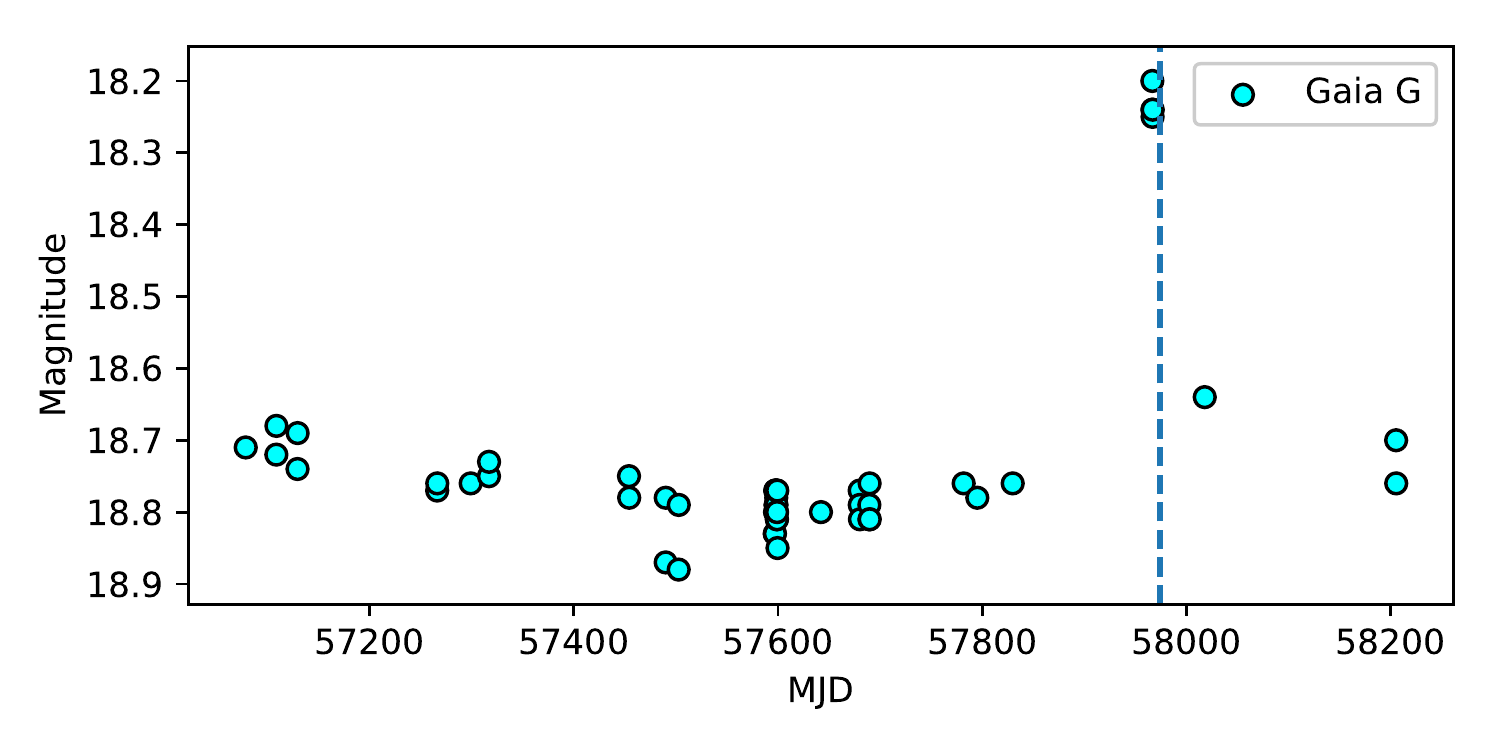}
\caption{Lightcurve of Gaia17byk. The dashed line indicates the date when we observed the source.}
\label{fig:Gaia17byk_lc}
\end{figure}

\subsection{Gaia17bzc}
Gaia17bzc (also known as AT2017fxl) was detected on 2017 August 3 \citep{Delgado2017vii}. The alert was in response to an increase in brightness ($\sim0.5$\,mag) of a red source residing in the Galactic Centre. We observed this source six days post-alert and measured a polarisation of $P = 6.86(\pm0.64)\%$ in \textit{Z} band. Unfortunately we were unable to obtain a magnitude measurement for our observation but Figure \ref{fig:Gaia17bzc_lc} highlights our time of observation.

\begin{figure}
\includegraphics[width=8.5cm]{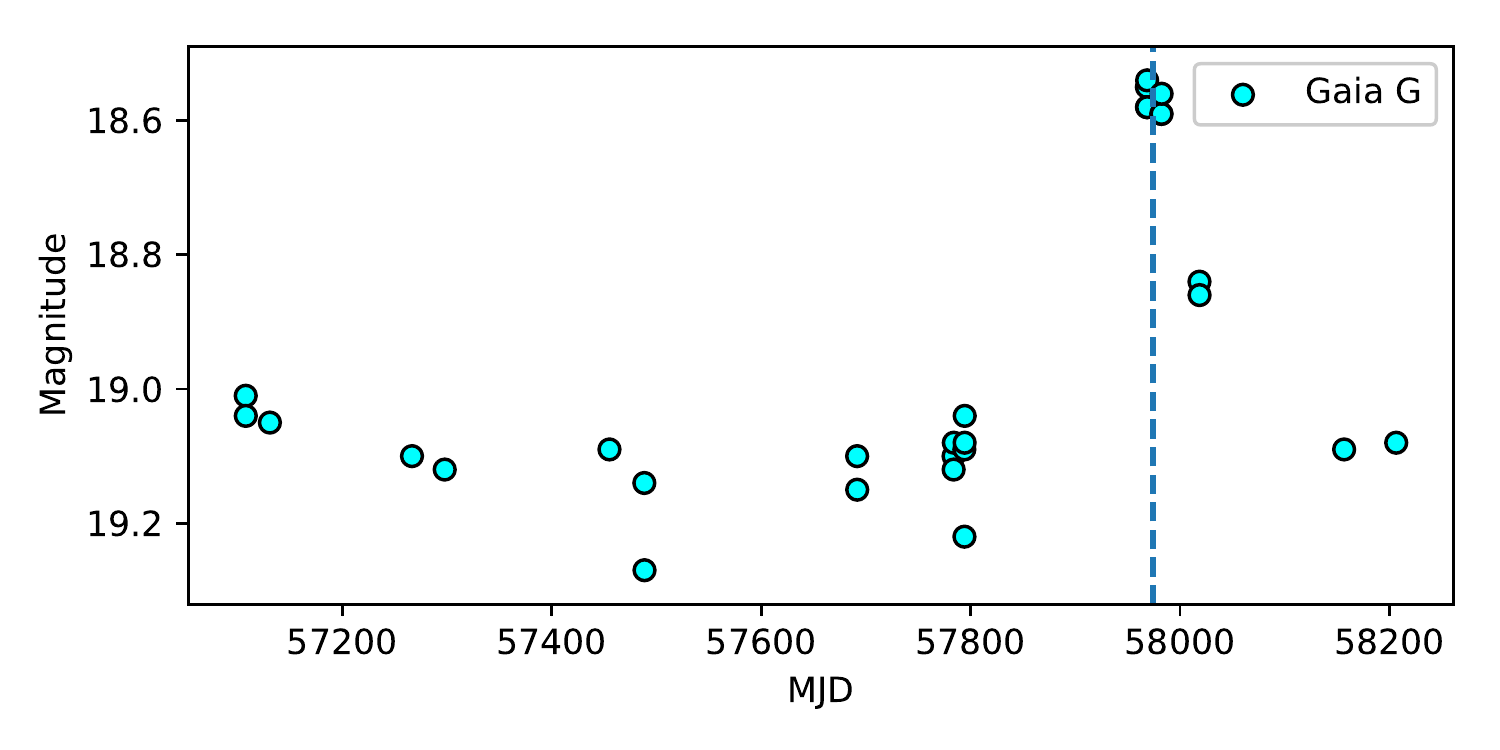}
\caption{Lightcurve of Gaia17bzc. The dashed line indicates the date when we observed the source.}
\label{fig:Gaia17bzc_lc}
\end{figure}

\subsection{GX 304-1}
GX 304-1 (also known as 4U 1258-61) is a well known High Mass X-ray Binary (HMXB). Initially observed as an X-ray source with period of 272 s \citep{Huckle1977,McClintock1977} a coincident optical counterpart was later discovered \citep{Mason1978}, the secondary is a Be star. GX 304-1 periodically enters phases on increased X-ray emission (see \citealt{Manousakis2008} and \citealt{Jenke2012} for examples). We responded to a recent alert of increased activity on 2016 May 17 \citep{Nakajima2016} and observed the source approximately five weeks post-alert. We measured a polarisation of $P = 6.80(\pm0.16)\%$, $P = 6.17(\pm0.45)\%$ and $P = 6.80(\pm0.08)\%$ in \textit{V}, \textit{B} and \textit{R} bands respectively. Previous observations suggest that the polarisation measurements of this source are dominated by Galactic dust scattering \citep{Mason1978}. 

\subsection{MASTER OT J023819.81-521134.1}
MASTER OT J023819.81-521134.1 was detected in the galaxy PGC 009998 on 2017 August 7 \citep{Balanutsa2017}. Although the source was initially suspected to be a SN, the trigger was reported to be likely due to high AGN activity within the galaxy \citep{Stanek2017a}. We observed the source approximately eight hours post-alert and measured a polarisation of $P = 0.66(\pm0.20)\%$ in \textit{V} band. We calculate $E(B-V) = 0.03$ at the source position corresponding to $P_{\rm Gal, dust} \leq 0.27\%$. 

\subsection{MASTER OT J220727-053121.8}
MASTER OT J220727 (also known as AT2016ecw, Gaia16arv and SN2016ecw) was discovered on 2016 June 16 \citep{Shurpakov2016}. The source was classified as a type Ia SN via follow-up spectroscopic observations \citep{Blagorodnova2016}. We observed the source three days post-alert and measured a polarisation of $P = 1.06(\pm0.34)\%$ in \textit{V} band.  We calculate $E(B-V) = 0.06$ at the source position corresponding to $P_{\rm Gal, dust} \leq 0.54\%$. This result is consistent with $P \sim 0.3\%$ expected for broadband measurements of type Ia SNe.

\subsection{OGLE16aaa}
OGLE16aaa is an optical transient that was detected on 2016 January 2 \citep{Wyrzykowski2016}. Three months of follow-up observations were conducted by \citet{Wyrzykowski2017}. The source exhibited a shallow temporal increase in brightness over several weeks followed by an equally shallow decline and spectra of the source determined the presence of very broad He II and H$\alpha$ lines leading to the classification of a TDE. We observed the source approximately five months post detection and measured a polarisation of $P = 1.81(\pm0.42)\%$ in \textit{V} band. This level of polarisation is significantly lower than previously observed relativistic TDEs (i.e. \citealt{Wiersema2012a}). We calculate $E(B-V) \sim 0.02$ at the source position corresponding to $P_{\rm Gal, dust} \lesssim 0.18\%$. We compared the brightness of the host nucleus of our observation to that of the host, pre-TDE with a fixed aperture of one arcsec. We found that the magnitude of both images within the aperture were consistent, suggesting that no TDE contamination of the host light was present at the time we observed the source. 

\subsection{P13 NGC 7793}
P13 NGC 7793 was first discovered as an ULX source in NGC 7793 \citep{Fabbiano1992} and later postulated to be an X-ray binary containing a $\sim15\msun$ black hole and spectral type B9Ia donor star with a period of $\sim64$\,days \citep{Motch2014}. Recent observations have shown that the source exhibits X-ray pulsations at a period of $\sim0.42$\,s \citep{Furst2016,Israel2017}, showing that the compact object is a neutron star. We responded to an alert of the source rebrightening on 2016 May 20 \citep{Soria2016} and observed the source approximately one month post-alert measuring a polarisation of $P < 6.54\%$ in \textit{V} band. To our knowledge, this is the first optical polarimetry of a ULX. The measured optical intensity is dominated by the companion star, but some emission mechanisms may lead to observable optical polarisation in sources like this, such as strongly beamed optical jet emission or irradiation of the companion star.

\subsection{PG 1553+113}
PG 1553+113 (also known as HESS J1555+111) is a well known BL Lac object first discovered in the Palomar-Green survey \citep{Green1986}. The source periodically enters outburst and produces very high energy emission (see \citealt{Aharonian2006,Abdo2010,Aleksic2015}). We observed the source in response to a report of recent activity on 2016 April 27 \citep{Kapanadze2016}. We observed the source approximately seven weeks post-alert and measured a polarisation of $P = 5.15(\pm0.09)\%$, $P = 5.26(\pm0.13)\%$ and $P = 4.78(\pm0.07)\%$ with angles of $\theta = 31.50(\pm0.49)$\,deg, $\theta = 31.55(\pm0.73)$\,deg and $\theta = 30.62(\pm0.43)$\,deg in the \textit{V}, \textit{B} and \textit{R} bands respectively. Our measurements are significantly higher than the $\sim2\%$ polarisation reported by \citep{Andruchow2011} in the $B$ and $R$ bands. This suggests that the source exhibits intrinsic polarimetric variability, possibly caused by a change in internal structure. They also report a polarisation angle of $\sim 127-128$\,deg compared to our values of $\sim 31-32$\,deg, a shift of $\sim 95-100$\,deg. Further polarimetric follow-up of this source is recommended. We also calculate $E(B-V) = 0.04$ at the source position corresponding to $P_{\rm Gal, dust} \leq 0.36\%$.

\begin{figure}
\includegraphics[width=8.5cm]{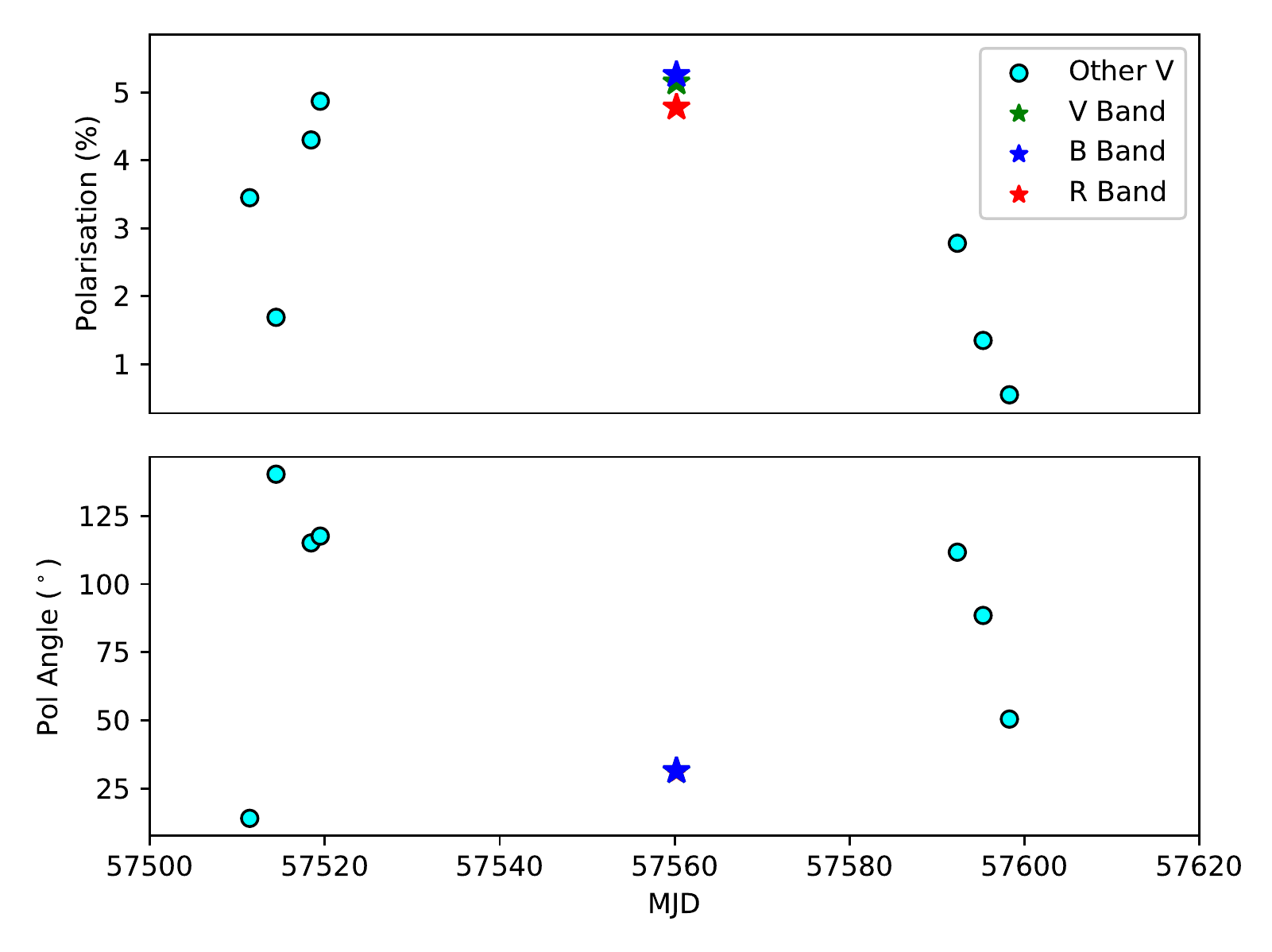}
\caption{Plot of PG 1553+113 showing the temporal evolution of the degree of polarisation and polarisation angle. The 'Other V' legend label refers to data taken from the Steward Observatory spectropolarimetric monitoring project.}
\label{fig:pg1553_plot}
\end{figure}

\subsection{PKS 1510-089}
PKS 1510-089 is a well known Blazar that periodically enters outburst producing very high energy emission (for examples see \citealt{Wu2005,DAmmando2011}). Previous optical polarimetry campaigns have shown that the polarisation of the source varies significantly over time-scales of $\sim$ few days \citep{Marscher2010}. We observed the source three times during our observing runs in response to reports of increased activity both in the NIR and at high energies \citep{DeNaurois2016,Mirzoyan2016,Carrasco2016i}. We observed the source on 2016 June 19, 20 and 22 and measured decreasing polarisation levels of $P = 8.76(\pm0.16)\%$, $P = 3.14(\pm0.16)\%$ and $P = 1.94(\pm0.35)\%$ and polarisation angles of $\theta=155.77(\pm0.54)$ deg, $\theta=138.55(\pm1.49)$ deg and $\theta=142.03(\pm5.08)$ deg in \textit{V} band. We calculate $E(B-V) = 0.09$ at the source position corresponding to $P_{\rm Gal, dust} \leq 0.81\%$. Our results confirm that the level of polarisation of PKS 1510-089 varies significantly on short time-scales, with only a small change of polarisation angle measured.

\begin{figure}
\includegraphics[width=8.5cm]{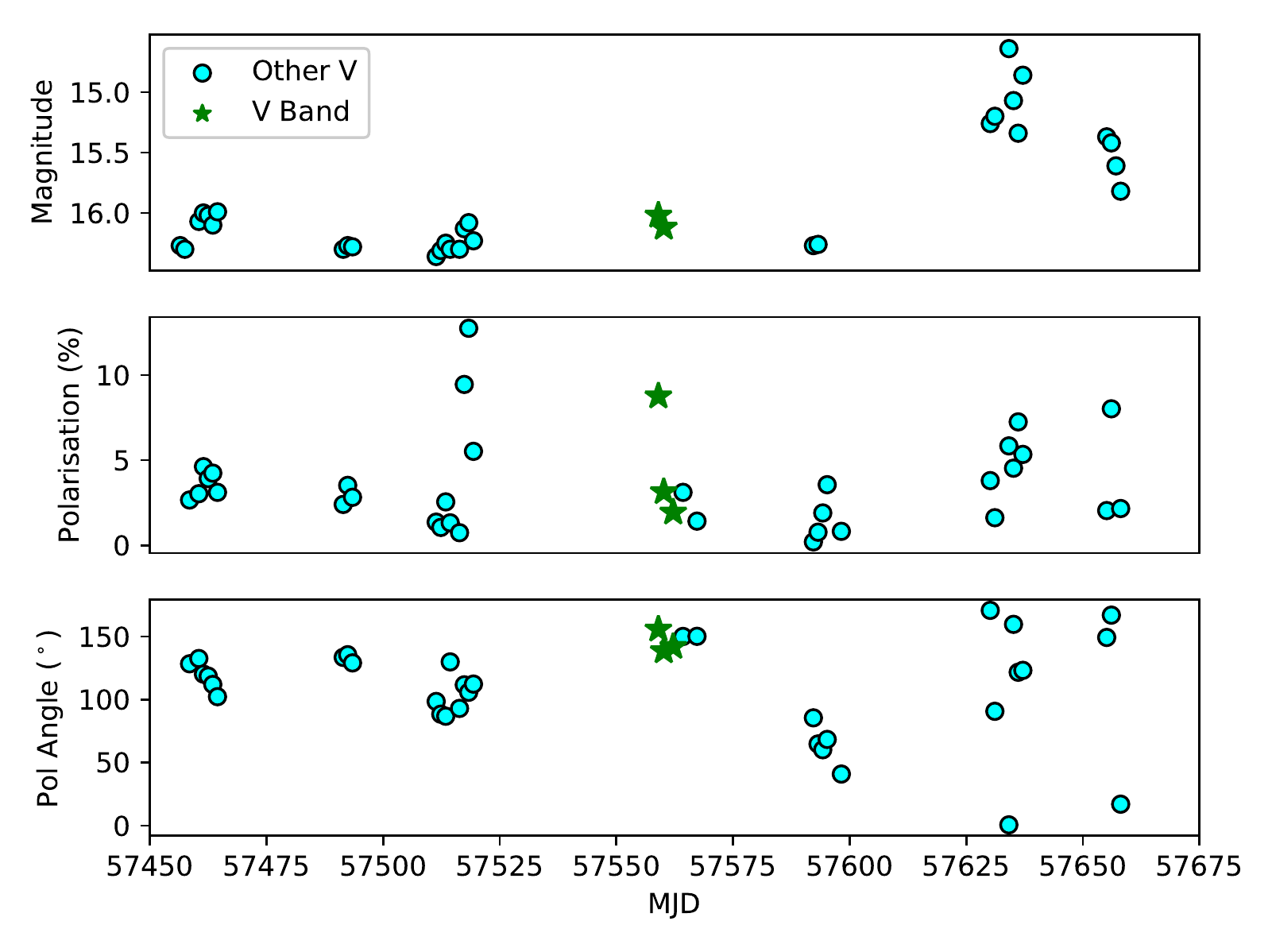}
\caption{Plot of PKS 1510-089 showing the temporal evolution of source brightness, degree of polarisation and polarisation angle. The 'Other V' legend label refers to data taken from the Steward Observatory spectropolarimetric monitoring project.}
\label{fig:pks1510_plot}
\end{figure}

\subsection{PKS 2023-07}
PKS 2023-07 is a well known Blazar that periodically enters outburst with some very high energy emission (i.e. \citealt{Vercellone2008,Gasparrini2009}). We responded to a recent increase in gamma-ray emission detected on 2016 April 13 \citep{Verrecchia2016} and observed the source approximately nine weeks post-outburst measuring a polarisation of $P = 7.36(\pm0.35)\%$ in \textit{V} band. We calculate $E(B-V) = 0.03$ at the source position corresponding to $P_{\rm Gal, dust} \leq 0.27\%$.

\subsection{PS16cnz}
PS16cnz (also known as AT2016cnm) was discovered on 2016 May 23 \citep{Chambers2016ii}. We observed the source approximately four weeks post-alert and measured a polarisation of $P < 0.60\%$ in \textit{V} band.

\subsection{PS16crs}
PS16crs (also known as AT2016cor and SN2016cor) was discovered in SDSS J154431.47+161814.9 on 2016 May 25 and was classified as a type Ia SN via follow-up spectroscopic observations \citep{Zhang2016}. We observed the source approximately three weeks post-alert and measured a polarisation of $P \leq 3.72\%$ in \textit{V} band.

\subsection{PS16ctq}
PS16ctq (also known as AT2016cut) was discovered on 2016 June 11 \citep{Chambers2016iii}. We observed the source nine days post alert and measured a polarisation of $P \leq0.50\%$ in \textit{V} band.

\subsection{PS16cvc}
PS16cvc (also known as AT2016cxb, MASTER OT J211223+144645.1 and SN2016cxb) was discovered on 2016 June 19 \citep{Chambers2016ii}. The source was classified as a type Ia SN via follow-up spectroscopic follow-up \citep{Tomasella2016}. We observed the source twice, approximately 24 and 96 hours post-alert and measured polarisations of $P < 0.71\%$ and $P < 0.74\%$ in \textit{V} band on both nights, respectively. This low level of continuum polarisation is consistent with previous results in the literature.  

\subsection{SXP 15.3}
SXP 15.3 (also known as RX J$0052.1-7319$ and MASTER OT J211223+144645) is a well known Pulsar/X-ray binary with a period of 15.3\,s that resides in the Small Magellanic Cloud \citep{Lamb1999,Covino2001}. The secondary star is a Be star, dominating the received optical emission. The source is known to periodically go into outburst and exhibits long term optical variability (i.e. see \citealt{Galache2008} and \citealt{Rajoelimanana2011} and references there-in). We responded to a recent alert of the source re-brightening on 2017 July 25 \citep{Kennea2017} and observed the source approximately 12 days post-alert, measuring a polarisation of $P < 1.45\%$ in \textit{Z} band.

\subsection{XTE J1709-267}
XTE J1709-267 (also known as RX J1709.5-2639) is a neutron star low mass X-ray binary, first detected by \textit{ROSAT} \citep{Voges1999}. The source has been observed extensively in a number of wavelengths (i.e. \citealt{Jonker2004}). We responded to a recent outburst on 2016 May 31 \citep{Nakahira2016} and observed the source approximately three weeks post-alert measuring a polarisation of $P < 2.00\%$ in \textit{V} band. We published the photometry results via an ATel \citep{Wiersema2016}. Low-mass X-ray binaries frequently show linearly polarised infrared emission, generally attributed to a transient or steady jet, which in rare cases can also be detected at optical wavelengths (for a review see \citealt{Russell2018}), particularly during outbursts. We set a strict limit on any optical jet emission for this source.

\bsp
\label{lastpage}
\end{document}